\documentclass{aa}  

\usepackage{graphicx}
\usepackage{txfonts}
\usepackage{subcaption}
\usepackage{enumitem}
\usepackage{xcolor}
\usepackage{pdfpages}

\usepackage{natbib,twoopt}
\usepackage[breaklinks=true]{hyperref} 
\bibpunct{(}{)}{;}{a}{}{,}             
\makeatletter
  \newcommandtwoopt{\citeads}[3][][]{\href{http://adsabs.harvard.edu/abs/#3}%
    {\def\hyper@linkstart##1##2{}%
     \let\hyper@linkend\@empty\citealp[#1][#2]{#3}}}
  \newcommandtwoopt{\citepads}[3][][]{\href{http://adsabs.harvard.edu/abs/#3}%
    {\def\hyper@linkstart##1##2{}%
     \let\hyper@linkend\@empty\citep[#1][#2]{#3}}}
  \newcommandtwoopt{\citetads}[3][][]{\href{http://adsabs.harvard.edu/abs/#3}%
    {\def\hyper@linkstart##1##2{}%
     \let\hyper@linkend\@empty\citet[#1][#2]{#3}}}
  \newcommandtwoopt{\citeyearads}[3][][]%
    {\href{http://adsabs.harvard.edu/abs/#3}
    {\def\hyper@linkstart##1##2{}%
     \let\hyper@linkend\@empty\citeyear[#1][#2]{#3}}}
\makeatother

\begin{document} 

   \title{A model grid for the reflected light from transition disks}

   \author{J. Ma
          \inst{1}
          \and
          H.M. Schmid \inst{1}\fnmsep
          }

   \institute{Institute for Particle Physics and Astrophysics, 
          ETH Zurich, Wolfgang Pauli Strasse 17, CH-8093 Zurich\\
              \email{jma@phys.ethz.ch}
             }

   \date{Received --; accepted --}

  \abstract
{The dust in protoplanetary disks is an important ingredient in planet formation and can be investigated with model simulations and quantitative imaging polarimetry of the scattered stellar light. }   
{This study explores circumstellar disks with calculations for the intensity and polarization of the reflected light. We aim to describe the observable radiation dependencies on parameters in order to constrain the dust scattering properties and the disk geometry.}
{The photon scattering and absorption by the disk are calculated with a Monte Carlo method for a grid of simple, rotationally symmetric models approximated at each point by a plane-parallel dusty atmosphere. The adopted geometry is described by a strongly illuminated inner wall of a transition disk with inclination $i$, a constant wall slope $\chi$, and an angular wall height $\alpha$. Dust scattering parameters are the single scattering albedo $\omega$, the Henyey-Greenstein scattering phase function with the asymmetry parameter $g$, and the maximal fractional polarization $p_{\rm max}$ induced by the scattering. First, the results for the reflectivity, the polarized reflectivity, and the fractional polarization of a plane-parallel surface element are calculated as functions of the incidence angle and the escape direction of the photons and as functions of the scattering parameters. Integration over all escape directions yields the surface albedo and the fraction of radiation absorbed by the dust. Second, disk images of the reflected intensity and polarization are calculated, and the appearance of the disk is described for various parameter combinations. The images provide many quantitative radiation parameters for a large range of model calculations, which can be compared to observations. These include the disk integrated intensity $\overline{I}/I_\star$, azimuthal polarization $\overline{Q}_\varphi/I_\star$, the polarization aligned with the apparent disk axes $\overline{Q}/I_\star$, the quadrant polarization parameters $Q_{xxx}$ and $U_{xxx}$, the disk-averaged fractional polarization $\langle p_\varphi \rangle$ or $\langle p_Q \rangle$, but also the front-to-back intensity ratio $I_{180}/I_{000}$ or the maximum fractional scattering polarization ${\rm max}(p_\varphi)$.} 
{The results of our simple disk models reproduce well the measurements for $I/I_\star$, $Q_\varphi/I_\star$, and $\langle p_\varphi \rangle$ reported for well-observed transition disks. They describe the dependencies of the scattered radiation on the disk geometry and the dust scattering parameters  in detail. Particularly strong constraints on disk properties can be obtained from certain diagnostic quantities: 
for example the fractional polarization $\langle p_\varphi \rangle$ or ${\rm max}(p_\varphi)$ depend predominantly on the dust-scattering parameters $\omega$ and $g$; for disks with
well-defined inclination, ratios of the quadrant polarization parameter depend  mainly on the scattering asymmetry $g$ and the wall slope $\chi$; wavelength dependencies of $I/I_\star$ and $Q_\varphi/I_\star$ can mostly be attributed to the wavelength dependence of the dust scattering parameters $\omega(\lambda)$, $g(\lambda)$, and $p_{\rm max}(\lambda)$; and the ratio between the scattered and thermal light of the disk roughly constrains the disk reflectivity ${\cal{R}}$ and the single scattering albedo of the dust $\omega$.}
{This computational investigation of the scattered radiation from transition disks shows well-defined dependencies on model parameters and the results can therefore be used as a diagnostic tool for the analysis of quantitative measurements, specifically in constraining or even determining the scattering properties of the dust particles in disks. Collecting and comparing such information for many systems is required to understand the nature of the scattering dust in planet-forming disks. }


   \maketitle
%

\section{Introduction}

Dust is an important ingredient of the planet formation in protoplanetary disks. Dust particles can be observed over a large wavelength range as strong absorbers, scatterers, or emitters of radiation and the data show a large diversity of geometric morphologies of these dust tracers for different disks, and even within a given disk. For example, in the millimeter range  one can see  the thermal emission of the large dust particles settled in the cold disk midplane; in the infrared (IR) range we see the thermal emission of the dust illuminated and heated by the central star; and in the visual and near-IR spectral range  we can observe the scattered light from the small dust particles located at the disk surface. 
To understand the evolution of dust particles in protoplanetary disks and their role in the planet-formation process we need to investigate these different dust components and combine the complementary information from all these observations \citep[e.g.,][]{Andrews15}.

This work simulates with model calculations the photometric and polarimetric properties of the stellar light scattered by the inner wall of transitions disks. The simulations should be useful for comparisons with high-resolution and high-contrast observations and provide better constraints on the dust particle properties and geometric parameters of disks. The calculated disk models adopt a dust scattering phase matrix which depends on the particle albedo, scattering asymmetry parameter, and the induced polarization by the scattering particles. 
These parameters are indicative of the particle size distribution and the presence of specific dust types such as high-albedo icy grains, low-albedo carbon-rich particles, or porous aggregates \citep[e.g.,][]{Draine84,Shen09,Kolokolova2010,Min2016,Tazaki19,Tazaki21}.
Such a characterization will improve our understanding of the evolution of the dust in protoplanetary disks and the resulting composition of the forming planets. 

Transition disks are a special class of protoplanetary disks characterized by a large, strongly depleted central cavity \citep[e.g.,][]{Calvet05,Espaillat14}. 
They are relatively easy targets for observation of the scattered light with high-resolution observations because they produce a strong signal at the inner wall of an extended disk $\gtrapprox 10$~AU which can be separated for systems in nearby star-forming regions with a distance of $\approx 100-150$~pc. For example, the intensity of the scattered light has been measured for several disks by the Hubble Space Telescope \citep[e.g.,][]{Grady05,Ardila07,Mulders13,Debes13}.
Particularly well suited for the disk detection are adaptive optics (AO) instruments at large ground-based telescopes equipped with imaging polarimeters, which allow the scattered and therefore polarized radiation of the disk  to be  separated from the very strong but typically unpolarized radiation of the central star \citep[e.g.,][]{Schmid22}. 
For this reason, there exist many observations of the polarized flux from disks taken with differential polarimetric imaging (DPI) \citep[e.g.,][]{Quanz11,Hashimoto12,Rapson15,Stolker16,Monnier17,Garufi17,Avenhaus18}. 
If the instrumental polarization and image convolution effects can be calibrated, then these data provide quantitative measurements of the polarized intensity of the disk. In favorable cases, both the intensity and the polarized intensity can be derived \citep[e.g.,][]{Perrin09,Monnier19,Hunziker21,Tschudi21}. Such measurements are particularly suitable for comparisons with model calculations for the investigation of dust scattering properties. \par
Model calculations for geometrically thin disks that treat multiple scattering and polarization by dust were first presented by \citet{Whitney92} long before such disks could be imaged. More popular were calculations for extended dusty envelopes as observed with seeing-limited observations \citep[e.g.,][]{Bastien88,Whitney93,Fischer94}.
Disk modelling obtained a strong boost from progress in high-resolution imaging, which revealed that many protoplanetary disks are geometrically thin and show structures such as cavities, gaps, and spirals. Many simulated disk images for the scattered light were presented to prove that observed disk morphologies are in good qualitative agreement with the proposed three-dimensional disk geometry \citep[e.g.,][]{Pinte08,Murakawa10,Dong12,deJuanOvelar13,JangCondell13,Pohl15,Dong16,Monnier17,JangCondell17}.
In these studies, the dust scattering properties were usually not the focus and the fractional polarization or the wavelength dependence of the scattered light was not investigated. 
There exist only a few disk studies where the dependence of the scattering polarization and intensity on dust parameters are investigated. Systematic radiative transfer calculations are presented by \citet{Tazaki19}, who, for a fixed disk geometry,  model the dependencies between the disk-integrated color of the scattered intensity and polarized intensity, and the disk averaged fractional polarization in order to constrain ---with observations--- the size and porosity of dust particles.
This work has the same goal but we aim to provide a more general overview of the dependencies of observational parameters on the dust scattering parameters and on the disk geometry.

In this paper, we present  a grid of simple transition disk models that describe the disk geometry with three parameters, the slope $\chi$, the angular height $\alpha$ of the inner disk wall, and the disk inclination $i$. The dust scattering properties are defined by the three parameters, single scattering albedo $\omega$, scattering asymmetry $g$, and induced fractional polarization $p_{\rm max}$ for a scattering angle of 90$^{\circ}$. \par
Such simple models are ideal for exploring the dependencies of the polarization and intensity of the scattered radiation for a large range of dust-scattering parameters, for particles with very low or very high albedos, with isotropic or strongly forward-scattering phase functions, and particles producing low or high fractional polarization. 
The results of this model grid should include roughly the range of observed intensity and polarization for transition disks in order to constrain the disk parameters. This investigation paves the way to identifying  observational parameters that have significant diagnostic potential for the analysis of the geometry of transition disks and the scattering particles. The calculations will also clarify how the models should be improved in order to better determine the properties of the observed disks.

The following section describes our model calculations and the parameters used for our model grid. Section 3 gives an overview of the reflectivity, the polarized reflectivity, and the dust absorption of a plane-parallel surface for the adopted dust-scattering parameters. The dependencies of the scattered radiation from transition disks on the model parameters are described in Sect.~4, and in 
Sect.~5 we present an investigation of the diagnostic potential of the presented results for the characterization of the dust and the disk geometry with observations. In Sect.~6, we summarize our results and assess the limitations of our interpretation of observational results stemming from the simplicity of our model simulations  compared to real disks which are typically
much more complex.

\section{Model description}\label{section: model description}

\begin{figure}
    \centering
    \includegraphics[width=0.48\textwidth]{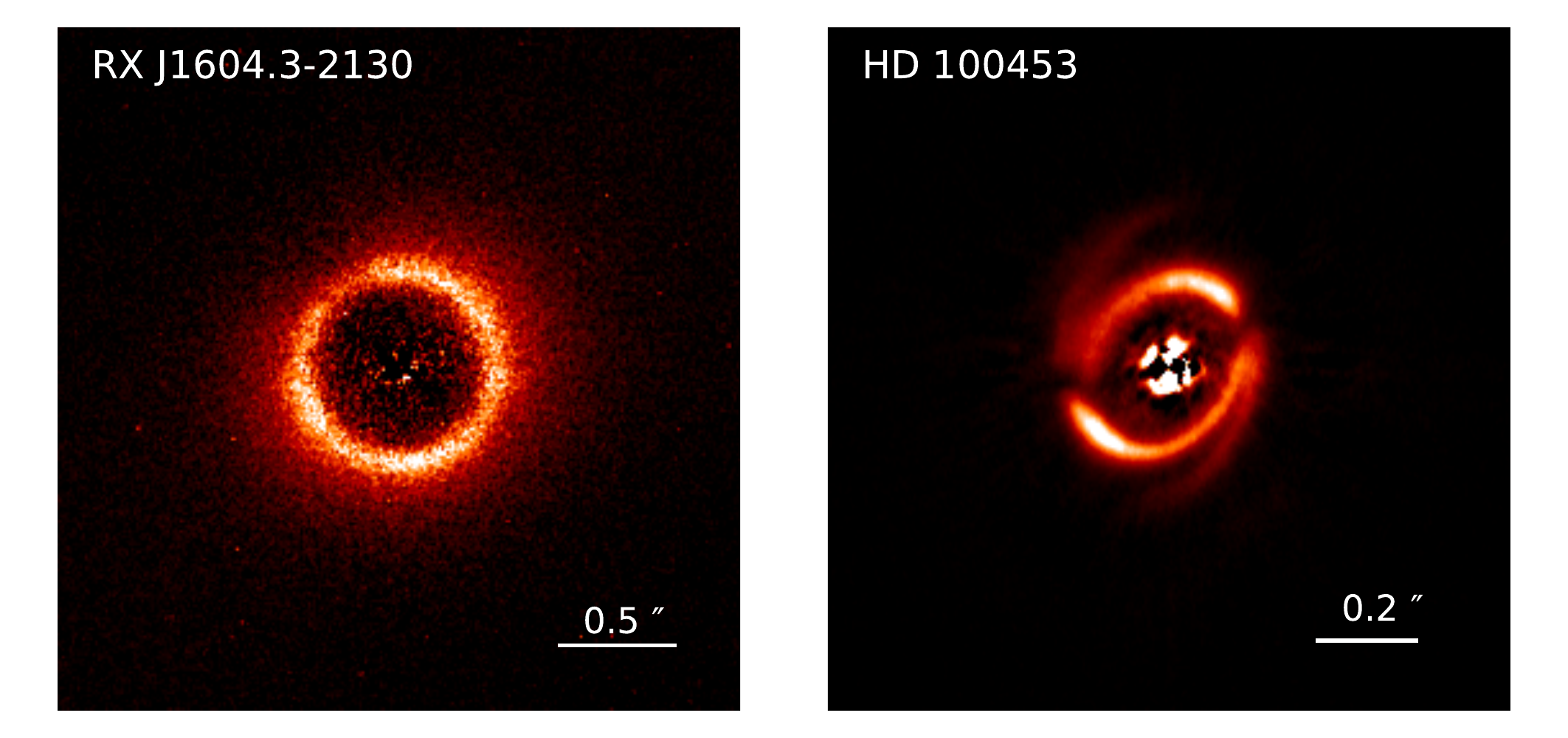}
    \caption{Examples of the $Q_\varphi$ image for the bright walls in the transition disks around RX J1604.3-2130 and HD 100453 (R-band observations with SPHERE/ZIMPOL taken from the ESO data archive).}
    \label{fig:disk-example}
\end{figure}
In many transition disks, narrow ring-like emission from the inner wall surrounding the large central cavity dominates the scattered light, as shown in Fig.~\ref{fig:disk-example} for a pole-on disk and an inclined disk with spiral features and shadows from a hot, unresolved dust disk located close to the star. 
Detailed descriptions can be found for RXJ1604.3-2130 \citep[e.g.,][]{Mayama12}, HD 100453 \citep[e.g.,][]{Benisty17}, HD 142527 \citep[e.g.,][]{Canovas13}, PDS 70 \citep[e.g.,][]{Keppler18}, HD 169142 \citep[e.g.,][]{Bertrang18}, HD 34700A \citep{Monnier19} and others.  
The bright signal in polarized intensity from these inner walls can be measured with relatively high precision. Often, it is also easy to derive a brightness ratio between the front and back sides of the disks, and our models help to understand such ratio measurements. 
For some strongly illuminated disks, it is even possible to extract the intensity of the scattered light and derive the fractional polarization of the scattered light \citep[e.g.,][]{Monnier19,Hunziker21,Tschudi21}. Based on these observational data, it seems reasonable to simulate the frequently observed ring-like disk morphology of transition disks with a simple geometry for the bright inner disk wall in order to capture the main parameter dependencies for the reflected intensity and polarization  with our model simulations.

\subsection{Disk geometry}\label{section: disk geometry}
\begin{figure}
    \centering
    \vspace{-0.3cm}
    \begin{subfigure}[!ht]{\hsize}
        \centering
        \includegraphics[width = \textwidth]{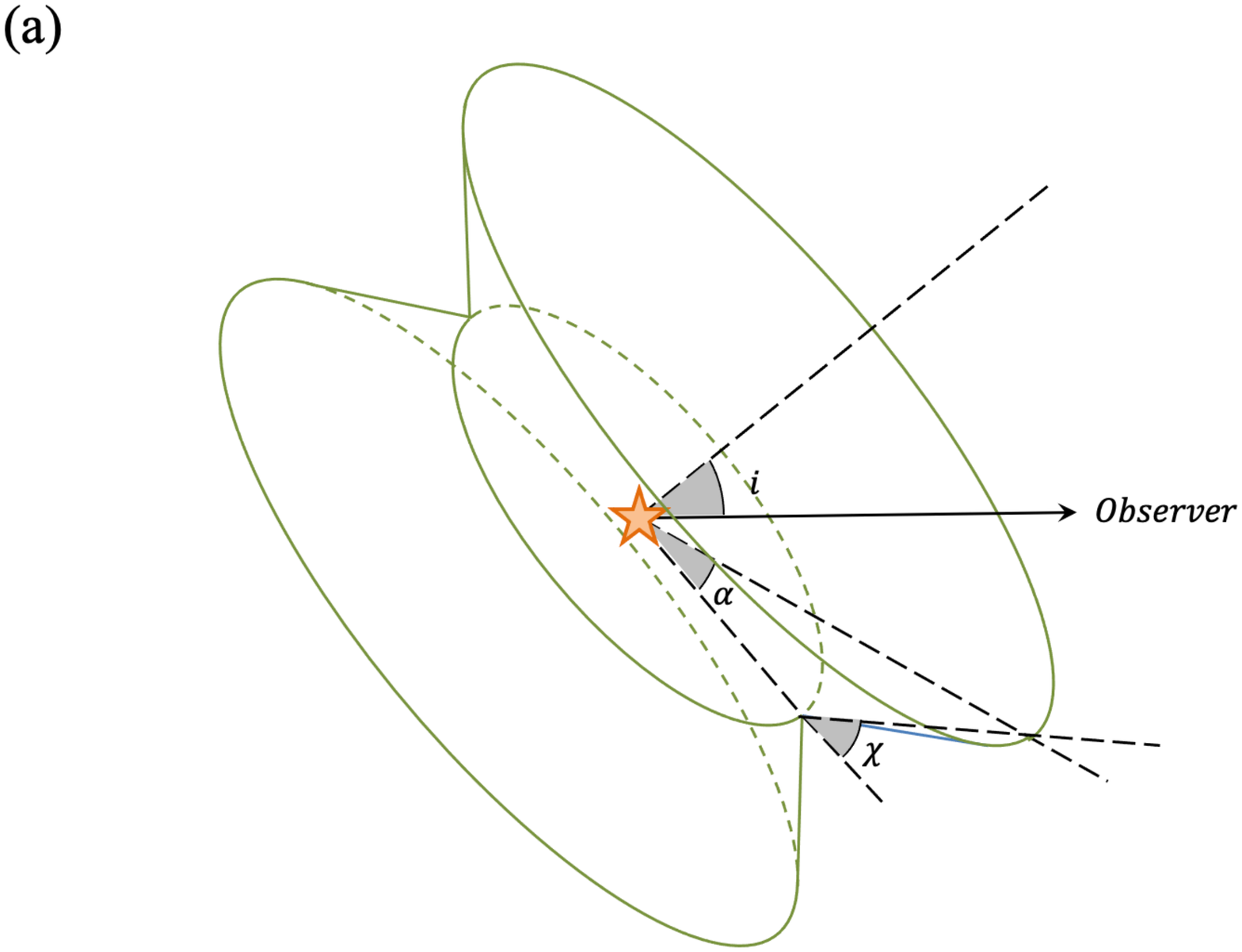}
        \vspace{-1.5cm}
        \label{fig:geometry}
    \end{subfigure}
    \begin{subfigure}[!ht]{\hsize}
        \centering
        \includegraphics[width = \textwidth]{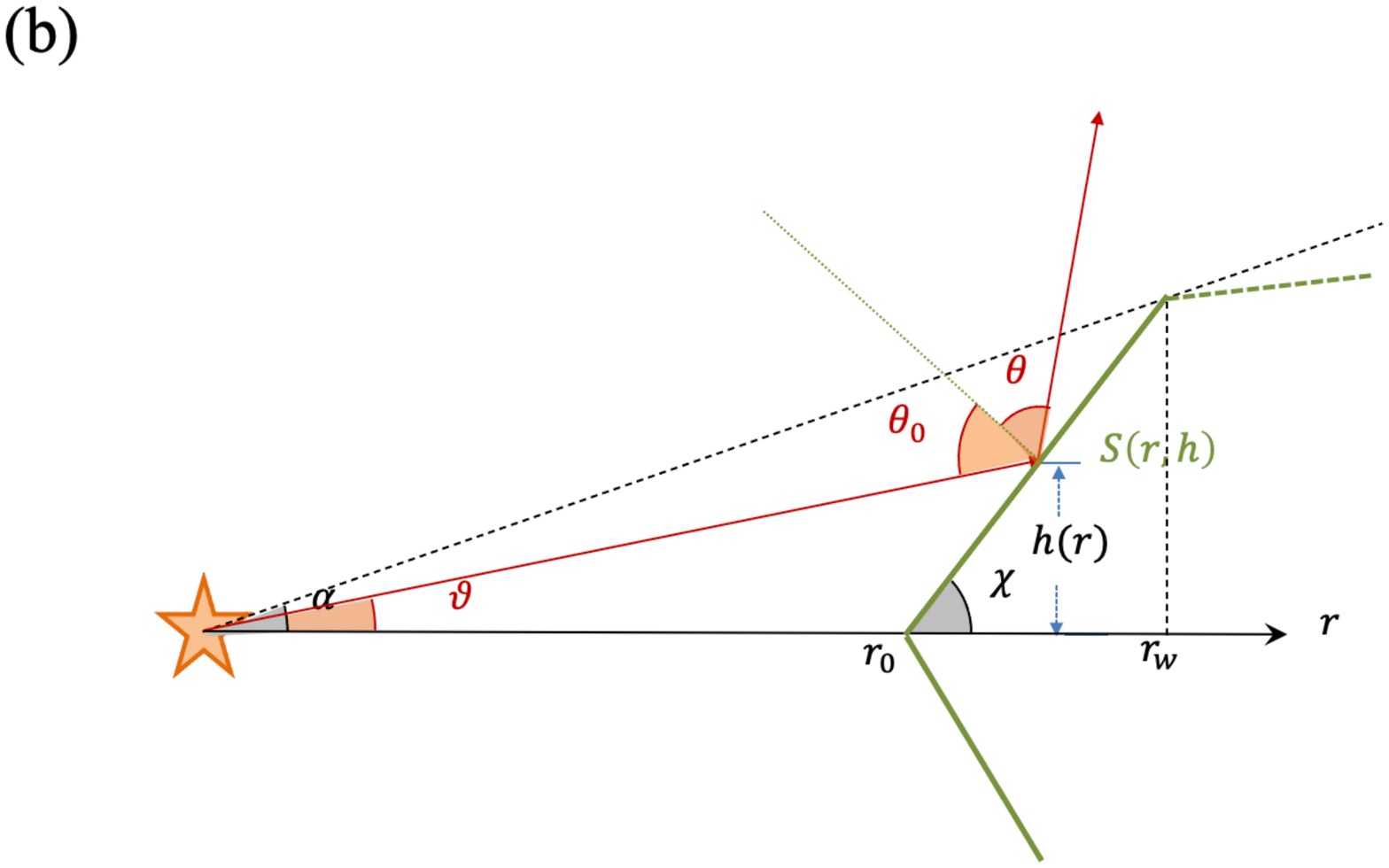}
        \label{fig:section}
    \end{subfigure}
    \vspace{-1.6cm}
    \caption{Geometry of the transition disk model. Panel (a) Sketch of the rotationally symmetric, inclined disk model with inclination $i$, wall slope $\chi,$ and angular height of the disk wall $\alpha$ . 
    Panel (b) Corresponding vertical disk cross-section with the midplane disk distance to the star $r$ and the height $h$ for the surface point $S(r,h)$, and radii $r_0$ and $r_w$ for the inner radius and the rim radius of the wall, respectively. Emitted photons from the star are characterized by the angular direction $\vartheta$, while $\theta_0$ and $\theta$ are the polar incidence and emergence angles from surface point $S$ for the special case of scattering in the principal plane.} 
    \label{fig:diskgeometry}
\end{figure}  
    
\begin{figure}
    \centering 
    \vspace{-0.5cm}
    \includegraphics[width=0.5\textwidth]{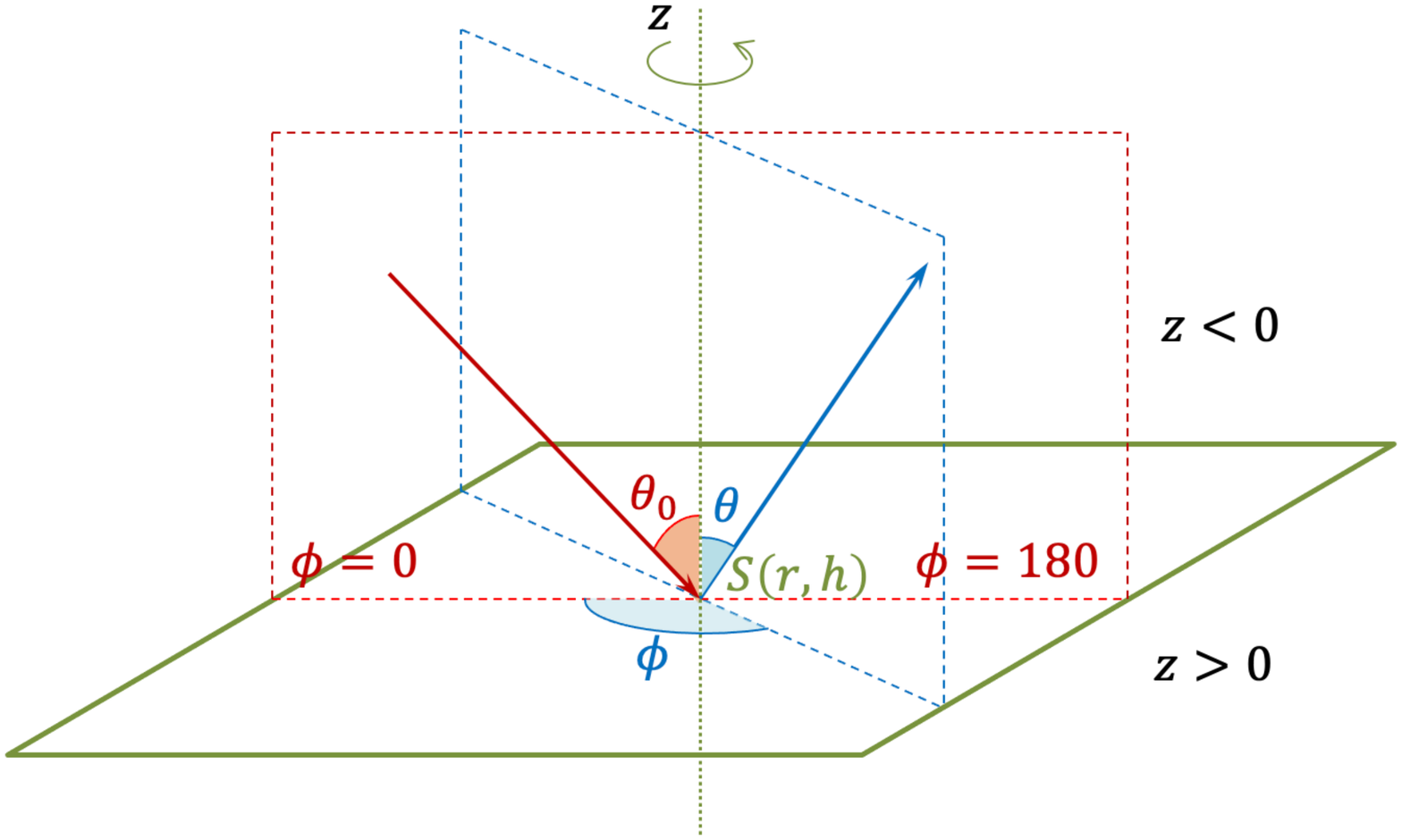}
    \vspace{-0.9cm}
    \caption{Plane parallel surface element $S(r,h)$ in polar coordinates. Photons always enter in the principal plane $\phi_0=0^\circ$, as indicated by the red arrow in the red dotted plane, and are assumed to escape at the same point $S$ in an upward direction with angle $(\theta,\phi)$, as shown by the blue arrow in the blue dotted plane.}
    \label{fig:coordinate}
\end{figure}

Our transition disk model is defined according to Fig.~\ref{fig:diskgeometry}(a) by an inclined, rotationally symmetric geometry with an inner wall with a vertical cross-section as shown in Fig.~\ref{fig:diskgeometry}(b). The height of the disk surface $h(r)$ from the disk midplane as a function of the radius $r$ is defined by an inner radius $r_0$ and a wall with a constant slope $\chi$ or 
\begin{equation}
    h(r)= \pm (r-r_0) \, {\rm tan}\,\chi \quad {\rm for}\quad r_0\leq r\leq r_w\,,
\label{Eq:hofr}
\end{equation}
where $r_w$ is the midplane radius of the wall rim with an angular height $\alpha$ with respect to the disk midplane. 
There must be $\alpha<\chi$ for an inner disk wall with a finite radial extent. The disk surface for $r>r_w$ has a slope smaller than $\alpha$ and is therefore not illuminated by the star.
The radius of the wall rim is defined by $r_w=r_0(1-\tan\alpha/\tan\chi)^{-1}$ and the corresponding wall height is $h_w=h(r_w)= \pm r_w\,{\rm tan}\,\alpha$. The positive and negative signs stand for the wall section above and below the disk midplane, referred to as the upper and lower part, respectively. 
 
The illuminated wall is described by the two parameters $\chi$ and $\alpha$, while $r_0$ is only a scaling factor for the disk size. For a given angular height $\alpha$,  the same relative amount of stellar radiation interacts with the disk independent of $\chi$. 
For our model grid, we choose a broad range for the wall slopes $\chi$ from 12.5$^\circ$ to 82.5$^\circ$. 
For the wall height $\alpha$, we consider values of between $5^\circ$ and $20^\circ$ but use mostly just $\alpha = 10^\circ$ ($h_w/r_w=0.176$) because many results for the reflected intensity and polarized intensity simply scale as $\propto\sin\alpha \propto \alpha$ for $\alpha<20^\circ$. This comfortably covers the typical range of transition disks \citep{Takami14, Woitke19, Avenhaus18}.

The disk can be seen by an observer under different inclinations from pole-on $i=0^\circ$ to edge-on $i=90^\circ$. The illuminated back wall or inner wall of the far side in our disk images  is always up from the star and the inner wall on the near side or front wall is below the star as if we are looking onto the disk. This also defines the apparent major and minor disk axes for inclined disks, which are horizontal and vertical, respectively. 

Depending on the inclination $i$ and the wall slope $\chi,$ the upper front side wall and the upper and lower back side walls are visible, partially visible, or not visible. The upper back side wall is fully visible near the minor axis for $i<90^\circ-\alpha$, and can even be visible or partly visible for slightly larger $i$. 

The front side wall is only visible if $i <90^\circ-\chi$ and it disappears for $i=90^\circ-\chi$, while the lower backside wall starts to become (partially) visible for $i >90^\circ-\chi$. The lower backside disappears behind the disk for nearly edge-on configurations $i\gtrapprox 90^\circ-\alpha$. The exact borderline depends on the unconstrained outer $r>r_w$ disk geometry.

\subsection{Illumination and surface reflectivity} 
Each disk element is treated like a plane-parallel, optically thick surface where photons are scattered until they escape or are absorbed. It is assumed that the photons enter and escape at the same point $S(r,h)$; see Fig.~\ref{fig:coordinate}.
The plane-parallel surface is described by a polar coordinate system with the surface normal as polar axis, where the incoming photon propagates in a perpendicular plane, often also called the principal scattering plane, with the polar incidence angle $\theta_0$ and the azimuthal incidence angle $\phi_0=0^\circ$ \citep[e.g.,][]{Chandrasekhar60,vandeHulst80}.
The incidence angle $\theta_0(r)$ at point $S$ for a disk with given $r_0$ and $\chi$ is
\begin{equation}
    \theta_0=90^{\circ} -\chi +\arctan \frac{(r-r_0)\tan\chi}{r}\,.
\end{equation}
The incident angle is $\theta_0 = 90^\circ - \chi$ at the midplane $r=r_0$ and gradually increases along the wall slope until it reaches $\theta_0 = 90^\circ -\chi + \alpha$ at the rim $r=r_w$.

The incident flux $F_0$ at point $S(r,h)$ per surface area is given by the stellar spectral luminosity $L_\lambda$, the distance to the star and the incidence angle
\begin{equation}
    F_0(S(r,h)) = \frac{L_\lambda}{4\pi (r^2+h^2(r))}\, \cos\theta_0\,.
\label{eq:influx}
\end{equation}

The distribution of the reflected radiation is described by the polar $\theta$ and azimuthal $\phi$ angles 
\begin{itemize}
    \item $I(\theta_0,\phi_0=0^\circ,\theta,\phi)$ for the intensity, and 
    \item $Q(\theta_0,\phi_0=0^\circ,\theta,\phi)$ and $U(\theta_0,\phi_0=0,\theta,\phi)$ for the polarized intensity.
\end{itemize}
In our models, we  only consider the Stokes parameters $Q$ and $U$ for linear polarization and disregard scattering processes producing circular polarization. $Q$ and $U$ for the escaping photons are defined in a plane perpendicular to the photon propagation direction given by a blue arrow in Fig.~\ref{fig:coordinate}. Thereby, the positive $Q$-direction is also perpendicular to the vertical blue plane with orientation $\phi$ and $\phi+180^\circ$, while the negative $Q$-direction is in this plane. The positive and negative $U$-directions are rotated with respect to $Q$ by $45^\circ$. \citep[see][]{Schmid92}. 
For the principal plane $\phi=0^\circ$ or $180^\circ$, the integrated linear polarization for many multiple scattered photons escaping from point $S(r,h)$ is, for symmetry reasons, either perpendicular or parallel or $\pm Q$ to the principal plane, while $U=0$.

\subsection{Dust scattering}
The reflectivity at a given point on the disk surface is treated as photon random walk with a Monte Carlo code. The parameters describing the interaction with the dust are the single scattering albedo $\omega$, a scattering asymmetry parameter $g$, and the polarization $p_{\rm max}$ introduced by scattering with an angle of 90$^\circ$.
The scattering albedo is defined as
\begin{equation}
    \omega = \dfrac{\sigma_{\rm sca}}{\sigma_{\rm sca}+\kappa_{\rm abs}}
    \label{EqOmega}
,\end{equation}
where $\sigma_{\rm sca}$ and $\kappa_{\rm abs}$ stand for scattering and absorption cross-sections, respectively. The scattering albedo $\omega$ is defined for  $[0, 1]$ and $\omega = 1$ represents a purely scattering particle while $\omega = 0$ is a purely absorbing particle without scattering.

The angular distribution of the scattered photons is described by the Henyey-Greenstein (HG) phase function \citep{Henyey41}:
\begin{equation}
F_{\rm HG}(\theta_s,g) =\frac{1}{4\pi}\dfrac{1-g^2}{(1+g^2-2g\cos\theta_s)^{\frac{3}{2}}}\,,
\end{equation}
where $\theta_s$ is the scattering angle with respect to the incident direction.
The asymmetry parameter $g \in ]-1,1[$ describes the predominant scattering direction $g = \langle\cos\theta\rangle$, where $g=0$ stands for isotropic scattering, $g>0$ means more forward scattering, and $g<0$ more backward scattering.

The fractional polarization introduced by the dust scattering $p_{\rm sca}(\theta_S)$ is approximated by an angle dependence which is identical to Rayleigh scattering but with a scaling factor $p_{\rm max} \leq 1:$
\begin{equation}
    p_{\rm sca}(\theta_s,p_{\rm max}) = p_{\rm max}\dfrac{cos^2\theta_s-1}{\cos^2\theta_s+1}
\label{Eqpol}
,\end{equation}
which accounts for depolarization effects of dust scattering when compared to the dipole-type Rayleigh scattering. Hereafter, we refer to the polarized scattering phase function 
\begin{equation}
    F_{\rm HGpol}(\theta_s,g,p_{\rm max}) = F_{\rm HG}(\theta_s,g)\cdot p_{\rm sca}(\theta_s,p_{\rm max}) 
\label{EqHGpol}
\end{equation} 
as the HG$_{\rm pol}$ function. In this description of a single scattering event, the angle distribution of the polarized intensity for unpolarized incoming light  depends only on $g$, and the maximum fractional polarization $p_{\rm max}$ introduced by scatterings with $\theta_s=90^\circ$ \citep[e.g.,][]{Schmid21}. 
The polarization of the reflected light from a multiple scattering surface changes significantly with $p_{\rm max}$, but the scattered intensity depends only slightly on $p_{\rm max}$ as described for example by \citet{Chandrasekhar60} for Rayleigh scattering.
The adopted description for the scattering phase function is simple and versatile, however it disregards more complex features of dust scattering, such as a back-scattering maximum for $\theta_s$ near $180^\circ$ as expected for larger particles or deviations from the polarization function given in Eq.~\ref{Eqpol} as observed for the dust in the debris disk HR~4796A \citep{Milli19,Arriaga20}.

\subsection{Monte Carlo simulations}
A widely used tool for calculations of the scattered intensity and the scattering polarization are Monte Carlo simulations which follow the random walk of photons \citep[e.g.,][]{Witt77}. In astronomy, many codes are available which use similar procedures to those described here for the dust scattering in circumstellar disks \citep[e.g.,][]{Pinte06,Min09,Whitney13}.

\paragraph{Scattering in a plane-parallel surface.} 
In a first step, we investigate the reflected intensity $I(\theta_0,0,\theta,\phi)$ and the Stokes parameters $Q(\theta_0,0,\theta,\phi)$, $U(\theta_0,0,\theta,\phi)$ from a homogeneous, semi-infinite, plane parallel dust  atmosphere for different incidence angles $\theta_0$ and dust parameters ($\omega,g,p_{\rm max}$). 
In a rotationally symmetric disk, the stellar illumination of a surface element always occurs in the principal plane and is described by the polar angle $\theta_0$ and the azimuthal angle $\phi_0=0^\circ$.
The incident photons have a random polarization angle as expected for unpolarized incident radiation.
The vertical optical depths for the photon interactions $\tau_z$ are calculated from random distributions of the free path lengths $\delta\tau$ and the photon propagation direction. 
The vertical optical depth $\tau_z$ is zero above the surface and is positive for deeper layers. After one or several scattering events, the photon may escape in a direction $(\theta,\phi)$ with a polarization angle $\gamma$. 

Our Monte Carlo simulations are based on the code described in \citet{Schmid92}, which was compared thoroughly with analytic calculations for isotropic and Rayleigh scattering \citep{Chandrasekhar60,Abhyankar70,Abhyankar71}. To this code, we added the Henyey-Greenstein-type dust scattering phase function with polarization. 
The procedure for calculating the scattering angles $(\theta_s,\phi_s)$ and the polarization angle change $\gamma_s$ of the photon includes the following steps. 
First the polar scattering angle $\theta_s$ is drawn from the probability distribution function ${\cal{P}}$ for the Henyey-Greenstein function using a rejection method: 
\begin{equation}
    {\cal{P}}(\theta_s) = \dfrac{1}{4\pi}\int_0^{\theta_s}\dfrac{1-g^2}{(1+g^2-2g\cos\theta)^{\frac{3}{2}}}\sin\theta d\theta
\end{equation}
Then we draw an evenly distributed random number $p_0$ in the interval $[0,1]$ and compare it to $p_{max}$. If $p_0>p_{max}$, then we adopt non-polarizing scatterings and therefore equally distributed azimuthal scattering angles $\phi_s$ and polarization angle $\gamma_s$. Otherwise, $\phi_s$ and $\gamma_s$ are evaluated with a probability distribution as for Rayleigh scattering, as described by \citet{Schmid92}.
The escaping photons are sampled in direction bins defined by the central angles $\theta_i=2.5^\circ+i\cdot 5^\circ$ and $\phi_j=j\cdot 10^\circ$. One bin covers the solid angle $\Omega_{ij}=|\sin\theta_i|\cdot (\pi/18) \cdot (\pi/36)$. 
The reflectivity $I(\theta,\phi)$ in a given bin is the ratio between the collected photons $N_{ij}$ and the incident photons $N_{\rm tot}$ scaled by the bin size $I=N_{ij}/(N_{\rm tot}\Omega_{ij})$. The relative statistical errors are $(N_{ij})^{-1/2}$ and they are large for small $\theta_i$ or small solid angle bins which collect fewer photons. 

\paragraph{Scattering in a circumstellar disk}
In our Monte Carlo simulations of the circumstellar disks, the photons are released with random polarization orientations from an isotropically emitting central star with a polar angle distribution $D(\vartheta)\propto \cos{\vartheta}$. For our calculations, $D(\vartheta)$  is restricted to the range $\vartheta \in [0,\alpha]$, where interaction with the inner disk wall will occur. The photon then penetrates the disk  at point $S(r,h(r))$   where dust scattering or absorption can occur as described above for the plane-parallel surface. The incidence angle $\theta_0(r,h) = 90^{\circ}-\chi+\vartheta$, radial distance from the star $r=r_0\sin^2\chi(1+\cot\chi \cot(\chi-\vartheta)),$ and $h(r)$ (Eq.~\ref{Eq:hofr}) follow from the adopted disk parameters $r_0$, $\chi$, and $\alpha$. 

The direction and polarization angle of the photon escaping from $S(r,h)$ are defined by $(\theta,\phi,\gamma)$ for the plane-parallel surface. Because the adopted disk is rotationally symmetric, these parameters define the azimuthal angle $\varphi$ of the surface region and the apparent inclination of the disk for which this escaping photon can be observed. In other words, each escaping photon characterized by $(r,\theta,\phi,\gamma)$ in a given disk geometry $(r_0,\chi,\alpha)$ can be converted into a photon escaping from the cylindrical coordinates $r,h,\varphi$ with a polar inclination angle $i$, or from a corresponding Cartesian point $x(i),y(i)$ in the sky plane image corresponding to a disk seen under inclination $i$. Also, the polarization direction of the photon must be rotated from $\gamma$ in the plane parallel system to $\gamma'$ in the system of the observer. The trigonometric formulas for these transformations are straightforward but remain complex. 

For the resulting intensity $I(x,y,i)$ of the scattered radiation, the photons are added together for each $(x(i),y(i))$ bin using inclination bins with a width of $5^\circ$ defined by the central angles $i_k = 2.5^\circ + k\cdot 5^\circ$.
For the Stokes $Q$ and $U$ parameters, the direction of the polarization for each photon $\gamma'_j$ must also be taken into account according to
\begin{equation}
    Q =\sum_j \cos (2 \gamma'_j) \quad {\rm and} \quad U =\sum_j \sin (2 \gamma'_j)\,.   
\end{equation}
After the calculation of the random walk of sufficiently large numbers of photons, the distribution of the scattered intensity $I(x,y,i)$ and the polarized intensities $Q_(x,y,i)$ and $Q_(x,y,i)$ for the reflected light from the disk can be established for each inclination bin. 

Various tests for the Monte Carlo code were carried out, including simulations and comparisons with analytic results for disks with a perfect Lambert surface, analyzing disks with isotropic scattering, or the polarization obtained for single scattered photons for $p_{\rm max} =1$, and many more. 

\section{Calculations for a plane parallel surface}
\label{Sect:plpar}
The reflectivity $I(\theta,\phi)$ and the Stokes parameters $Q(\theta,\phi)$ and $U(\theta,\phi)$ of the reflected radiation for a given point $S(r,h)$ on the disk wall depend on the illumination angle $\theta_0$ and the dust scattering parameters $\omega$, $g,$ and $p_{\rm max}$. This section gives a brief overview of the scattered light from a plane parallel atmospheric surface to support the discussion and interpretation of the results of the scattered radiation from an entire disk. Previous results for Henyey-Greenstein scattering in a semi-infinite atmosphere are described in \citet{Dlugach74} and are tabulated in \citet{vandeHulst80}, but without considering the scattering polarization.
Appendix~\ref{AppSurface} provides an overview of the parameter dependencies for the reflected intensity and polarization for an atmospheric surface with dust scattering as described by Eqs.~(\ref{EqOmega}) to (\ref{EqHGpol}).

\begin{figure}[]
    \resizebox{0.9\hsize}{!}{\includegraphics{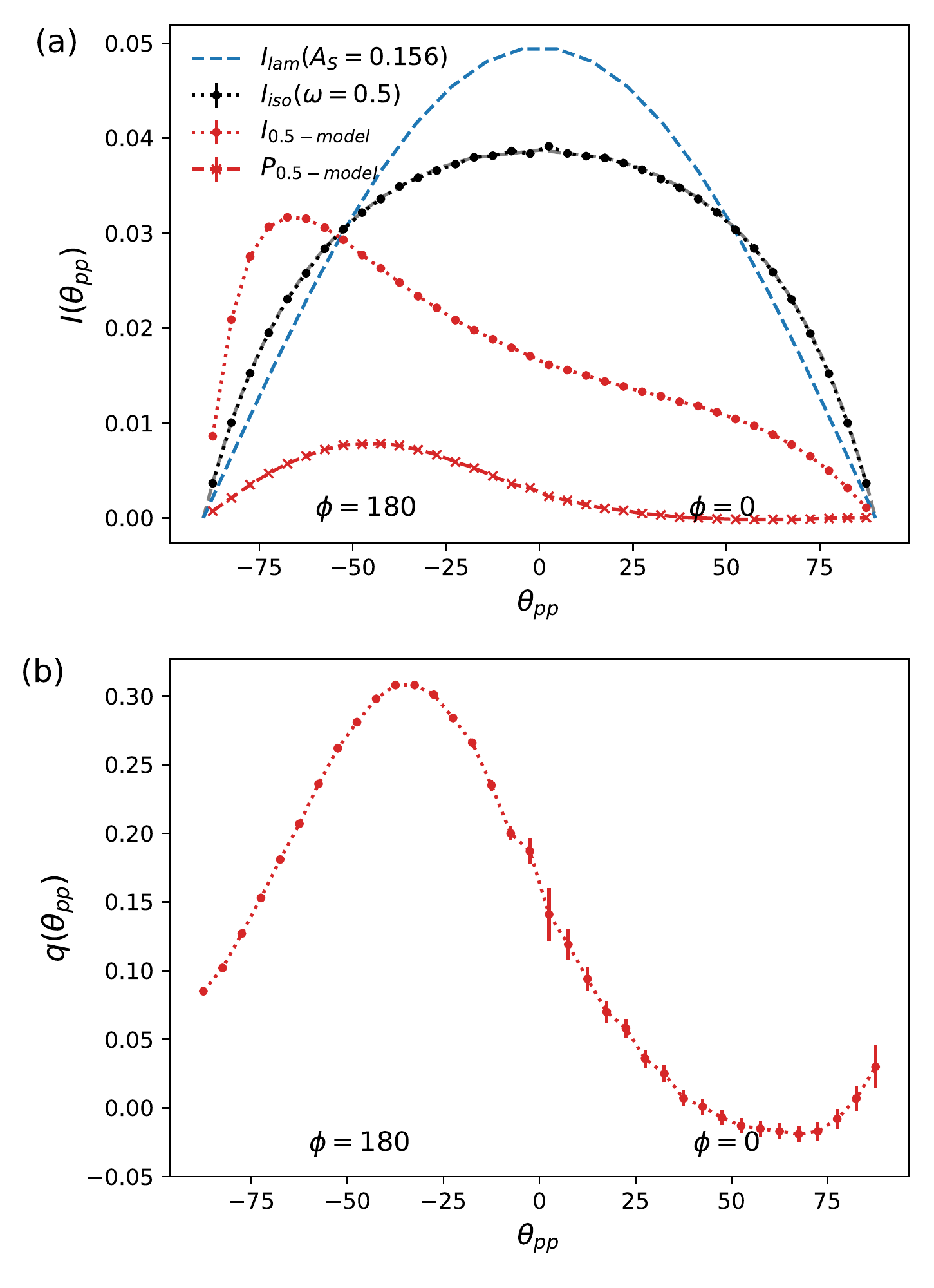}}
    \vspace{-0.4cm}
    \caption{Reflection in the principal plane of the 0.5-model. Panel (a): Reflected intensity $I(\theta_{\rm pp})$ and polarized intensity $P(\theta_{\rm pp})$ for the 0.5-model ($\theta_0=57.5^{\circ}$, $\omega=0.5$, $g=0.5$, $p_{\rm max}=0.5$) are shown in red. The analytic solution and simulations of $I_{\rm iso}(\theta)$ for the reflectivity of a surface with isotropically scattering particles $g=0$ with $\theta_0=57.5^{\circ}$, $\omega=0.5$, $p_{\rm max}=0.5$ is shown in black. 
    The analytic result $I_{\rm Lam}(\theta)\propto \cos\theta$ for a Lambertian surface with a surface albedo $A_S=0.156$ is in blue. 
    Panel (b): Fractional Stokes parameter $q(\theta_{\rm pp})$ for the 0.5-model.}
    \label{fig:05model}
\end{figure}

\subsection{The reference model ("0.5-model").} \label{section: reference model}

To simplify our discussion we introduce a reference model with fixed incidence angle $\theta_0=57.5^{\circ}$ and dust scattering parameters $\omega=0.5$, $g=0.5$, and $p_{\rm max}=0.5$, which we also refer to as the 0.5-model. The selected dust scattering values are quite typical for interstellar dust grains for wavelengths of around $1~\mu$m \citep{Draine03}. Appendix~\ref{AppSurfacePara} presents results for other model parameters and compares them to this reference case.

Further, we focus our description on the reflection in the principal plane which is the scattering plane perpendicular to the surface. For this plane, we can plot one-dimensional functions for $I(\theta_{\rm pp})$ (see Fig.~\ref{fig:05model}), but also $Q(\theta_{\rm pp})$ or the fractional polarization $p(\theta_{\rm pp})$ and other parameters. The angle $\theta_{\rm pp}$ for the escaping radiation in the principal plane is defined for the range $-90^\circ$ to $+90^\circ$, where $\theta_{\rm pp}=+\theta$ for $\phi=0^\circ$ and $\theta_{\rm pp}=-\theta$ for $\phi=180^\circ$, which are for $\theta_0\gg 0^\circ$ the backward reflection side and the forward reflection side, respectively.

The scattering in the principal plane describes for an axisymmetric circumstellar disk the reflected intensity and polarization for the front and back side along the minor axis. 
Of course, the $\phi$-dependence of the reflectivity is important for other regions of the disk and this is discussed in Appendix~\ref{AppSurfacePhi} for the reference model. 

\paragraph{The reflected intensity in the principal plane.}
The reflected intensity of the 0.5-model in the principal plane $I(\theta_{\rm pp})$ is compared in Fig.~\ref{fig:05model} with the analytic solution $I_{\rm iso}(\theta)$ for isotropic $(g=0)$ scattering with $\omega=0.5$ and $\theta_0=57.5^\circ$ as calculated by \citet{Chandrasekhar60} and with our Monte Carlo code, and $I_{\rm Lam}(\theta)$ for a gray, Lambertian surface with surface albedo $A_S$. The intensity distributions for $I_{\rm iso}$ and $I_{\rm Lam}$ are rotationally symmetric with respect to the surface normal and produce no scattering polarization. The surface albedo $A_S$ of the Lambert model was set to 0.156, because this yields the same surface albedo (equivalent to the term reflectivity ${\cal{R}}$) as the $\omega=0.5$ isotropic scattering model.

The forward scattering dust of the 0.5-model plotted in Fig.~\ref{fig:05model} produces a clear asymmetry with respect to $\theta_{\rm pp}=0^\circ$ and for most viewing angles much less reflected flux than isotropic scattering, because photons are scattered predominantly further "down" into deep layers where photon absorptions are frequent \citep[][]{Mulders13}.
A Lambertian surface behaves like $I_{\rm Lam}(\theta)\propto \cos\theta$ and this is equivalent to a surface brightness (SB), which is independent from the observers viewing angle $\theta$. The isotropic scattering case $I_{\rm iso}(\theta_{\rm pp})$ in Fig.~\ref{fig:05model} produces a higher surface brightness for large angles $\theta>60^\circ$ when compared to $\theta<60^\circ$. For the forward scattering dust of the 0.5-model, there exists a strong difference between the "bright" forward scattering direction $\theta_{\rm pp}<-45^\circ$ and the "faint" backward scattering direction $\theta_{\rm pp}>45^\circ$. 
For example, the intensity ratio between $I(-67.5^\circ)$ and $I(+67.5^\circ)$ is about a factor of 4.

\paragraph{The polarization in the principal plane.}
In the principal scattering plane, the fractional polarization of the reflected radiation can be described by the fractional Stokes parameter $q=Q/I;$ because of reasons of  symmetry, the reflected radiation can only be polarized perpendicular or parallel to this plane. Therefore, the fractional polarization is $p(\theta_{\rm pp})=|q(\theta_{\rm pp})|$ and $u(\theta_{\rm pp})=0$. The polarization $q(\theta_{\rm pp})$ shows a strong maximum perpendicular to the scattering plane around $\theta_{\rm pp}=-30^\circ$ (Fig.~\ref{fig:05model}(b)). This angle for the maximum polarization $\theta_{\rm pp}^{\rm maxp}$ is about $90^\circ$ away from the incidence angle $\theta_0=57.5^\circ$ and represents the polarization dependence for the adopted Rayleigh-like phase curve for the fractional polarization (Eq.~\ref{Eqpol}). 
The polarization of the reflected light is dominated by photons escaping after the first scattering event, while the polarization of multiple scattered photons is randomized and therefore they add mainly to the "unpolarized" intensity and lower the fractional polarization. The $q(\theta_{\rm pp}) = q(\theta,0^\circ:180^\circ)$ curve shows a weakly negative section around the back-scattering directions $\theta_{\rm pp}=50^\circ-80^\circ$. This is a well-known higher order scattering effect \citep[e.g.,][]{vandeHulst80}, which for dust (or haze) scattering in Jupiter and Titan leads to strong "negative" limb polarization effects \citep{Schmid11,McLean17,Bazzon14}.

Also included in Fig.~\ref{fig:05model}(a) is the polarized flux $P(\theta_{\rm pp})=p(\theta_{\rm pp})\times I(\theta_{\rm pp})$ of the reflected radiation. The maximum of $P(\theta_{\rm pp})$ lies between the maxima of $p(\theta_{\rm pp})$ and $I(\theta_{\rm pp})$.
High-contrast observations of disks with differential polarimetric imaging provide often only the signal for the polarized flux $P(\theta_{\rm pp})$ which differs very significantly from the reflected intensity of a Lambert surface or an isotropically scattering atmosphere, and even from the reflected intensity of a surface with Henyey-Greenstein scattering particles.

\subsection{Scattering parameter dependencies}
The 0.5-model represents only one combination of model parameters, and Appendix~\ref{AppSurface} provides a more detailed description of  the dependencies of the intensity $I(\theta,\phi)$ and polarizations such as $Q(\theta,\phi)$, $U(\theta,\phi),$ and $p(\theta,\phi)$ on the incidence angle $\theta_0$, the scattering albedo $\omega$, asymmetry $g$, and polarization $p_{\rm max}$. Here, we summarize some key results.

The single scattering albedo $\omega$ has a very important impact on the amount of reflected intensity. A low single scattering albedo significantly reduces  the multiple scattered photons, which contribute more "randomized" or unpolarized reflected light, while the strongly polarizing single scattered photons are less suppressed. \\
The scattering asymmetry parameter $g$ can greatly change the angular distribution of the reflected intensity $I(\theta_{\rm pp})$. The reflected intensity  also depends significantly on $g$ because the incoming stellar photons are scattered for large $g$ predominantly into deeper layers where absorption is very likely. Interestingly, the fractional polarization of the scattered radiation $p(\theta_{\rm pp})$ depends only slightly on the scattering asymmetry parameter. \\
The parameter $p_{\rm max}$ of the polarization phase function has of course a strong effect on fractional polarization $p(\theta,\phi)$ of the reflected radiation
but only a small impact on the reflected intensity.

\begin{figure}[]
    \centering
    \resizebox{0.9\hsize}{!}{\includegraphics{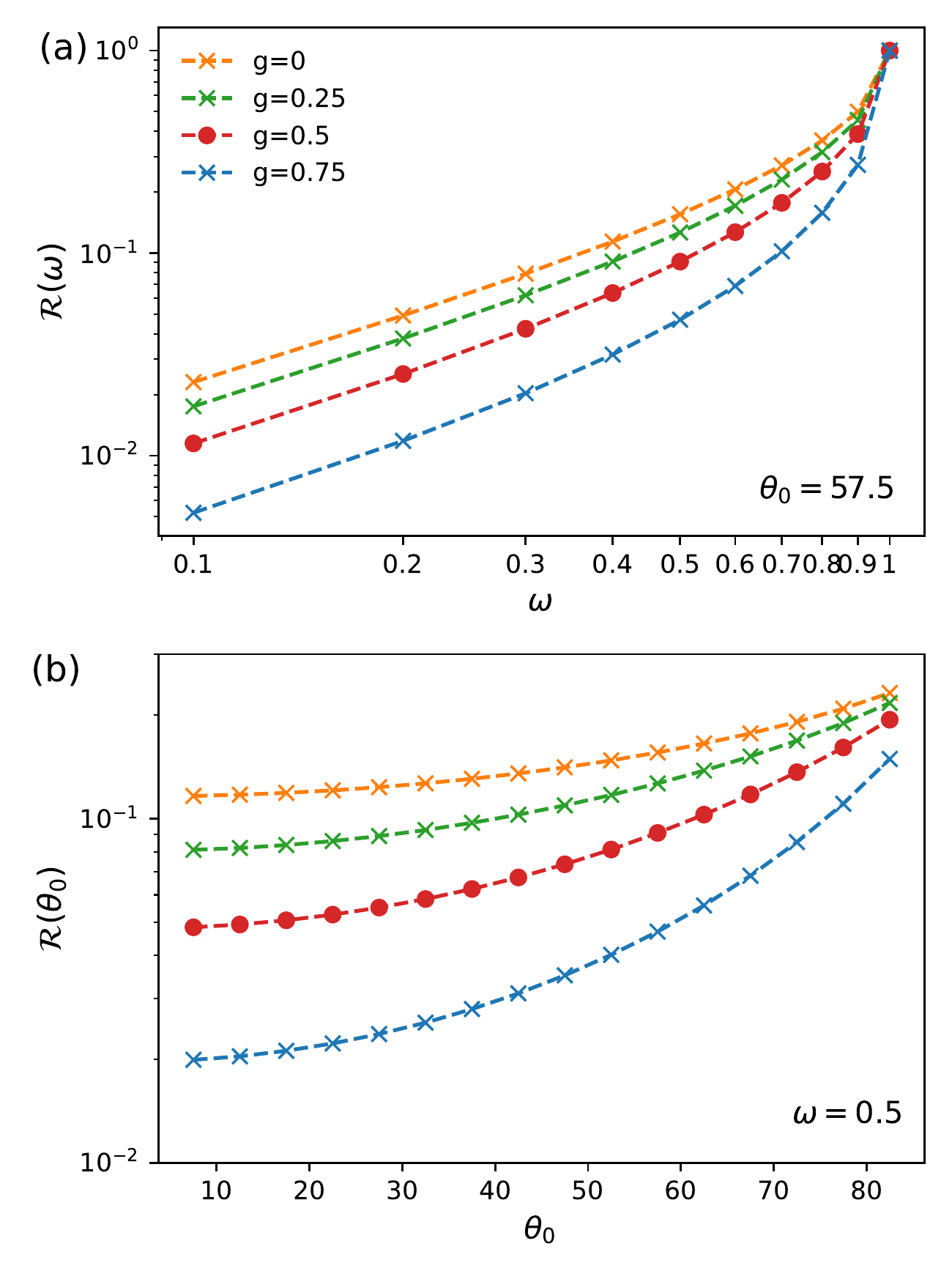}}
    \caption{Integrated reflectivity ${\cal{R}}$ for different asymmetry parameters $g$ as functions of $\omega$ (panel a) and $\theta_0$ (panel b). }
    \label{fig:reflectivity}
\end{figure}

\paragraph{Absorption and surface albedo.}
A very interesting value from the plane parallel surface calculations is the integrated reflectivity ${\cal{R}}(\theta_0,\omega,g)$ or surface albedo according to
\begin{equation}
    {\cal{R}} = \frac{1}{F_0}\int_0^{90^\circ} \int_0^{360^\circ} I(\theta,\phi) \sin\theta\, {\rm d}\phi{\rm d}\theta\,.
\end{equation}
This integrated reflectivity ${\cal{R}}$ does not depend on the viewing directions $\theta$ and $\phi$ and also the small dependence on $p_{\rm max}$ can be neglected. This allows us to characterize the reflectivity of a surface of dust described by the HG scattering phase function in a simple way by ${\cal{R}}(\theta_0,\omega,g)$. Values of ${\cal{R}}$ for the models discussed in this section are also included in Table~\ref{tab:param}.\\
The radiation absorbed by the dust is directly related to the reflectivity ${\cal{R}}$ by the relation
\begin{equation}
    {\cal{K}}(\theta_0,\omega,g) = 1 - {\cal{R}}(\theta_0,\omega,g) \,.
\label{Eq-kappa}
\end{equation}
The total radiation energy absorbed by the dust, integrated for all wavelengths, can be roughly estimated from the thermal IR emission of the dust. Therefore, ${\cal{R}}$ is an important parameter for the energy budget of circumstellar disks. 

The dependence of the integrated reflectivity ${\cal{R}}$ on the single scattering albedo $\omega$, the asymmetry parameter $g,$ and the incidence angle $\theta_0$ is shown in Fig.~\ref{fig:reflectivity}. The reflectivity is high for small $g$ (and would be very high for negative $g$ not considered in this work), for grazing incidence $\theta_0\rightarrow 90^\circ$, and in particular for high scattering albedo $\omega$. For $\omega=1$, there is no absorption ${\cal{K}}=0$ and all radiation is reflected from the surface ${\cal{R}}=1$. \\
The reflectivity for the 0.5-model is at a level of 9~\%, indicating that about $\approx 90~\%$ or most of the radiation falling onto this type of dust surface is absorbed. 

\section{Calculations for transition disk models}
The geometry of the simple three-dimensional, rotational symmetric disk models used in this work are described by three parameters, namely disk inclination $i$, angular wall height $\alpha$, and wall slope $\chi$. The scattering at each point $S(r,h)$ is treated as for a plane-parallel surface. 

The primary results of the model calculation are images for the scattered light intensity $I(x,y)$ and Stokes polarization parameters $Q(x,y)$ and $U(x,y)$.
The central star is at $x_0=0$, $y_0=0$ and the $x$-axis is aligned with the major axis of the inclined disk and the more distant part of the minor axis is in the positive $y$-direction. The orientation of Stokes $Q$ and $U$ are aligned with the $x,y$-coordinates, and $Q$ is positive for polarization in the $y$-direction.

For the polarization of a circumstellar disk, the polarized flux $P=(Q^2+U^2)^{1/2}$ can also be described by
\begin{equation}
    P=(Q_\varphi^2+U_\varphi^2)^{1/2}\,,
\end{equation}
where $Q_\varphi$ and $U_\varphi$ are the azimuthal polarization components defined with respect to the star according to the description of \citet{Schmid06} for the radial Stokes parameters $Q_r$, $U_r$, and using $Q_\varphi = -Q_r$ and $U_\varphi = -U_r$.
The angle of the scattered polarization of a circumstellar disk is to first order in azimuthal direction $Q_\varphi$, and $U_\varphi\approx 0$ is very small, meaning that the polarized flux can be approximated by $P(x,y)\approx Q_\varphi(x,y)$ and the fractional polarization by $p(x,y)\approx p_\varphi(x,y)=Q_\varphi(x,y)/I(x,y)$. The $Q_\varphi(x,y)$ parameter confers the advantage that it is not biased by the statistical noise of the Monte Carlo calculations or the photon noise in observations. 

To quantify the reflected radiation, we use reflected intensity $\overline{I}/I_\star$, the Stokes $Q$ intensity $\overline{Q}/I_\star$, and the azimuthally polarized intensity $\overline{Q}_\phi$, which are disk-integrated values relative to the intensity of the central star. There is always $\overline{U}=0$ in our models because of the adopted disk symmetry. For the disk-averaged fractional polarization, we calculate the ratios $\langle p_\varphi\rangle=\overline{Q}_\varphi/\overline{I}$ and $\langle p_Q\rangle = \overline{Q}/\overline{I}$ which are flux-weighted averages. 

\begin{figure*}[!ht]
    \centering
    \begin{subfigure}[b]{\hsize}
        \centering
        \includegraphics[width=17cm]{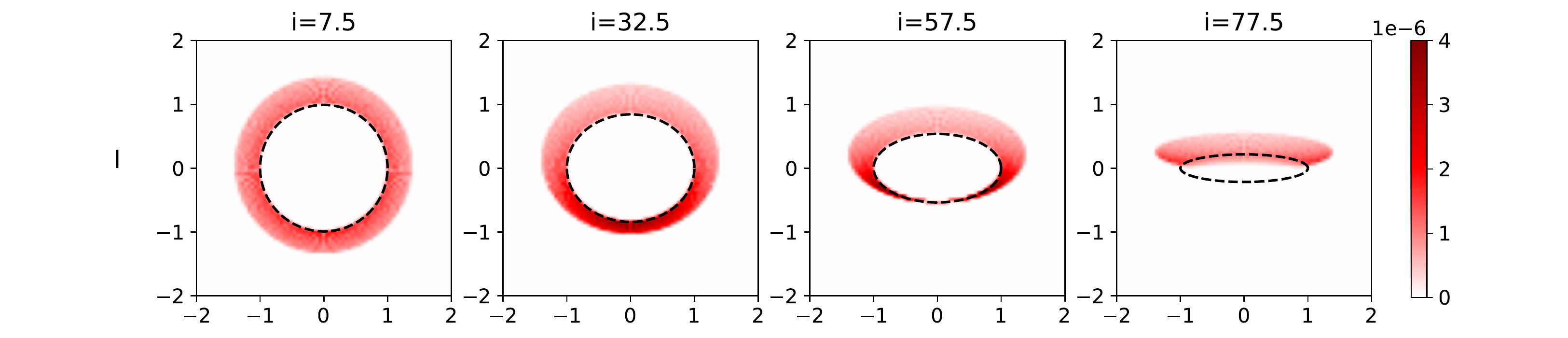}
        \vspace{-0.2cm}
    \end{subfigure}
    \begin{subfigure}[b]{\hsize}
        \centering
        \includegraphics[width=17cm]{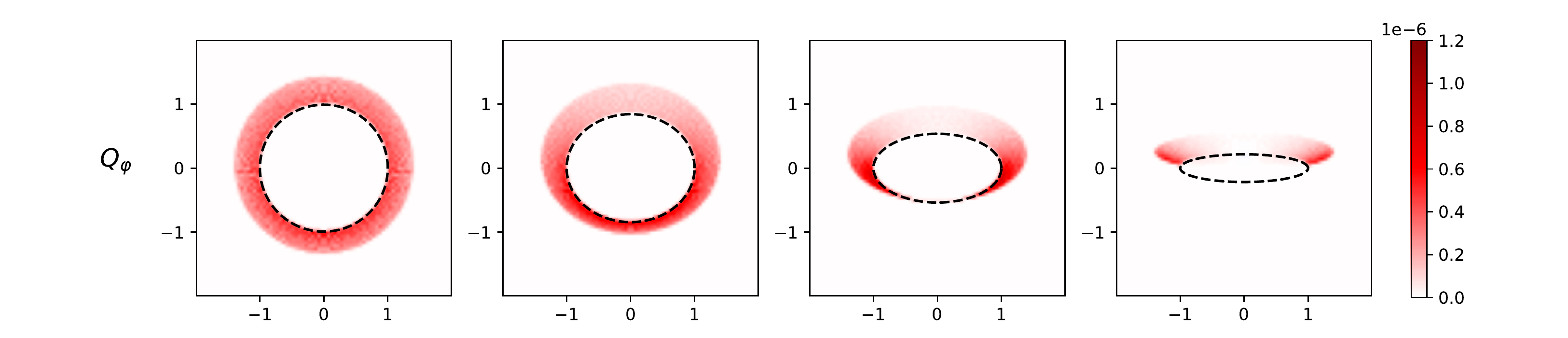}
        \vspace{-0.2cm}
    \end{subfigure}
    \begin{subfigure}[b]{\hsize}
        \centering
        \includegraphics[width=17cm]{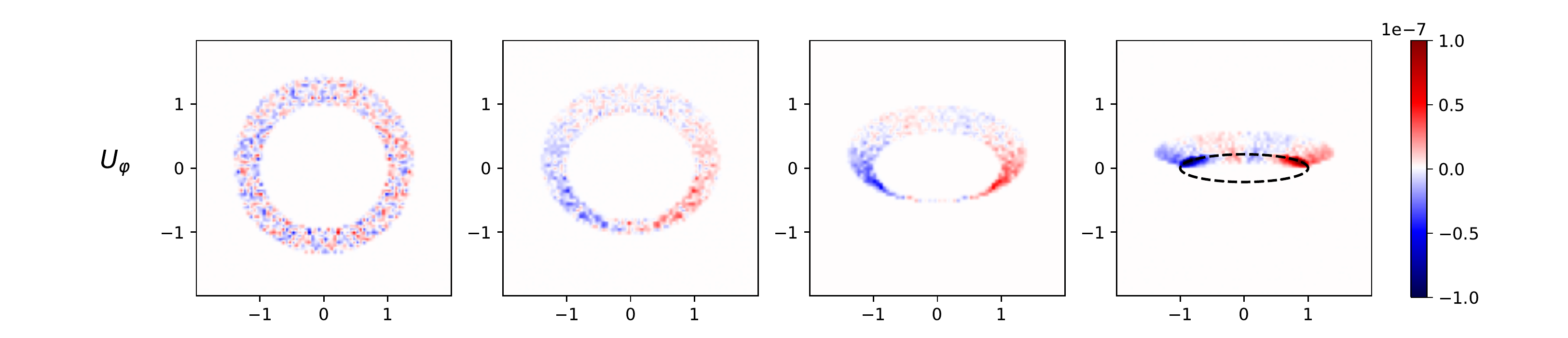}
        \vspace{-0.2cm}
    \end{subfigure}
    \begin{subfigure}[b]{\hsize}
        \centering
        \includegraphics[width=17cm]{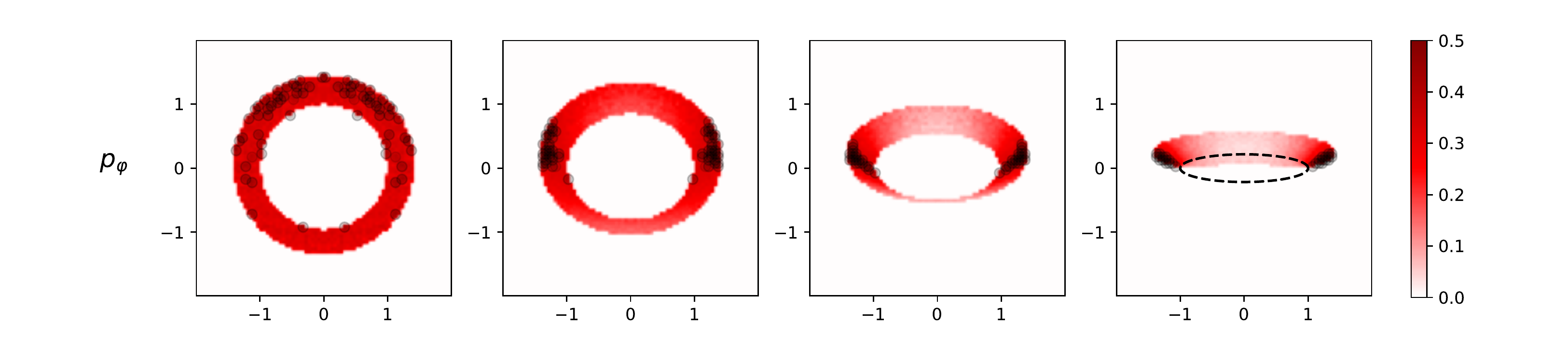}
        \vspace{-0.2cm}
    \end{subfigure}
    \begin{subfigure}[b]{\hsize}
        \centering
        \includegraphics[width=17cm]{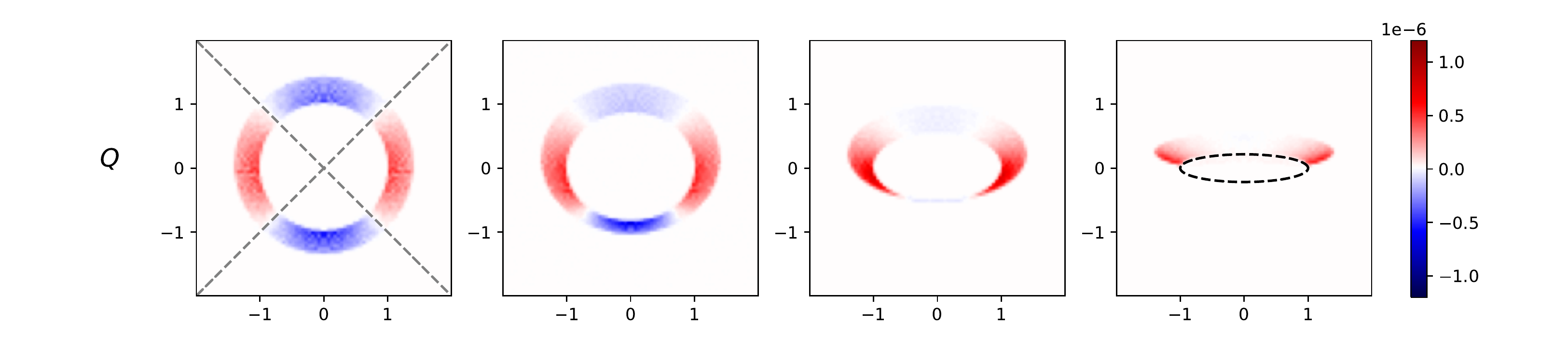}
        \vspace{-0.2cm}
    \end{subfigure}
    \begin{subfigure}[b]{\hsize}
        \centering
        \includegraphics[width=17cm]{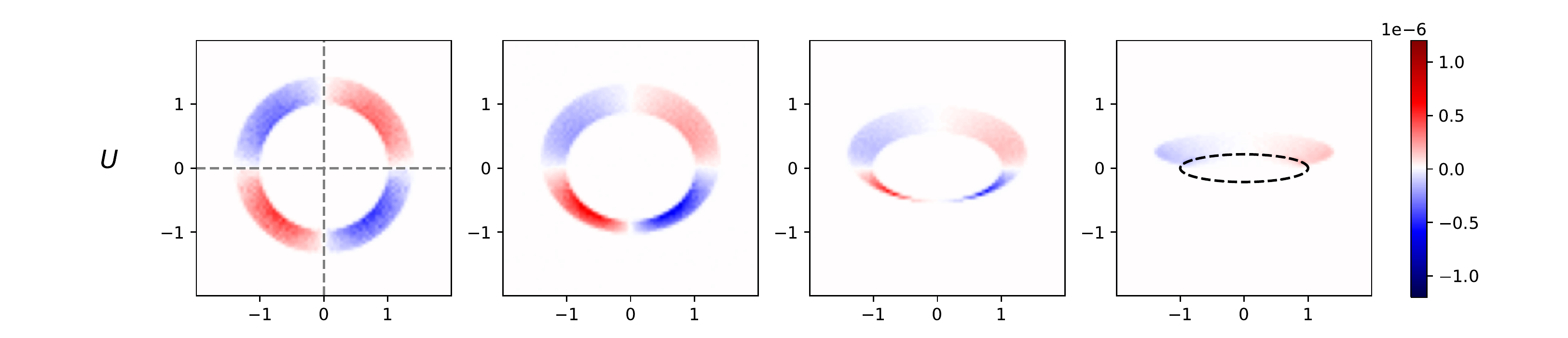}
        \vspace{-0.2cm}
    \end{subfigure}
    \caption{Images of radiation parameters for the 0.5-reference model for four inclinations $i$. The dashed ellipses show $r_0 = 1$ in the disk midplane and the dashed straight lines show the splitting for $Q_{xxx}$ and $U_{xxx}$. For $p_\varphi$, the locations with $p_{\varphi}\ge \max(p_{\varphi})$ are marked with gray transparent points. We note the changed color scale for the weak and therefore noisy $U_\phi$ signal. \vspace{-0.1cm} 
    }
    \label{fig:05-2d}
\end{figure*}

In addition, we use the quadrant polarization parameters $Q_{000}$, $Q_{090}$, $Q_{180}$, $Q_{270}$ for Stokes $Q$ and $U_{045}$, $U_{135}$, $U_{225}$, $U_{315}$ for Stokes $U$ for the characterization of the azimuthal dependence of the polarization. The indices give the position angle of the Stokes quadrant for which the $Q$ or $U$ signal is integrated, where $Q_{000}$ is the backside quadrant, $Q_{090}$ the left quadrant and so on as indicated in Fig.~\ref{fig:05-2d}. Basic symmetry properties of the polarization quadrant parameters $Q_{xxx}$ and $U_{xxx}$ and relations with respect to Stokes $Q$ and $U$ or the azimuthal polarization $Q_\varphi$ are described in \citep{Schmid21}. 

\subsection{Reference model (0.5-model) for transition disks}
The geometry for the reference model for the transition disks is defined by a wall slope of $\chi=32.5^\circ$ and an angular wall height of $\alpha=10^\circ$. This corresponds to incidence angles of $\theta_0(r_0)=57.5^\circ$ near the midplane and $\theta_0(r_w)=67.5^\circ$ at the wall rim. For the dust, we again adopt the scattering parameters $\omega=0.5$, $g=0.5,$ and $p_{\rm max}=0.5$, so that the three-dimensional 0.5-model disk model corresponds precisely to the plane parallel 0.5-model. 

Along the disk minor axis, the scattering occurs in the principal plane and the calculated reflected radiation is very similar to the results for the scattering in plane-parallel surface atmosphere. For surface regions at other disk azimuth angles, the reflectivities depend on $\theta$ and $\phi$ and cannot be compared directly to the presented plane parallel surface results. 
For regions near the major axis, the scattering angle $\theta_s$ is always close to $90^\circ$ and therefore the polarization signal at these positions is strong. 

\subsubsection{Images of the reference model for transition disks}

The two-dimensional images $I(x,y)$, $Q_\varphi(x,y)$, $U_\varphi(x,y)$, $p_\varphi(x,y)$, $Q(x,y)$, and $U(x,y)$ for the 0.5-model are shown in Fig.~\ref{fig:05-2d} for the inclinations $i=7.5^\circ,32.5^\circ,57.5^\circ$, and $77.5^\circ$. 
For polar viewing angles $i\la 10^\circ$, the images for the intensity $I(x,y)$, azimuthal polarization $Q_\varphi(x,y)$, and fractional polarization $p_\varphi(x,y)$ are close to axisymmetric. For intermediate
$i\approx 20^\circ - 50^\circ$, the disk intensity $I(x,y)$ shows a high surface brightness for the front wall because of the strongly ($g=0.5$) forward scattering HG function. For higher inclinations ($i\ge 57.5^\circ$), the front wall becomes invisible and the maximum brightness moves to regions close to the major axis of the projected disk. For high inclination $i=77.5^\circ$, only the scattering from the back wall remains visible. For $i> 57.5^\circ$,  the lower side of the back wall  also contributes to the signal. 

 For higher inclination, the azimuthally polarized intensity $Q_\varphi(x,y)$ shows a reduced signal compared to $I(x,y)$ near the minor axis, where the reflection is dominated by forward and backward scattering which produces only little polarization.
The relative level of $Q_\varphi(x,y)$ with respect to $I(x,y)$ is shown in the $p_\varphi(x,y)$ images. For all plotted disk inclinations, there is ${\rm max}(p_\varphi(x,y))\approx 30~\%$ because there is always a surface region with a scattering angle close to $\theta_S\approx 90^\circ$ where the fractional polarization of the scattered radiation is close to the maximum. The disk location where $\theta_S\approx 90^\circ$ depends strongly on disk inclination (Fig.~\ref{fig:05-2d}) and therefore the photon escape angles $\theta,\,\phi$  also change, but this has only a small effect on the resulting ${\rm max}(p_\varphi)$ value (see Figs.~\ref{fig:05model2} and \ref{fig:hg_dep}).

The calculations show a weak but significant $U_\varphi(x,y)$ signal for inclined disks (Fig.~\ref{fig:05-2d}). This indicates that the position angle of the scattering polarization can deviate (slightly) from a strictly azimuthal orientation as a result of multiple scatterings. This can happen everywhere on the disk except the minor axis, because the scattered light is not escaping in the principal plane (see also Fig.~\ref{fig:05model2}). This effect was described by \citet{Canovas15}, who report a relative $U_\varphi/Q_\varphi$ signal of up to 4.5~\% for their optical thick disk model. The parameters of our "0.5-reference model" are different but our results qualitatively confirm the effects described by \citet{Canovas15}, including the quite subtle distribution of positive and negative $U_\varphi$ regions on the disk. 
For $i=32.5^\circ$, the $U_\varphi$ signal is mostly negative on the left ($x<0$) and positive on the right side ($x>0$) of the disk ring and the ratio $U_\varphi/Q_\varphi$ for the right side is $+1.4~\%$. The corresponding values for $i = 7.5^{\circ}$, $57.5^{\circ}$, and $77.5^{\circ}$ are $0.6\%$, $2.5\%,$ and $10\%$, respectively. For high inclination there are clearly positive and negative $U_\varphi$ parts on one disk side, and this more complex pattern is particularly strong for the lower back-side surface visible in the $i=77.5^\circ$ case. 

$Q_\varphi$ and $U_\varphi$ are vector components and even a signal of $|U_\varphi|/Q_\varphi\approx 10~\%$ means that the polarization angle deviates only $6^\circ$ from the azimuthal direction and $Q_\varphi \approx 0.99~(p\times I)$ and therefore we disregard in the following description of the disk polarization the $U_\varphi$ signal. 
However, if $U_\varphi$ could be measured accurately, this would provide important information for circumstellar disks. Nevertheless, the investigation of this topic is beyond the scope of this paper. 

Figure~\ref{fig:05-2d} also shows  the disk-aligned Stokes $Q(x,y)$ and $U(x,y)$ polarization for different $i$. The polarization quadrants $Q_{xxx}$ and $U_{xxx}$ are indicated and they are used for the quantitative characterization of the azimuthal dependence of the polarization according to \citet{Schmid21}.
For pole-on disks, the absolute value is the same for all quadrants and the integrated Stokes parameters $\overline{Q}$ and $\overline{U}$ are zero because positive and negative quadrants cancel each other. For inclined disks, the polarization quadrants show  clear front--back asymmetries which are apparent in the flux ratios for $Q_{180}/Q_{000}$ or $U_{135}/U_{045}$. These asymmetries are caused by the scattering asymmetry $g$ and the disk geometry and are discussed in the following subsection. For high-inclination systems, the $Q$-signal is strongly concentrated in the positive $Q_{090}$ and $Q_{270}$ quadrants.

\begin{figure*}[!ht]
    \centering
    \resizebox{0.8\hsize}{!}{\includegraphics{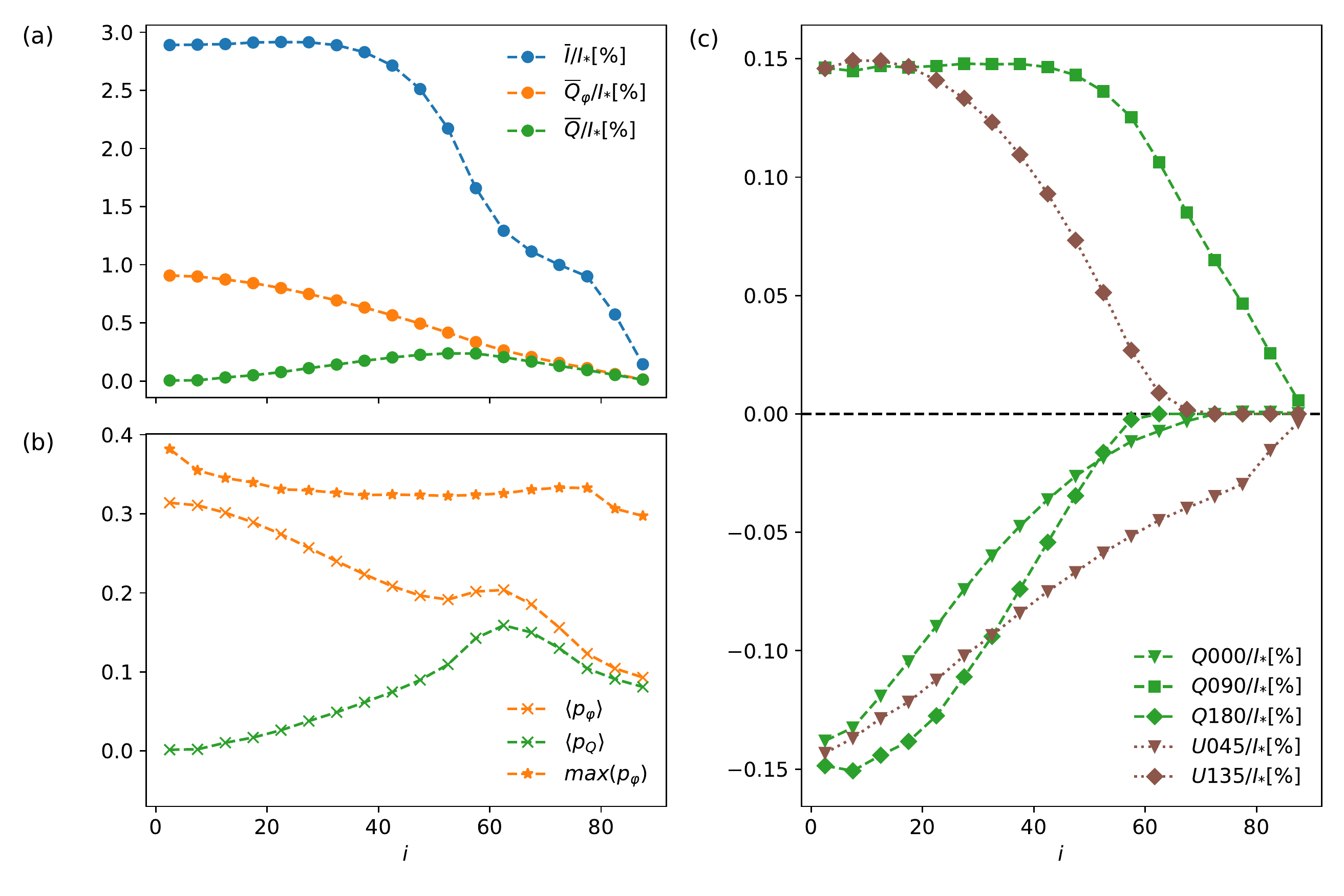}}
    \caption{Radiation parameters for the 0.5-model as functions of disk inclination $i$.
Panel    (a): Disk integrated intensity $\overline{I}(i)/I_\star$, azimuthal polarization intensity $\overline{Q}_\varphi(i)/I_\star$, and Stokes $Q$-intensity $\overline{Q}(i)/I_\star$ expressed relative to the stellar intensity. Panel 
    (b): Disk-averaged fractional polarization $\langle p_\varphi(i)\rangle$ and $\langle p_Q(i)\rangle$ of the scattered radiation, and the maximum fractional azimuthal polarization measured in the disk image ${\rm max}(p_\varphi(i,x,y))$.}
    \label{fig:fluxvsiref}
\end{figure*}

\subsubsection{Radiation parameters for the 0.5-model} \label{section: param for ref}
The radiation parameters $\overline{I}(i)/I_\star$,  $\overline{Q}_\phi(i)/I_\star$, $\overline{Q}(i)/I_\star$, and $Q_{xxx}(i)/I_\star$ or $U_{xxx}(i)/I_\star$ for our transition disk models are first presented in detail in Fig.~\ref{fig:fluxvsiref} for the reference case. 
Relative quantities with respect to the intensity of the central star and as a function of inclination are given because observations often provide such relative measurements and a comparison with our models assumes implicitly that the stellar emission is isotropic without additional absorption or emission components being present like "disturbing" hot dust between the star and the observer or between the star and the disk. 

We use as additional parameters the averaged fractional polarizations $\langle p_\varphi(i)\rangle$ and $\langle p_Q(i)\rangle$ and the maximum value for the fractional polarization ${\rm max}(p_\varphi(i,x,y))$ for the scattered light. Calculated model values for $i=7.5^\circ$, $32.5^\circ$, $57.5^\circ$, and $77.5^\circ$ are given in Table~\ref{tab:diskresults} for the reference disk case (column 3) and many other transition disks with different model parameters. 

\paragraph{Disk-integrated intensity and polarized intensity.}
According to Fig.~\ref{fig:fluxvsiref}(a), the relative disk intensity is for a pole-on orientation at the level of $\overline{I}_{\rm obs}(0)/I_\star=2.9\%$. This value is for $i<40^\circ$ almost independent of the inclination because the drop in brightness for the back side for enhanced $i$ is compensated by the forward scattering gain on the front side. 
For $i>40^\circ$, the bright front side starts to disappear; but for $i>60^\circ$, the lower back side wall below the disk midplane becomes visible and therefore the $\overline{I}(i)/I_\star$ curve has a shoulder around $i=70^\circ$. For large $i>70^\circ$, significant parts of the back-side wall are also hidden by the unilluminated surface of the front-side disk, and there is a step drop $\overline{I}(i)/I_\star\rightarrow 0$ for $i\rightarrow 90^\circ$.

The curve for the azimuthal polarization $\overline{Q}_\phi(i)/I_\star$ shows a steady decrease from the maximum at $i=0^\circ$ towards $i=90^\circ$. 
For $i=0^\circ$ or pole-on disks, the scattering occurs everywhere under the polarimetrically very favourable scattering angle of about $\theta_S\approx 90^\circ$. For increasing inclination, the induced polarization is reduced for disk areas on the front and back side, because forward and backward scattering conditions produce less polarization. 
Only the regions close to the major axis keep a strong polarization signal  for increased $i$  because there the scattering angle is $\theta_S\approx 90^\circ$ for all $i$.

For Stokes $\overline{Q}(i)/I_\star$ , the curve starts for $i=0^\circ$ at zero, because the positive and negative quadrant contributions compensate each other for pole-on disks. 
For larger $i,$ the negative $Q$-features near the minor axis diminish and the two positive $Q$-features near the major axis become dominant so that $\overline{Q}\rightarrow \overline{Q}_\varphi$ for high inclination.

\paragraph{Fractional polarization of the scattered light.}
Figure~\ref{fig:fluxvsiref}(b) shows the disk averaged fractional polarizations, which are the ratios $\langle p_\varphi(i)\rangle=\overline{Q}_\varphi(i)/\overline{I}(i)$ and $\langle p_Q(i)\rangle=\overline{Q}(i)/\overline{I}(i)$ of the parameters given in panel (a). The azimuthal polarization $\langle p_\varphi(i)\rangle$ is about 31\% for pole-on disks, decreases for increasing $i$ like $\overline{Q}_\varphi(i)$, and reaches a plateau around $i\approx 40^\circ$ when the strong intensity contributions from the front side start to disappear. The plateau with $\langle p_\varphi(i)\rangle\approx 20\%$ extends to about $i\approx 60^\circ$ and the shape of this plateau depends on the changing visibility and obscuration of the back wall. The fractional Stokes Q polarization $\langle p_Q(i)\rangle$ is zero at $i=0^\circ$, increases with inclination, and joins the curve $\langle p_\varphi(i)\rangle$  for $i\ga 60^\circ$
, because $\overline{Q}_\varphi$ and $\overline{Q}$ are dominated for high $i$ by the signal of the two disk regions near the major axis.

The same plot includes also determinations for the maximal fractional polarization of the disk ${\rm max}(p_\varphi(i,x,y))$ as measured in the two-dimensional $p_\varphi$ images. Interestingly, this value is almost constant around ${\rm max}(p_\varphi(i))\approx 33~\%$, because for all inclinations  there exist certain disk surface regions where the scattering angle is close to $90^\circ$ (Fig.~\ref{fig:05-2d}). 

\paragraph{Azimuthal dependence of the disk polarization.} 
The images for $Q_\varphi$, $Q,$ and $U$ in Fig.~\ref{fig:05-2d} show a very strong dependence of the azimuthal distribution of the scattering polarization on the disk inclination. 
For the characterization of the azimuthal distribution, we use the quadrant polarization parameters $Q_{000}$, $Q_{090}$, $Q_{180}$, $Q_{270}$, and $U_{045}$, $U_{135}$, $U_{225}$, and $U_{315}$\citep[see][]{Schmid21}, which measure the Stokes $Q$ and Stokes $U$ polarization fluxes of the disk in the individual positive and negative quadrants indicated in Fig.~\ref{fig:05-2d} for $Q$ and $U$ and $i=7.5^\circ$. 
Axisymmetric disk models provide five independent quadrant values $Q_{000}$, $Q_{090}$, $Q_{180}$, $U_{045}$, and $U_{135}$, because some parameters are redundant, that is, $Q_{090} = Q_{270}$, $U_{045} = - U_{315}$, and $U_{135} = - U_{225}$ , as a result of the disk symmetry.

Figure~\ref{fig:fluxvsiref}(c) shows the dependence of the normalized quadrant polarization parameters $Q_{xxx}/I_\star$ and $U_{xxx}/I_\star$ on the disk inclination for the reference model. For pole-on, axisymmetric disks all quadrant parameters have the same value $Q_{xxx}=U_{xxx}=\overline{Q}_\varphi/(2\pi)$. For higher $i,$ first the backside quadrants $Q_{000}$ and $U_{045}$ weaken, because of inefficient backward reflectivity for forward scattering dust and the unfavorable scattering angle $\theta_s \gg 90^\circ$ for the production of strongly polarized radiation. The signals in the front-side quadrants $Q_{180}$ and $U_{135}$ remain stronger than for the backside quadrants, because they profit from the strong ($g=0.5$) forward scattering effect. For larger inclination, the illuminated front-side wall disappears as soon as $i$ approaches $i\approx 90^\circ - \chi$, which is $\approx 57.5^\circ$ for the reference model.  
For larger $i,$ only regions near the major axis $Q_{090}$ and $U_{270}$ or $U_{045}$ and $U_{315}$ have a favorable scattering angle $\theta_s \approx 90^\circ$, are still not hidden by the front side and produce a relatively strong polarization signal. 

\subsection{Disk appearance for different model parameters}

\begin{figure}
    \centering
    \begin{subfigure}[b]{\hsize}
        \centering
        \includegraphics[width=\textwidth]{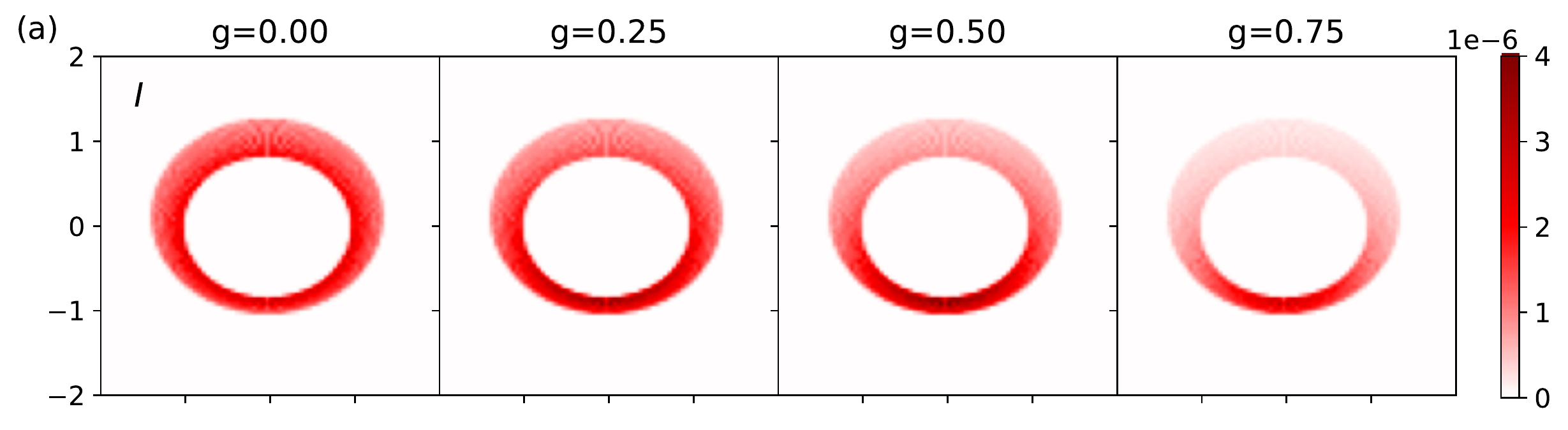}
    \end{subfigure}
    \begin{subfigure}[b]{\hsize}
        \centering
        \includegraphics[width=\textwidth]{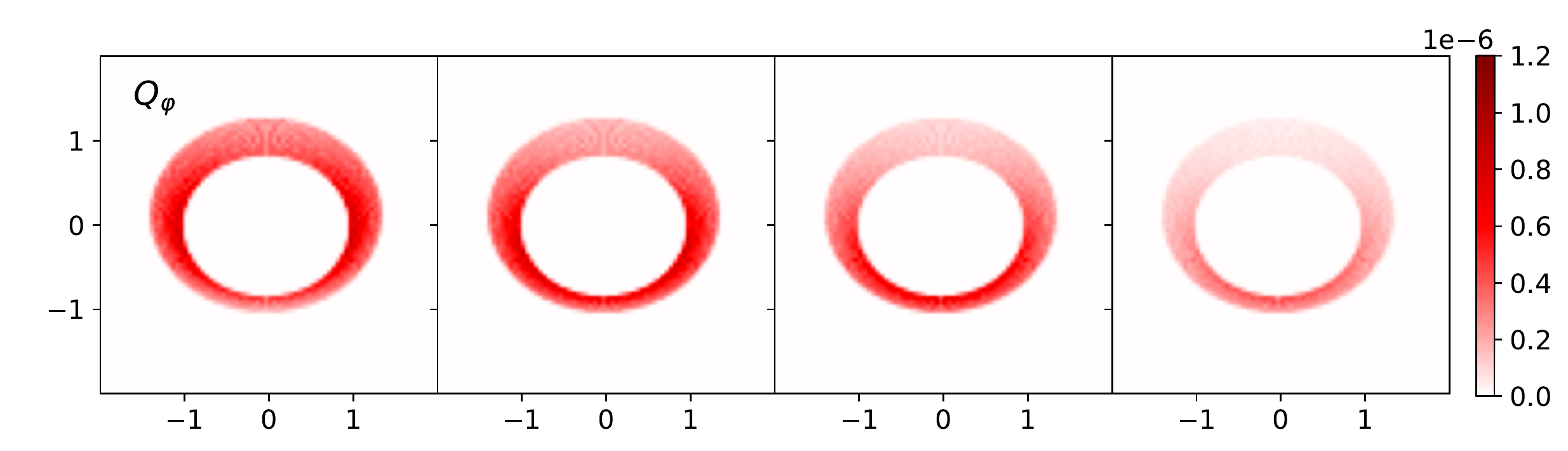}
    \end{subfigure}
    \begin{subfigure}[b]{\hsize}
        \centering
        \includegraphics[width=\textwidth]{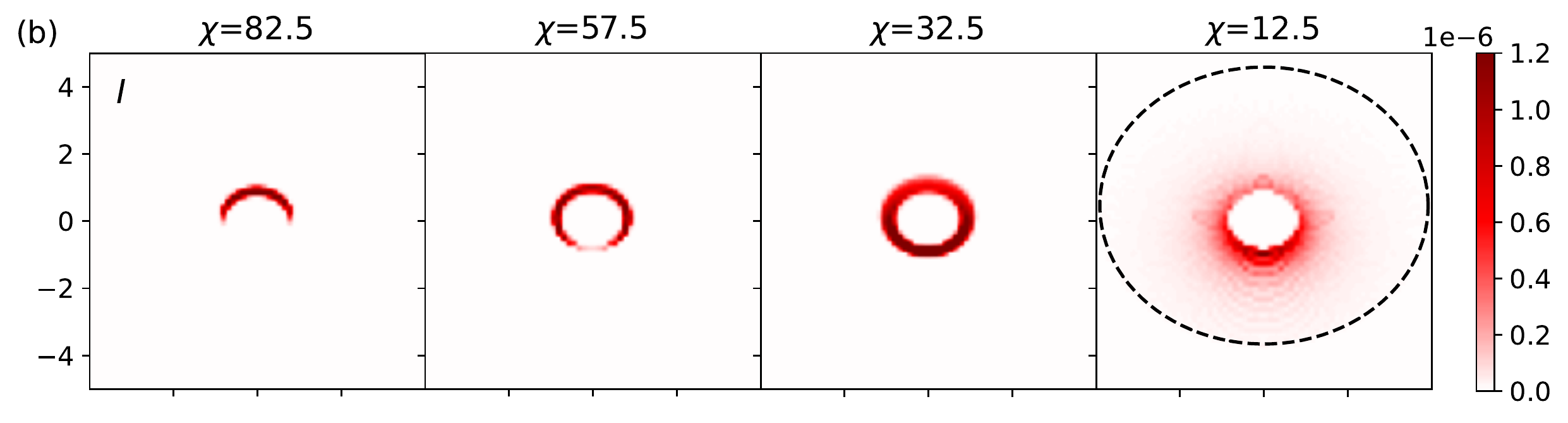}
    \end{subfigure}
    \begin{subfigure}[b]{\hsize}
        \centering
        \includegraphics[width=\textwidth]{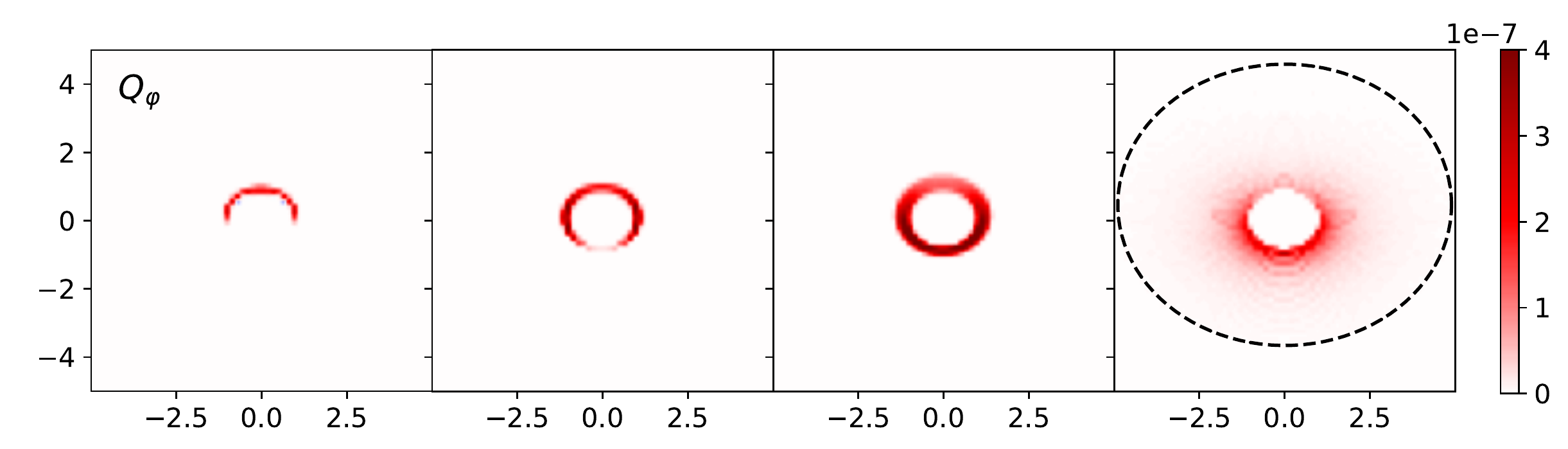}
    \end{subfigure}
    \begin{subfigure}[b]{\hsize}
        \centering
        \includegraphics[width=\textwidth]{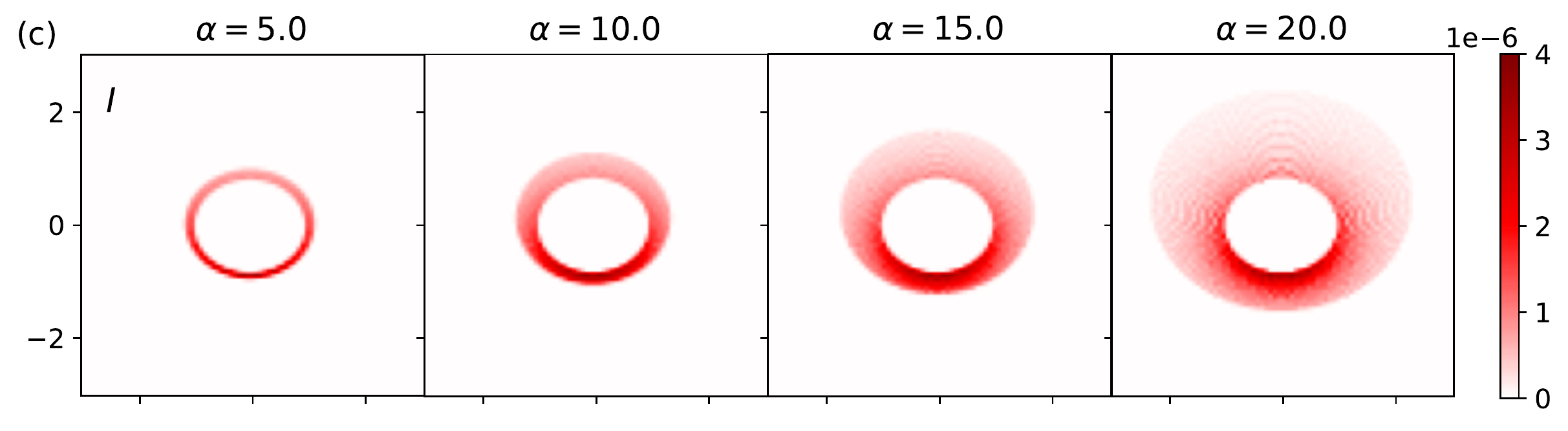}
    \end{subfigure}
    \begin{subfigure}[b]{\hsize}
        \centering
        \includegraphics[width=\textwidth]{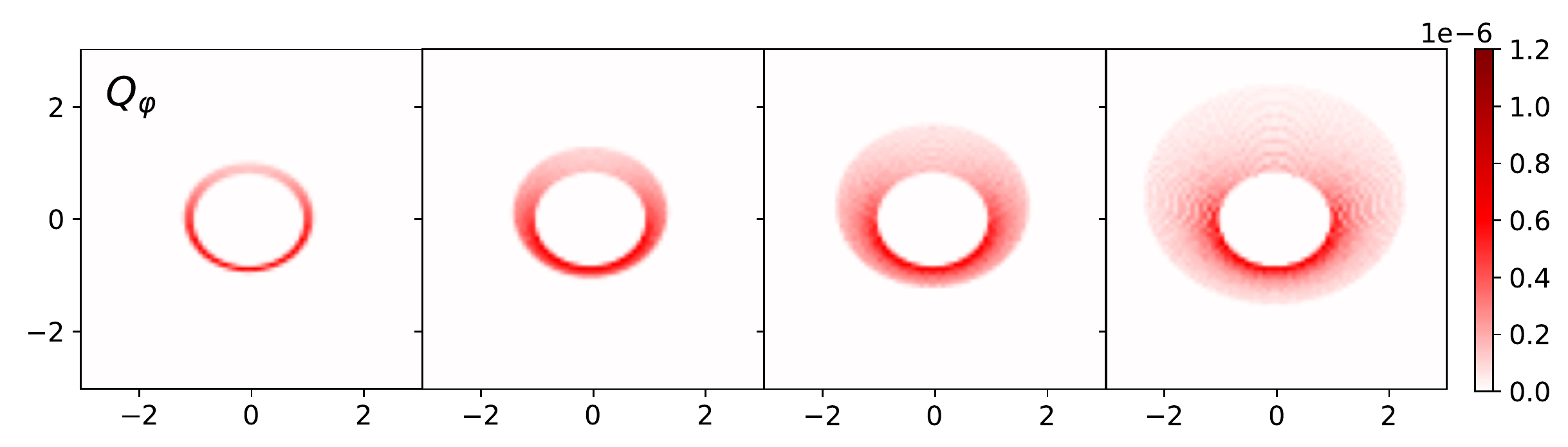}
    \end{subfigure}
    \vspace{0.2cm}
    \caption{Reflected intensity $I(x,y)$ and azimuthal polarization $Q_\varphi(x,y)$ for 
    (a) the scattering asymmetry parameters $g=0.0,0.25,0.5,$ and 0.75 and fixed parameters $\omega=0.5$, $p_{\rm max}=0.5$, $i=32.5^\circ$, $\chi=32.5^\circ$, and $\alpha=10^\circ$. The case $g=0.5$ is the reference model; 
    (b) the disk wall slopes $\chi = 82.5^\circ,57.5^\circ,32.5^\circ$, and $12.5^\circ$. The case $\chi=32.5^\circ$ is the reference model. The dashed ellipses in $\chi=12.5^\circ$ show the outer edge of the disk; and 
    (c) the disk wall heights $\alpha=5^\circ,10^\circ,15^\circ$ and $20^\circ$ and fixed parameters $\omega=0.5$, $g=0.5$, $p_{\rm max}=0.5$, $i=32.5^\circ$, and $\alpha=10^\circ$. The case $\alpha=10^\circ$ is the reference model. }
    \label{fig:disk-morphology}
\end{figure}

The model parameters $i$, $\chi$, and $\alpha$ for the disk geometry and $\omega$, $g$, and $p_{\rm max}$ for the dust scattering  all influence the resulting intensity and polarization of the disk to a greater or lesser degree. Figure~\ref{fig:05-2d} is a good example of the strongly changing geometric appearance of a disk seen under different inclinations.
The scattering albedo $\omega$  strongly defines the scattered intensity and $p_{\rm max}$ the fractional polarization, but the overall morphology of the disk images is not greatly changed by these parameters. 
However, the disk appearance depends significantly on the scattering asymmetry parameter $g$ and the geometry of the disk wall described by the wall slope $\chi$ and wall height $\alpha$. 

The scattering asymmetry $g$ can strongly change the relative brightness between the front- and backside as illustrated in Fig.~\ref{fig:disk-morphology}(a) for disks with $i=32.5^\circ$. The wall slope is $\chi=32.5^\circ$ and therefore $\theta_0$ between $57.5^\circ$ and $67.5^\circ$ and the back wall is seen under a perpendicular viewing angle $\theta=0^\circ$, while the front wall is seen for $\theta=-65^\circ$. Along the minor axis, the signal from the reflected light corresponds at $r_0$ to the principal plane calculations $I(\theta_{\rm pp})$ in Fig.~\ref{fig:hg_dep}. The reflectivity depends on the $g$-parameters and for isotropic scattering $g=0,$ the back wall of the disk is brighter in intensity $I$ and azimuthal polarization $Q_\varphi$ (Fig.~\ref{fig:disk-morphology}(b)) if the flux is integrated radially along the entire wall surface as in Fig.~\ref{fig:hg_dep}, where results are given for a unit surface area. Because the width of the projected back wall is about twice as large as for the projected front wall, the surface brightnesses for $I(x,y)$ and $Q_\varphi(x,y)$ are comparable for the two sides for the $g=0$ case. However, the front side clearly dominates for larger $g$ (see also Fig.~\ref{fig:hg_dep}) as observed in many disks.

The slope of the inner disk wall also has a very important impact on the front-to-back brightness ratio as can be seen in Fig.~\ref{fig:disk-morphology}(b).
For inclined disks $i=32.5^\circ$ and steep walls (e.g., $\chi=82.5^\circ$), the illuminated inner wall of the disk is only visible on the back side, but not on the front side because $i+\chi>90^\circ$. The front side is still barely visible for $\chi=57.5^\circ$ ($i+\chi=90^\circ$) but is clearly visible for moderately steep $\chi=32.5^\circ$ or flat $\chi=12.5^\circ$ disks. In these cases, the front side is much brighter than the back side because of the forward scattering. Indeed, flat disks with the same wall height of $\alpha=10^\circ$ are very extended. This can be inferred from the panels in the fourth column in Fig.~\ref{fig:disk-morphology}(b) when comparing them with the reference model in the third panel. 

The angular height $\alpha$ defines the total radiation received by the disk from the central star. A disk wall with a larger angular height will intercept more photons from the star and the reflected radiation will be enhanced. A higher wall height does not change the reflectivity for the inner disk part, but adds reflecting surface regions further out. For angular heights $\alpha\leq 20^\circ$ the reflected radiation $\overline{I}/I_\star$, $\overline{Q}_\phi/I_\star$, $\overline{Q}/I_\star$ scales roughly proportional to $\alpha$.

The strong brightness dependencies for the front side of the presented disk models is partly an artificial effect caused by the adopted simple disk geometry. Real disks have no sharp edge between a strongly illuminated inner wall and a dark, nonilluminated (self-shadowed) flat disk surface, and therefore the transition from one regime to the other will be much more gradual and not an abrupt switch, as is seen for our model results. For this reason, caution is advised when considering the model results for the brightness of the steep inner wall on the front side of the disk for the interpretation of observations. 

\begin{table*}[]
    \caption{Numerical results for our grid of disk models}
    \label{tab:diskresults}
    \centering
    \addtolength{\tabcolsep}{-1.2pt}    
    \resizebox{\textwidth}{!}{%
    \begin{tabular}{c c c c c c c c c c c c c c c c}
    \hline\hline
    \noalign{\smallskip}
 & Parm.  & Ref. & \multispan{2}{\hfil $\omega$-dep.\hfil} 
                         & \multispan{3}{\hfil$g$-dep.\hfil}
                                 &\multispan{2}{\hfil$p_{\rm max}$-dep.\hfil}
                                        & \multispan{3}{\hfil$\chi$-dep.\hfil} 
                                            & \multispan{3}{\hfil$\alpha$-dep.\hfil} \\
 & $\omega$ & $\omega_r$=$0.5$ & 0.20 & 0.80 & 
 $\omega_r$ & $\omega_r$ & $\omega_r$ & $\omega_r$ & $\omega_r$ & $\omega_r$ & $\omega_r$ & $\omega_r$  & $\omega_r$ & $\omega_r$ & $\omega_r$ \\ 
 & $g$      & $g_r=0.5$ & $g_r$ & $g_r$ & 0.00 & 0.25 & 0.75 & $g_r$ & $g_r$ 
            & $g_r$ & $g_r$ & $g_r$ & $g_r$ & $g_r$ & $g_r$\\
 & $p_{\rm max}$      
            & $p_r=0.5$ & $p_r$ & $p_r$ & $p_r$ & $p_r$ & $p_r$ & 0.20 & 0.80 & 
               $p_r$ & $p_r$ & $p_r$ & $p_r$ & $p_r$ & $p_r$ \\ 
 & $\chi$   & $\chi_r$=32.5$^\circ$ & $\chi_r$ & $\chi_r$ & $\chi_r$ & $\chi_r$ & $\chi_r$ 
 & $\chi_r$ & $\chi_r$ & $12.5^\circ$ & $57.5^\circ$ & $82.5^\circ$  & $\chi_r$ & $\chi_r$ & $\chi_r$\\
 & $\alpha$ & $\alpha_r=10^\circ$ &$\alpha_r$ &$\alpha_r$ &$\alpha_r$ &$\alpha_r$ &$\alpha_r$ &$\alpha_r$ &$\alpha_r$ &$\alpha_r$ &$\alpha_r$ &$\alpha_r$ & $5^\circ$& $15^\circ$ & $20^\circ$\\
\noalign{\smallskip}
\hline
para & $i$ \\
 $\overline{I}/I_\star [\%] $
 & 7.5  & 2.89 & 0.82 & 7.87 & 4.03 & 3.81 & 1.50 & 2.92 & 2.85 & 3.30 & 1.82 & 0.41 & 1.36 & 4.62 & 6.65\\
 & 32.5 & 2.89 & 0.84 & 7.48 & 3.69 & 3.52 & 1.68 & 2.90 & 2.87 & 3.79 & 1.25 & 0.73 & 1.35 & 4.66 & 6.71\\
 & 57.5 & 1.66 & 0.48 & 4.36 & 2.58 & 2.18 & 0.93 & 1.66 & 1.66 & 4.62 & 0.98 & 0.89 & 0.78 & 2.66 & 3.82\\
 & 77.5 & 0.90 & 0.24 & 2.66 & 2.25 & 1.51 & 0.38 & 0.89 & 0.91 & 2.06 & 0.86 & 0.80 & 0.46 & 1.08 & 0.87\\
 \noalign{\smallskip}
 $\overline{Q}_\varphi/I_\star [\%]$ 
 & 7.5  & 0.90 & 0.34 & 1.52 & 1.36 & 1.27 & 0.44 & 0.36 & 1.45 & 1.22 & 0.53 & 0.13 & 0.42 & 1.43 & 2.02\\
 & 32.5 & 0.69 & 0.26 & 1.17 & 1.00 & 0.92 & 0.36 & 0.28 & 1.12 & 1.04 & 0.31 & 0.10 & 0.33 & 1.10 & 1.56\\
 & 57.5 & 0.33 & 0.12 & 0.58 & 0.56 & 0.47 & 0.17 & 0.13 & 0.54 & 0.75 & 0.13 & 0.06 & 0.15 & 0.54 & 0.78\\
 & 77.5 & 0.11 & 0.04 & 0.21 & 0.26 & 0.18 & 0.05 & 0.04 & 0.18 & 0.38 & 0.05 & 0.04 & 0.06 & 0.14 & 0.13\\
\noalign{\smallskip}
 $-Q_{000}/\overline{Q}_\varphi$ 
 & 7.5  & 0.147  & 0.145 & 0.148 & 0.170 & 0.155  & 0.143 & 0.144  & 0.147  & 0.140  & 0.160 & 0.272 & 0.145  & 0.150 & 0.152\\
 & 32.5 & 0.086  & 0.081 & 0.092 & 0.151 & 0.110  & 0.076 & 0.085  & 0.087  & 0.067  & 0.147 & 0.431 & 0.079  & 0.094 & 0.104\\
 & 57.5 & 0.035  & 0.036 & 0.034 & 0.070 & 0.046  & 0.034 & 0.034  & 0.036  & 0.015  & 0.101 & 0.190 & 0.028  & 0.043 & 0.051\\
 & 77.5 & -0.005 & 0.005 & 0.025 & 0.001 & -0.008 & 0.004 & -0.005 & -0.004 & -0.003 & 0.016 & 0.030 & -0.011 & 0     & 0.018\\
  $-U_{045}/\overline{Q}_\varphi$ 
 & 7.5  & 0.152 & 0.153 & 0.154 & 0.170 & 0.160 & 0.148 & 0.154 & 0.152 & 0.146 & 0.162 & 0.252 & 0.149 & 0.154 & 0.156\\
 & 32.5 & 0.135 & 0.128 & 0.142 & 0.201 & 0.163 & 0.119 & 0.133 & 0.135 & 0.111 & 0.211 & 0.234 & 0.126 & 0.144 & 0.152\\
 & 57.5 & 0.153 & 0.141 & 0.171 & 0.222 & 0.182 & 0.136 & 0.154 & 0.154 & 0.083 & 0.273 & 0.161 & 0.134 & 0.173 & 0.195\\
 & 77.5 & 0.265 & 0.245 & 0.295 & 0.290 & 0.271 & 0.263 & 0.265 & 0.267 & 0.089 & 0.351 & 0.163 & 0.223 & 0.306 & 0.373\\
  $Q_{090}/\overline{Q}_\varphi$ 
 & 7.5  & 0.161 & 0.160 & 0.162 & 0.162 & 0.163 & 0.162 & 0.160 & 0.162 & 0.161 & 0.164 & 0.175 & 0.162 & 0.162 & 0.162\\
 & 32.5 & 0.213 & 0.212 & 0.217 & 0.219 & 0.223 & 0.204 & 0.215 & 0.213 & 0.205 & 0.248 & 0.193 & 0.212 & 0.214 & 0.214\\
 & 57.5 & 0.375 & 0.375 & 0.371 & 0.341 & 0.368 & 0.367 & 0.375 & 0.373 & 0.329 & 0.298 & 0.322 & 0.379 & 0.366 & 0.356\\
 & 77.5 & 0.422 & 0.424 & 0.420 & 0.404 & 0.421 & 0.414 & 0.422 & 0.421 & 0.478 & 0.348 & 0.402 & 0.441 & 0.382 & 0.303\\
  $U_{135}/\overline{Q}_\phi$ 
 & 7.5  & 0.166  & 0.167  & 0.165  & 0.149  & 0.159  & 0.170  & 0.167  & 0.165  & 0.170 & 0.154  & 0.063 & 0.167  & 0.164 & 0.162\\
 & 32.5 & 0.177  & 0.183  & 0.168  & 0.109  & 0.147  & 0.193  & 0.177  & 0.177  & 0.201 & 0.095  & 0.001 & 0.187  & 0.168 & 0.159\\
 & 57.5 & 0.079  & 0.090  & 0.066  & 0.033  & 0.055  & 0.102  & 0.082  & 0.080  & 0.176 & 0.003  & 0.002 & 0.096  & 0.065 & 0.052\\
 & 77.5 & -0.002 & -0.002 & -0.003 & -0.003 & -0.002 & -0.002 & -0.002 & -0.002 & 0.020 & -0.002 & 0.004 & -0.003 & 0     & 0    \\
  $-Q_{180}/\overline{Q}_\varphi$ 
 & 7.5  & 0.168 & 0.168 & 0.163 & 0.141 & 0.156 & 0.170 & 0.170 & 0.165 & 0.173 & 0.149 & 0.020 & 0.169 & 0.163 & 0.160\\
 & 32.5 & 0.135 & 0.144 & 0.125 & 0.062 & 0.096 & 0.166 & 0.137 & 0.136 & 0.167 & 0.017 & 0     & 0.144 & 0.126 & 0.117\\
 & 57.5 & 0.007 & 0.008 & 0.006 & 0.001 & 0.002 & 0.013 & 0.006 & 0.007 & 0.069 & 0     & 0     & 0.009 & 0.005 & 0.004\\
 & 77.5 & 0     & 0     & 0     & 0     & 0     & 0     & 0       & 0   & 0     & 0     & 0     & 0     & 0     & 0    \\
 \noalign{\smallskip}
   $\langle p_{\varphi} \rangle$ 
 & 7.5  & 0.31 & 0.41 & 0.19 & 0.34 & 0.33 & 0.29 & 0.12 & 0.51 & 0.37 & 0.29 & 0.33 & 0.31 & 0.31 & 0.31\\
 & 32.5 & 0.24 & 0.31 & 0.16 & 0.27 & 0.26 & 0.21 & 0.10 & 0.39 & 0.27 & 0.25 & 0.14 & 0.24 & 0.24 & 0.23\\
 & 57.5 & 0.20 & 0.26 & 0.13 & 0.22 & 0.22 & 0.18 & 0.08 & 0.32 & 0.16 & 0.14 & 0.07 & 0.20 & 0.20 & 0.21\\
 & 77.5 & 0.12 & 0.16 & 0.08 & 0.12 & 0.12 & 0.12 & 0.05 & 0.20 & 0.18 & 0.06 & 0.05 & 0.12 & 0.13 & 0.15\\
 \noalign{\smallskip}
 $\langle p_{Q} \rangle$ 
 & 7.5  & 0    & 0    & 0    & 0    & 0    & 0    & 0    & 0.01 & 0    & 0.01 & 0.02  & 0    & 0    & 0   \\
 & 32.5 & 0.05 & 0.06 & 0.03 & 0.06 & 0.06 & 0.04 & 0.02 & 0.08 & 0.05 & 0.08 & -0.01 & 0.05 & 0.05 & 0.05\\
 & 57.5 & 0.14 & 0.18 & 0.09 & 0.13 & 0.15 & 0.12 & 0.06 & 0.23 & 0.09 & 0.07 & 0.03  & 0.14 & 0.14 & 0.13\\
 & 77.5 & 0.10 & 0.14 & 0.07 & 0.09 & 0.10 & 0.10 & 0.04 & 0.17 & 0.18 & 0.04 & 0.04  & 0.11 & 0.10 & 0.09\\
   ${\rm max}(p_\varphi)$ 
 & 7.5  & 0.35 & 0.49 & 0.22 & 0.37 & 0.37 & 0.35 & 0.16 & 0.55 & 0.49 & 0.34 & 0.43 & 0.34 & 0.37 & 0.38\\
 & 32.5 & 0.32 & 0.44 & 0.21 & 0.36 & 0.35 & 0.32 & 0.14 & 0.53 & 0.42 & 0.30 & 0.35 & 0.32 & 0.34 & 0.36\\
 & 57.5 & 0.32 & 0.43 & 0.22 & 0.37 & 0.35 & 0.30 & 0.13 & 0.52 & 0.40 & 0.31 & 0.27 & 0.32 & 0.33 & 0.35\\
 & 77.5 & 0.33 & 0.42 & 0.23 & 0.38 & 0.37 & 0.29 & 0.13 & 0.53 & 0.40 & 0.26 & 0.21 & 0.33 & 0.34 & 0.36\\
 \noalign{\smallskip}
   $I_{180}/I_{000}$ 
 & 7.5  & 1.23 & 1.29 & 1.14 & 0.96 & 1.09 & 1.39 & 1.23 & 1.23 & 1.37 & 0.98 & 0.08 & 1.22 & 1.24 & 1.26\\
 & 32.5 & 1.93 & 2.37 & 1.37 & 0.65 & 1.13 & 3.20 & 1.93 & 1.95 & 3.00 & 0.16 & 0    & 1.91 & 1.98 & 2.04\\
 & 57.5 & 0.29 & 0.44 & 0.16 & 0.04 & 0.11 & 0.75 & 0.29 & 0.29 & 6.11 & 0    & 0    & 0.32 & 0.27 & 0.26\\
 & 77.5 & 0    & 0    & 0    & 0    & 0    & 0    & 0    & 0    & 0.97 & 0    & 0    & 0    & 0    & 0   \\
 \noalign{\smallskip}
 \hline
 \end{tabular}
 }
 \addtolength{\tabcolsep}{1.2pt}
 \tablefoot{Results are given for the reference model for $i=7.5^\circ,\, 32.5^\circ,\, 57.5^\circ$, and $77.5^\circ$, and models that deviate for one of the following parameters: scattering albedo $\omega$, asymmetry $g$, maximum scattering polarization $p_{\max}$, wall slope $\chi,$ and angular wall height $\alpha$. Numerical uncertainties for the calculated model results are at a level of about one or two units for the last digit of the given values.}
\end{table*}

\subsection{Quantitative disk model results}

The radiation parameters introduced for the reference model are also used for the investigation of the disk parameter dependencies for disks with different parameters. 
 Table~\ref{tab:diskresults} shows numerical values for a quantitative assessment. Results for the inclinations $i=7.5^\circ,\,32.5^\circ,\,57.5^\circ,$ and $77.5^\circ$ for 14 models are given including the reference model and models where one of the parameters $\omega$, $g$, $p_{\rm max}$, $\chi,$ or $\alpha$ deviates from the reference model. 

The quadrant polarizations are given as values relative to the integrated azimuthal polarization $(\pm)\, Q_{xxx}/\overline{Q}_\varphi$ or $(\pm)\,U_{xxx}/\overline{Q}_\varphi$. Because our disk models are rotationally symmetric, the absolute values for all relative quadrant parameters are for a pole-on view $i=0^\circ$ equal to $1/(2\pi)=0.159$ \citep[see][]{Schmid21}. The disk-averaged fractional polarization values $\langle p_\varphi\rangle$ and $\langle p_Q\rangle$ can be deduced from $\langle p_{\varphi}\rangle=\overline{Q}_\varphi/\overline{I}$ and $\langle p_{Q}\rangle=\langle p_{\varphi}\rangle (Q_{000}/Q_\varphi+2\,Q_{090}/Q_\varphi+Q_{180}/Q_\varphi)$. However, these are important observational parameters and  are therefore also listed. In addition, Table~\ref{tab:diskresults} includes the values for the maximum fractional polarization ${\rm max}(p_{\varphi})$, and the front-to-back intensity ratio $I_{180}/I_{000}$ for the comparison with the front-to-back polarization ratio $Q_{180}/Q_{000}$.

Numerical uncertainties in Table~\ref{tab:diskresults} are at the level of less than one to a few units for the last digit of the given values. Uncertainties are larger for small inclination $i$, because the solid angles per bin for the collection of escaping photons are smaller $(\propto \sin i)$ for pole-on viewing directions. The uncertainties can also be estimated from the jitter in the expectedly smooth curves in Fig.~\ref{fig:fluxvsiref}.

The maximum fractional polarization ${\rm max}(p_\varphi(x,y,i))$ occurs in the two-dimensional image at a disk location $(x,y)$ where the scattering angle is close to $\theta_S\approx 90^\circ$ and produces a high fractional polarization close to the maximum value for given scattering parameters (Fig.~\ref{fig:05-2d}). For a large range of disk inclinations $i,$ there exists a surface region where $\theta_S\approx 90^\circ$ and therefore ${\rm max}(p_\varphi(x,y,i))$ does not depend strongly on the disk inclination $i$, unlike the disk-averaged value $\langle p_\varphi(i)\rangle$ as illustrated in Fig.~\ref{fig:fluxvsiref}. 
This is confirmed by the values in Table~\ref{tab:diskresults} which also show that the dependencies of ${\rm max}(p_\varphi)$ on $\alpha$, $\chi,$ and $g$ are small. Therefore, the ${\rm max}(p_\varphi(x,y))$ parameter is very useful for constraining dust scattering parameters, in particular $\omega$ and $p_{\rm max}$.

For most of the models
in Table~\ref{tab:diskresults}, Appendix \ref{section: inclination-dep} provides curves of the radiation parameters as a function of disk inclination, as in the reference model in Fig.~\ref{fig:fluxvsiref}. These are meant to help with assessment of the dependencies of the reflected radiation on the disk parameters.

\section{Diagnostic parameters for dust scattering}

The calculated radiation parameters from the transition disk models can be compared with observational data in order to constrain the properties of real disks, in particular the dust scattering parameters $\omega$, $g,$ and $p_{\rm max}$. However, the model results show quite complex dependencies on the disk geometry and the dust scattering properties. Therefore, we investigate in this section the diagnostic potential of "observational" radiation parameters for constraining and deriving scattering parameters of the dust and refine geometric properties of the disk.

\subsection{Typical radiation parameters for transition disks}
\begin{figure}[]
    \centering
    \includegraphics[width=0.45\textwidth]{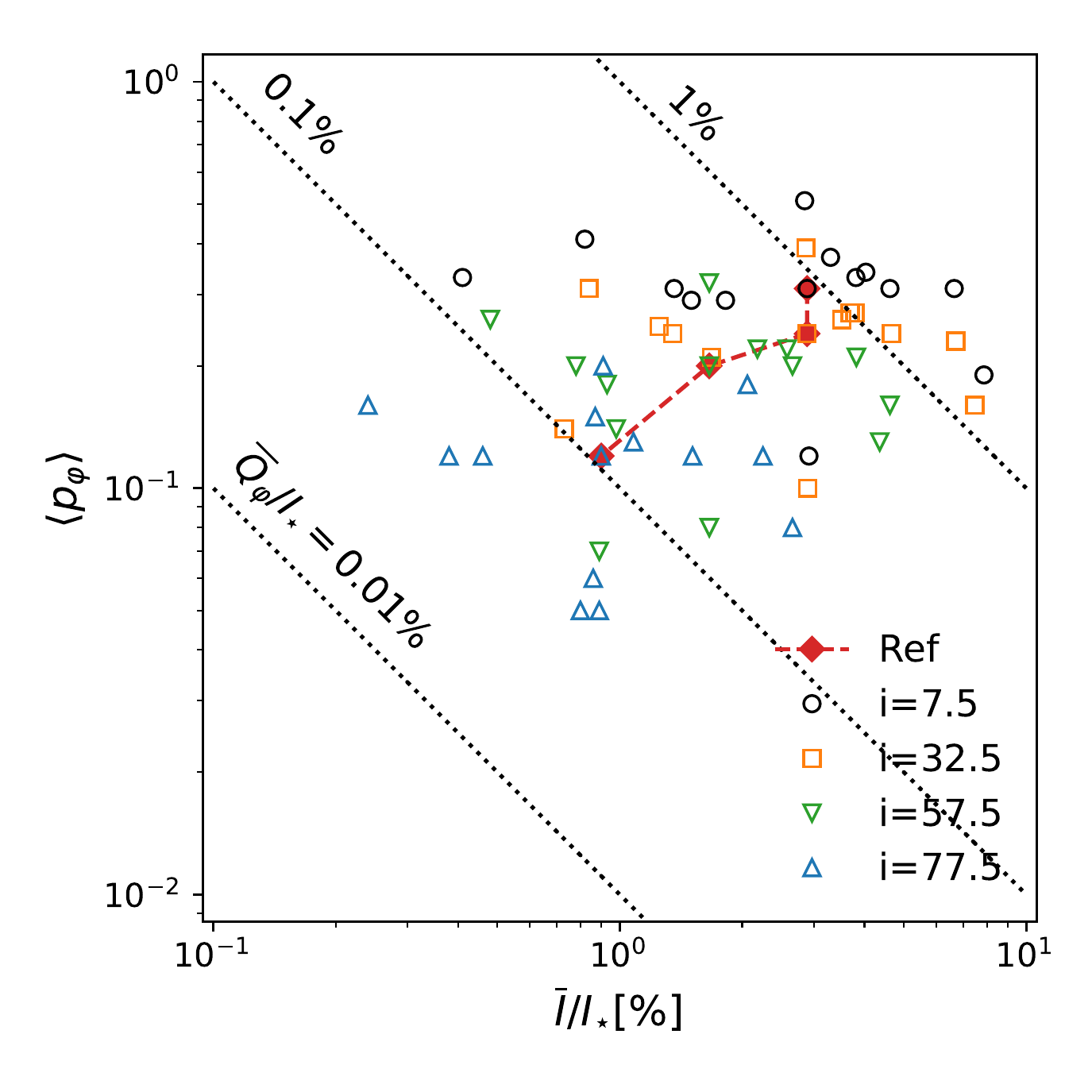}
    \caption{Distribution of the scattered intensity $\overline{I}/\overline{I}_\star$ and the fractional polarization $\langle p_\varphi\rangle$ (Y-axis) for all disk models given in Table~\ref{tab:diskresults}. The dotted lines indicate the constant values for the azimuthal polarization $\overline{Q}_\varphi/\overline{I}$. The colors indicate the inclination $i$ of the disks and the red line marks the results for the reference model.}
    \label{fig:DiagFluxPol}
\end{figure}

The model results for the radiation parameters $\overline{I}/I_\star$, $\overline{Q}_\varphi/I_\star$, and $\langle p_\varphi \rangle$ for $i=12.5^\circ$, $32.5^\circ$, $57.5^\circ$, and $77.5^\circ$ for the reference disk and 13 other models where one intrinsic disk parameter differs from the reference case are listed in Table~\ref{tab:diskresults}.
Figure~\ref{fig:DiagFluxPol} shows their distribution in the $\overline{I}/I_\star$--$\langle p_\varphi \rangle$ diagram and the points cover ranges of about a factor of 30 for $\overline{I}/I_\star$ and $\overline{Q}_\varphi/I_\star$ and about a factor of 10 for the fractional polarization $\langle p_\varphi \rangle$. Typical values for this selection of models are about $\overline{I}/I_\star\approx 2~\%$, $\overline{Q}_\varphi/I_\star\approx 0.4~\%$ and a fractional polarization of $\langle p_\varphi \rangle \approx 20~\%$, which are in rough agreement with measurements obtained for transition disks; for example, with the  $\overline{Q}_\varphi/I_\star$ values for transition disks compiled by \citet{Garufi17}, or the fractional polarization measurements of $\langle p_\varphi \rangle$ provided by \citet{Perrin09}, \citet{Monnier19}, \citet{Hunziker21}, and \citet{Tschudi21}.

The highlighted reference model is located in the middle of the point cloud and clearly shows  enhanced values for disk intensities, polarized intensities, and fractional polarization for low inclination $i$ and small values for high $i$. This pattern is shared by most disk models and therefore the parameters are systematically displaced towards the upper right for low $i$ and towards the lower left for high $i$, which are marked with different colors. 
The points for the reference models are surrounded by values from the other models with one parameter changed and this yields offsets in the $\overline{I}/I_\star$--$\langle p_\varphi \rangle$ plane by some amount, but typically by less than a factor of a few for the parameter range considered by our models. 

\subsection{Dependencies on model parameters}

\begin{table*}[!ht]
    \centering
    \caption{Dependence of the disk scattering intensity and polarization on model parameters} 
    \begin{tabular}{c c c c c c}
    \hline\hline
         & $X_r$ & $X_{r-1}$, $X_{r+1}$ & 
                   $\overline{I}/I_\star$ & 
                               $\overline{Q}_\varphi/I_\star$ & 
                                           $\langle p_\varphi \rangle$ \\
     \hline
    $\omega$ & 0.5 & 0.2,0.8 
                &  $1.92\pm0.52$ \tablefootmark{b} 
                    & $1.11 \pm 0.02$
                        & $-0.54 \pm 0.03$ \\
    $g$    & 0.5 & 0.25,0.75 
                & $-0.64 \pm 0.14$ 
                    & $-0.81 \pm 0.10$ 
                        & $-0.20 \pm 0.01$ \\
                                    
    $p_{\rm max}$ & 0.5 & 0.2,0.8 
                & $-0.01 \pm 0.01$ 
                    & $1.01\pm 0.01$ 
                        & $1.02 \pm 0.01$ \\    
\noalign{\smallskip}    
    $\alpha$ & 10 & 5, 20
                    & $1.16 \pm 0.06$ 
                        &  $1.12 \pm 0.04$
                             & $-0.02 \pm 0.01$ \\ 
   $\chi$     & 32.5 & 12.5, 57.5 
                    & $-0.65 \pm 0.08$ 
                            & $-0.76 \pm 0.03$
                                &$-0.07 \pm 0.09$\\
    $\cos i$    & $0.84$ & $0.99,\,0.61$ \tablefootmark{a}
                    & $0.12 \pm 0.06$
                        & $1.62 \pm 0.03$
                            & $1.51 \pm 0.08$\\
    \hline
    \end{tabular}
    \tablefoot{
    The gradients of the radiation parameters integrated intensity $\overline{I}/I_\star$, azimuthal polarization $\overline{Q}_\varphi/I_\star$, and averaged fractional polarization $\langle p_\varphi \rangle$ for the scattered light from transition disk with respect to the geometric parameters inclination $i$, angular wall height $\alpha$, wall slope $\chi,$ and the dust parameters scattering albedo $\omega$, asymmetry $g,$ and maximum scattering polarization $p_{\rm max}$.
    \tablefoottext{a}{The linear relation for $\overline{I}/I_{\star}$ is valid to $\cos i=0.61$ corresponding to $i = 42.5^\circ$.}
    \tablefoottext{b}{Here the linear relation does not fit well, so we use a power law instead: $\overline{I}/\overline{I}_r = \exp{[A'_\omega(\omega-\omega_r)/\omega_r]}$, $A'_\omega = 1.59\pm 0.04$}. 
        }
    \label{tab:diagnostic}
\end{table*}

The radiation parameters for the transition disks depend in various ways on the model parameters and we roughly  summarize these dependencies with linear relations $Y = aX +b$. We express the dependencies with a relative gradient $A_X$ at a reference point $X_r,\,Y_r$:
\begin{equation}
    \frac{Y(X)-Y_r}{Y_r} = A_X \, \frac{X-X_r}{X_r} \,,   
\label{Eq:gradient}
\end{equation}
where $Y$ represents radiation parameters like $\overline{I}/I_\star$, $\overline{Q}_\varphi/I_\star$, or $\langle p_\varphi \rangle$ and $X$ represents one of the model parameters for the dust scattering $\omega$, $g$, $p_{\rm max}$, or $\alpha$, $\cos i$, $\chi$ for the disk geometry. Also, 
$Y_r$ and $X_r$ are the corresponding values for the reference model and $A_X=a\,(X_r/Y_r)$ describes the relative gradient for the parameters of the reference model $Y_r$, $X_r$. Derived values for $A_X$ are summarized in Table~\ref{tab:diagnostic}. For example, if $\omega$ is enhanced by $20~\%$ from $\omega_r=0.5$ to $\omega=0.6$, $\langle p_\varphi \rangle$ decreases by $-0.54 \cdot 20~\% = -10.8~\%$ with respect to the reference model, taking the gradient $A_\omega[\langle p_\varphi \rangle]=-0.54$.

The gradients $A_X$ listed in Table~\ref{tab:diagnostic} are derived from the gradient between the two bracketing values $Y(X_{r-1})$ and $Y(X_{r+1})$ and the uncertainties represent the mean gradient difference for the points $Y(X_{r-1})$, $Y_r$ and $Y_r$, $Y(X_{r+1})$ with respect to $A_X$. The simple linear relation in Eq.~\ref{Eq:gradient} accurately describes the dependencies $Y(X)$ for many combinations of radiation parameters $Y$ and model parameters $X,$ as can be inferred from the small uncertainties indicated in Table~\ref{tab:diagnostic}. 

For $A_\omega[I/I_\star],$ a large uncertainty of $\pm 0.52$ or a strong deviation from a linear relation is obtained, because the reflected intensity has a dependence on the scattering albedo which is roughly exponential $\overline{I}/I_\star = c\,\exp(a\,\omega)$ in the range $\omega=0.2-0.8,$ which is similar to the results of the plane parallel case (e.g., Fig.~\ref{fig:reflectivity}(a)). This exponential function or the corresponding Taylor expansion can also be expressed with the relative gradient at the reference point $A'_X$: 
\begin{equation}
    \frac{Y(X)-Y_r}{Y_r} =\exp\Bigl(A'_X\frac{X-X_r}{X_r}\Bigr)-1 
    = Z + \frac{1}{2}Z^2 +\frac{1}{6}Z^3+ \cdots \,,
\end{equation}
where $Z=A'_X(X-X_r)/X_r$. For the gradient $A'_\omega = 1.59\pm 0.04$, a good fit is obtained for $\overline{I}/\overline{I}_r=\exp{[A'_\omega(\omega-\omega_r)/\omega_r}]$.

The inclination dependence of the polarized intensity $\overline{Q}(i)/I_\star$ and the fractional polarization $\langle p_\varphi(i) \rangle$ show a broad maxima for pole-on disks or $i\approx 0^\circ$ and a decrease similar to a $\cos i$-function for larger $i$. This can be well fitted with the linear function as in Eq.~\ref{Eq:gradient}, if $\cos i$ is used as model parameter $X$. 
However, for $X=\cos i,$ the obtained gradients $A_{\cos i}$ must be interpreted accordingly. First, a positive $A_{\cos i}$ means that the $Y(i)$ decreases with $i$ similar to $\cos i$. For the scattered intensity $\overline{I}/I_\star$, the gradient is close to zero and this is equivalent to a very flat dependence. Second, the large gradients of $A_{\cos i}\approx 1.5$ to $1.6$ for $\overline{Q}(i)/I_\star$ and $\langle p_\varphi(i) \rangle$ mean that there is still a relatively small deviation of $<|\pm 15~\%|$ from the exact $\cos i$-function, because only a very small $\cos i$-range around the reference value is sampled for the covered inclination range from $7.5^\circ$ to $42.5^\circ$. Third, the $i$-range for the fitting had to be restricted to $42.5^\circ$, because for larger $i$ the radiation parameters show bumps and strong changes due to the disappearance of the inner wall on the front side.

The dependencies given in Table~\ref{tab:diagnostic} apply for the reference case $Y_r(X_r)$, but should represent  the behavior of the disk radiation parameters described by the range $[X_{r-1},X_{r+1}]$ quite well; this range covers the region where the radiation functions $Y(X)$ ($\overline{I}/I_\star$, $\overline{Q}_\varphi/I_\star$, $\overline{Q}/I_\star$, $\langle p_\varphi \rangle$, and $\langle p_Q \rangle$) show smooth dependencies in Fig.~\ref{fig:integrate-param}.
This range excludes highly inclined disks $i>57.5^\circ$ or disks with steep inner walls $\chi>57.5^\circ$ where disk visibility and obscuration effects play an important role and produce strong and "bumpy" parameter dependencies.

The value for radiation parameters with two or more model parameters different from the reference value $Y(X_1,X_2)$ can be estimated from 
\begin{equation}
    \frac{Y(X_1,X_2)}{Y_r} \approx 
    \Bigl(A_1\frac{X_1-X_{1,r}}{X_{1,r}}+1\Bigr)\, \Bigl(A_2\frac{X_2-X_{2,r}}{X_{2,r}}+1\Bigr)\,.
\end{equation}
This assumes that the relative gradient $A_1$ for the $Y(X_1)$ remains roughly constant for small offsets along model parameter $X_2$ and vice versa. This approximation is certainly good for moderate offsets $X_1-X_{1,r}$ and $X_2-X_{2,r}$ from the reference model. 

It is clear from Table~\ref{tab:diagnostic} that the integrated intensity $\overline{I}/I_\star$, azimuthal polarization $\overline{Q}_\varphi/I_\star$, and fractional polarization $\langle p_\varphi \rangle$ depend on several model parameters. This creates ambiguities between the six model parameters involved and it is not straightforward to derive or constrain individual disk or dust parameters from the scattered radiation.

The following subsections discuss the different radiation parameters that are particularly useful for constraining the dust scattering parameters $\omega$, $g,$ and $p_{\rm max}$. Powerful diagnostic information can be obtained from radiation parameters that do not depend on disk wall height $\alpha$ or slope $\chi,$ because these are often poorly known geometric parameters of the disk. 

\subsection{Wavelength dependencies for the scattering parameters}
The most basic observational parameters for the scattered radiation from circumstellar disk are the disk-integrated intensity $\overline{I}/I_\star$ and the azimuthal polarization $\overline{Q}_\varphi/I_\star$. However, both parameters depend strongly on the disk geometry and the dust scattering parameters and it is difficult to disentangle the complex dependencies. 

An important approach to constrain dust scattering parameters is therefore the determination of the wavelength dependence for parameters of the reflected radiation, like $(\overline{I}/I_\star)_\lambda$, $(\overline{Q}_\varphi/I_\star)_\lambda$, or $Q_{180}/Q_\varphi$, because it seems reasonable to assume that the disk scattering geometry is identical or close to identical for different wavelengths. 
This means that color effects in the reflected light of circumstellar disks are predominantly produced by the wavelength dependence of the dust scattering parameters $\omega(\lambda)$, $g(\lambda),$ and $p_{\rm max}(\lambda)$. Colors for a radiation parameter $Y$ can be quantified as differences 
$\Delta Y/\Delta \lambda = (Y_{\lambda 2}-Y_{\lambda 2})/(\lambda_ 2-\lambda_1)$
or ratios 
$\Lambda Y/\Lambda \lambda = (Y_{\lambda 2}/Y_{\lambda 2})/(\lambda_ 2/\lambda_1)$.
This requires high-quality measurements of the same source for different wavelengths and such measurements or estimates have been obtained for a few transition disks, such as TW Hya \citep{Debes13}, HD~100546 \citep{Mulders13}, HD~135344B \citep{Stolker16}, HD~142527 \citep{Hunziker21}, or HD~169142 \citep{Tschudi21}. 

\begin{figure}
    \centering
    \includegraphics[scale=0.5]{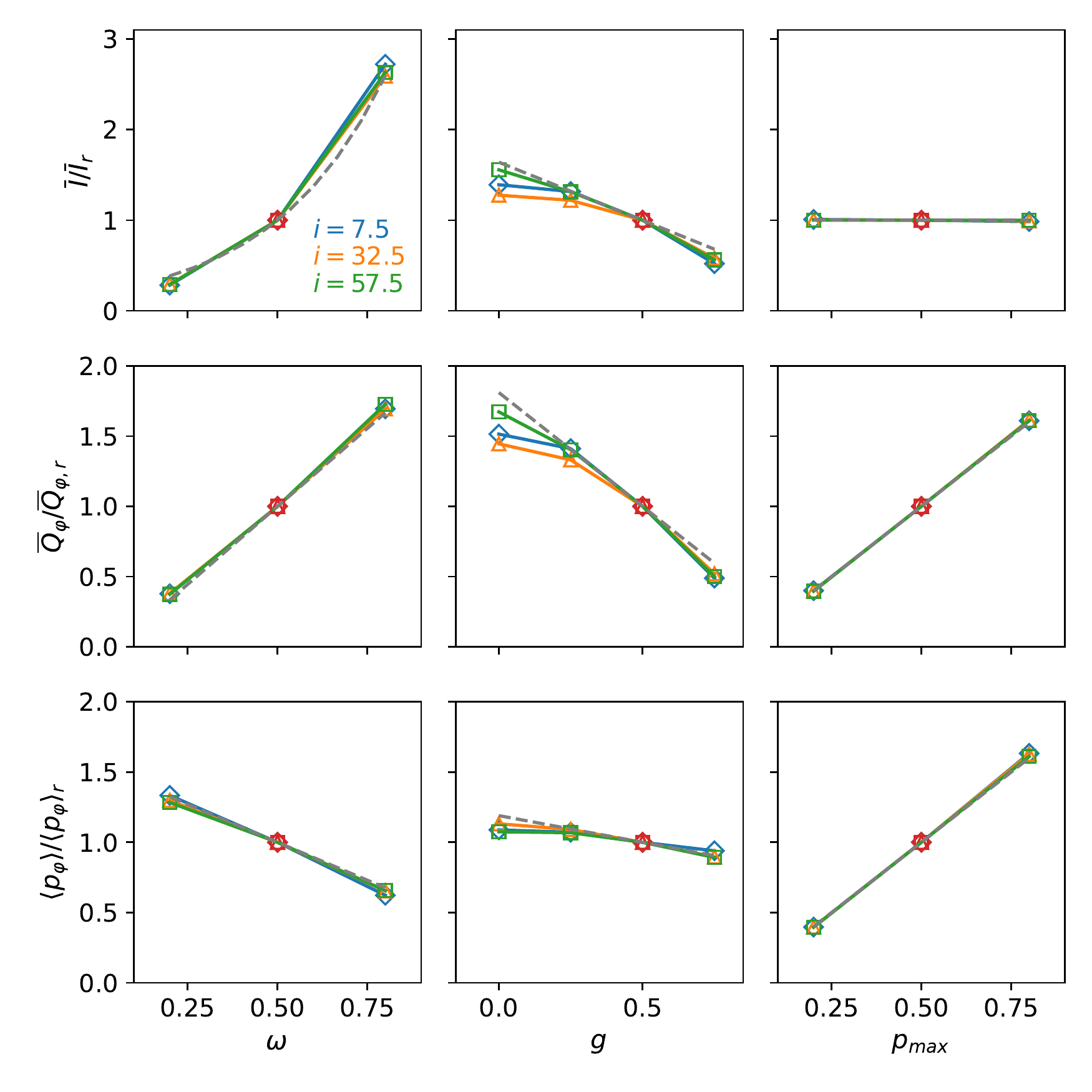}
    \caption{Relative change in disk integrated intensity $\overline{I}/\overline{I}_r$, azimuthal polarization $\overline{Q}/\overline{Q}_r$, and fractional polarization $\langle p_\varphi\rangle/\langle p_\varphi\rangle_r$  with respect to the reference model as a function of the dust scattering parameters $\omega$, $g,$ and $p_{\rm max}$. The dashed gray line is the fit from Table~\ref{tab:diagnostic}. In most panels the fitted lines coincide with the calculated results.}
    \label{fig:DiagRelative}
\end{figure}

The model results and the gradients $A_X(Y)$ given in Tables~\ref{tab:diskresults} and \ref{tab:diagnostic} provide the relative changes ($\Lambda (Y/Y_r)-1$) for the disk intensity $\overline{I}$, azimuthal polarization $\overline{Q}_\varphi$, and fractional polarization $\overline{p}_\varphi$ for parameters "near" the reference model as plotted in Fig.~\ref{fig:DiagRelative}. 
For disks with properties comparable to those of the reference model, an observed color like $\Lambda (\overline{Q}_\varphi/I_\star)$ for $\lambda_1$ and $\lambda_2$ can then be compared with these results to constrain the wavelength dependencies of the dust scattering parameters $\omega(\lambda)$, $g(\lambda)$, and $p_{\rm max}(\lambda)$ compatible with the measured color.

In several observational studies the transition disk was found to  reflect  radiation at longer wavelengths $\lambda>1~\mu$m more efficiently than that at shorter wavelengths $\lambda<1~\mu$m \citep[e.g.,][]{Mulders13,Stolker16,Hunziker21}. The diagrams in Fig.~\ref{fig:DiagRelative} indicate that this requires dust with either a larger single scattering albedo $\omega$ for longer wavelengths, a lower scattering asymmetry $g$, or a solution where the combined effect of both parameters can explain the observed color change.

\subsection{Fractional polarization} \label{section: fractional polarization}

\begin{figure}
    \centering
    \resizebox{\hsize}{!}{\includegraphics[]{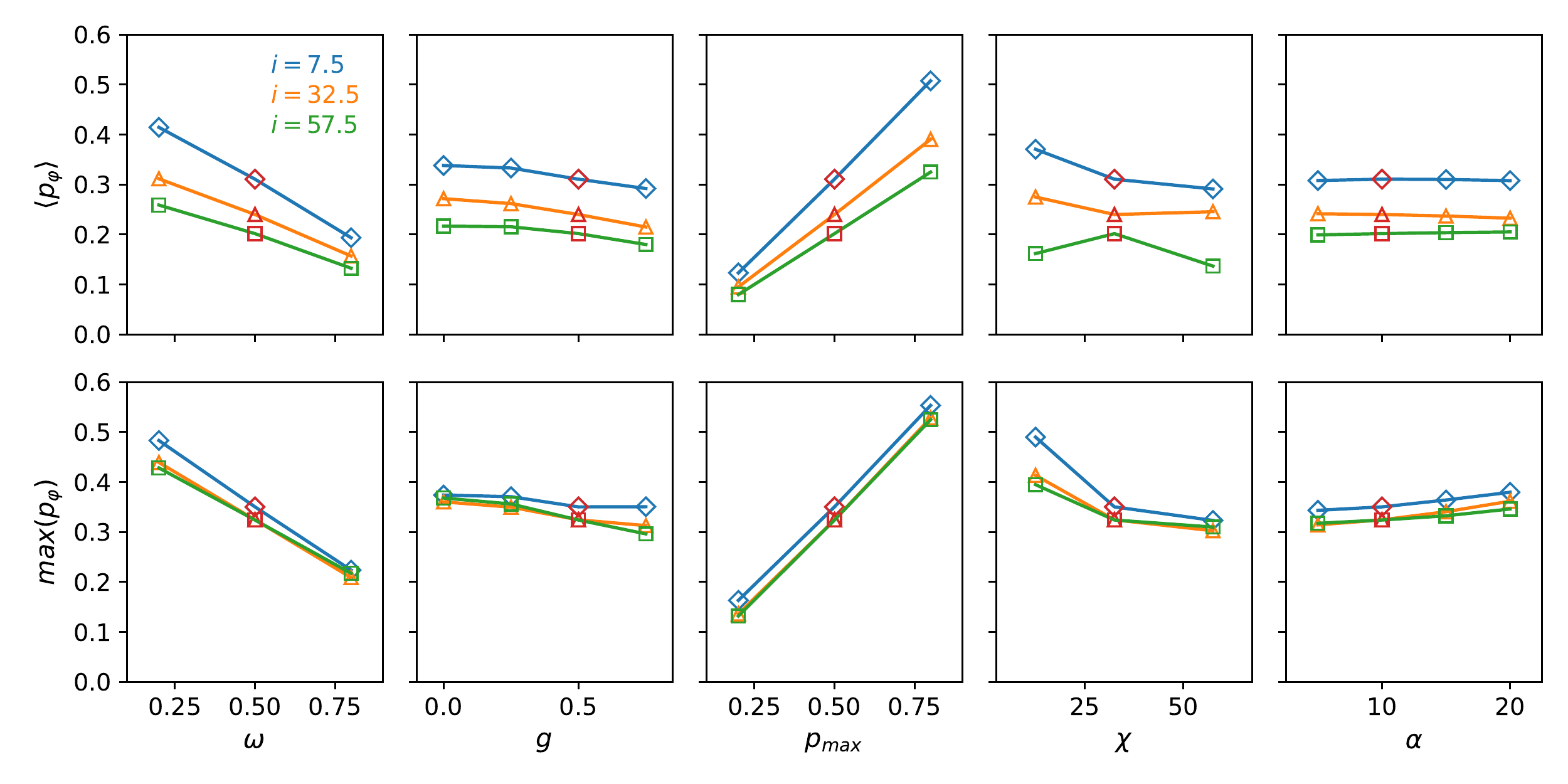}}
    \caption{Dependencies of the disk-averaged fractional polarization 
    $\langle{p}_\varphi\rangle$ and maximum fractional polarization ${\rm max}(p_\varphi)$ on the scattering parameters $\omega$, $g,$ and $p_{\rm max}$ and the geometric parameters $i$ and $\chi$ near the reference case.}
    \label{fig:Diagpol}
\end{figure}

The parameters for the fractional polarization $\langle {p}_\varphi \rangle$, $\langle {p}_Q \rangle$, or ${\rm max}(p_\varphi)$ are relative quantities between the polarization and intensity of the scattered radiation and therefore they are independent of the size of the scattering region; for our disk model they are almost independent of the  wall height $\alpha$. Figure~\ref{fig:Diagpol} shows the dependencies for the azimuthal polarization $\langle {p}_\varphi \rangle$ according to the results in Table~\ref{tab:diskresults} with dependencies as described by the gradients given in Table~\ref{tab:diagnostic}.

The fractional polarization $\langle p_{\varphi}\rangle$ shows strong dependencies on $\omega$ and $p_{\rm max}$, but is independent of $\alpha$, and is therefore ideal for constraining these two dust scattering parameters. 
Important for the fractional polarization is the scattering angle $\theta_S$ and therefore $\langle p_\varphi \rangle$ depends on the inclination $i$, which is fortunately often well known for disks. The wall slope $\chi$ has little but still some  influence on $\langle p_\varphi \rangle$ because it influences the relative contributions of scatterings with different $\theta_S$ to the flux-weighted average $\langle p_\varphi \rangle$. 

The dependence of the fractional polarization on $i$ and $\chi$ can be further reduced if only the maximum fractional polarization of a disk max($p_\varphi$) is considered as shown in the lower panel row of Fig.~\ref{fig:Diagpol}. This parameter selects the disk surface regions where the scattering angle is close to $\theta_S\approx 90^\circ$, which produces the highest possible fractional polarization for a surface with the given scattering parameters. Such regions exist for disks with high and low inclination, or steep and flat inner walls. 

An important aspect is that the determination of ${\rm max}(p_\varphi)$ requires only a measurement of $p_\varphi(x,y)\approx Q_\varphi(x,y)/I(x,y)$ for disk locations where the fractional polarization is close to the maximal fractional polarization. No full disk integrations for $Q_\varphi$ and $I$ are required. Therefore, the ${\rm max}(p_\varphi)$ may be obtained for disks where reliable $\langle p_\varphi \rangle$ measurements for an estimate of scattering parameters are not possible, because the disk deviates strongly from axisymmetry, because some disk regions are shadowed by an inner disk, or because it is observationally difficult to measure the disk intensity $I$ or polarized intensity $Q_\varphi$ well for all disk regions for the determination of integrated quantities $\overline{I}$ and $\overline{Q}_\phi$.
For inclined disks, ${\rm max}(p_\varphi)$ is predicted for locations near the major axis (see Fig.~\ref{fig:05-2d}) which are well separated from the central star and therefore favorable for accurate measurements. 

\begin{figure}
    \centering
    \resizebox{\hsize}{!}{\includegraphics[]{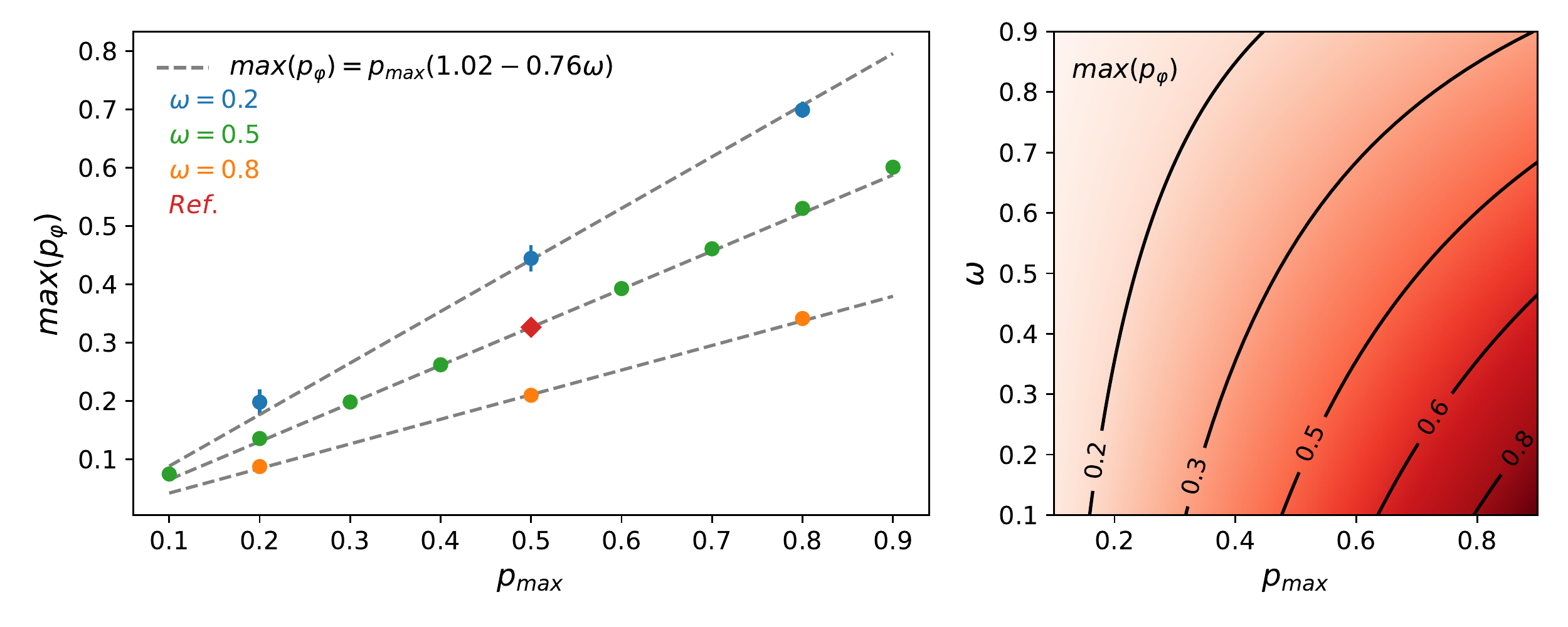}}
    \caption{Maximum fractional polarization $p_{\varphi}$ as a function of scattering parameter $p_{max}$ and $\omega$. The gray dashed lines represent $\omega=0.2, 0.5,$ and $0.8$ for the reference parameters $X_r:\,g=0.5,\,i=32.5^\circ,\,\chi=32.5^\circ$. The points near the reference case illustrate $p_{max}$ values for $X_{r-1}$ and $X_{r+1}$.}
    \label{fig:max_p}
\end{figure}

The parameter dependencies for ${\rm max}(p_\varphi)$ can be quite accurately approximated by two linear dependencies on $p_{\rm max}$ and $\omega $ while ignoring minor effects on $g$, $i$ , $\chi,$ and $\alpha$:
\begin{equation}
    \frac{{\rm max}(p_\varphi)}{{\rm max}(p_\varphi)_r} \approx \Big( 1 + A_{\rm pmax}\dfrac{p_{\rm max} - p_{\rm max,r}}{p_{\rm max, r}} \Big)\Big(1+ A_\omega\dfrac{\omega-\omega_r}{\omega_r}\Big)
\end{equation}
with the gradients $A_{p{\rm max}} = 1.00 \pm 0.03$ and $A_\omega = -0.59 \pm 0.01$ based on the values from Table~\ref{tab:diskresults}. Because of the strict proportionality ${\rm max}(p_\varphi)\propto p_{\rm max}$, and using the reference values for ${\rm max}(p_\varphi)_r$, $p_{\rm max,r}$, and $\omega_r$, we obtain an  even more simple dependence: 
\begin{equation}
    {\rm max}(p_\varphi) \approx p_{\rm max}(1.02 - 0.76\cdot\omega)\,.
\end{equation}
This relation is compared in Fig.~\ref{fig:max_p} with model results and the agreement is excellent for a large range of $p_{\rm max}$ and $\omega$ parameters. Thus, the fractional polarization $\langle p_\varphi \rangle$ or ${\rm max}(p_\varphi)$ of the scattered radiation from disks provides strong constraints on the possible value combination for the single scattering albedo $\omega$ and the maximum scattering polarization $p_{\rm max}$. If one of these parameters can be constrained with additional observations or theoretical considerations, then the other parameter is also well defined.

\subsection{Azimuthal distribution of the scattered polarization}
\label{section: azimuthal distribution}
\begin{figure}
    \centering
    \resizebox{\hsize}{!}{\includegraphics{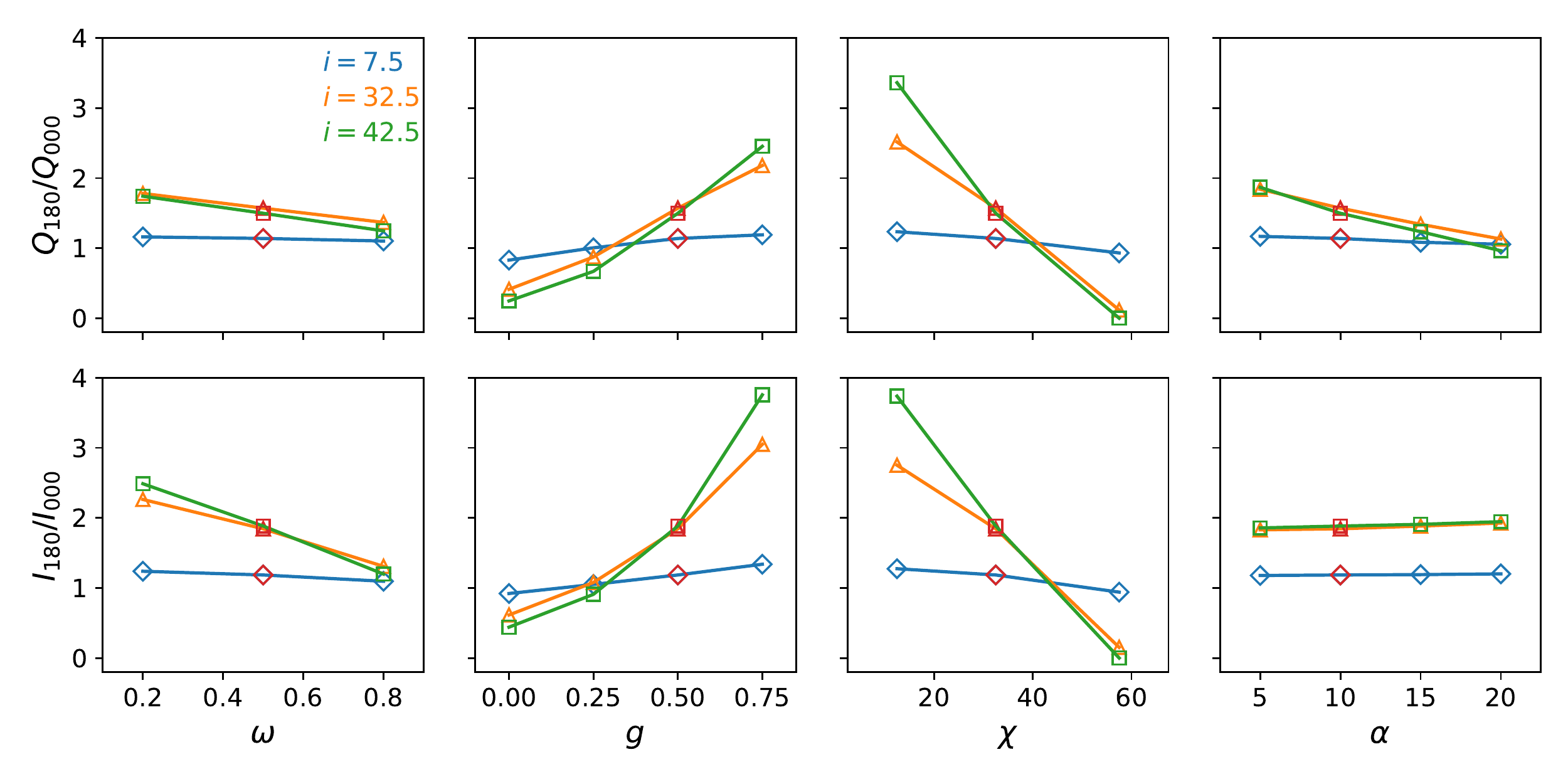}}
    \caption{Quadrant polarization ratios $Q_{180}/Q_{000}$ and $I_{180}/I_{000}$ as a function of $\omega$, $g$, $\chi,$ and $\alpha$. The largest inclination $i$ is restricted to $42.5^\circ$ to avoid effects related to the disappearance of the inner wall on the front side.}
    \label{fig:DiagAzi}
\end{figure}

\begin{figure}
    \centering
    \resizebox{\hsize}{!}{\includegraphics{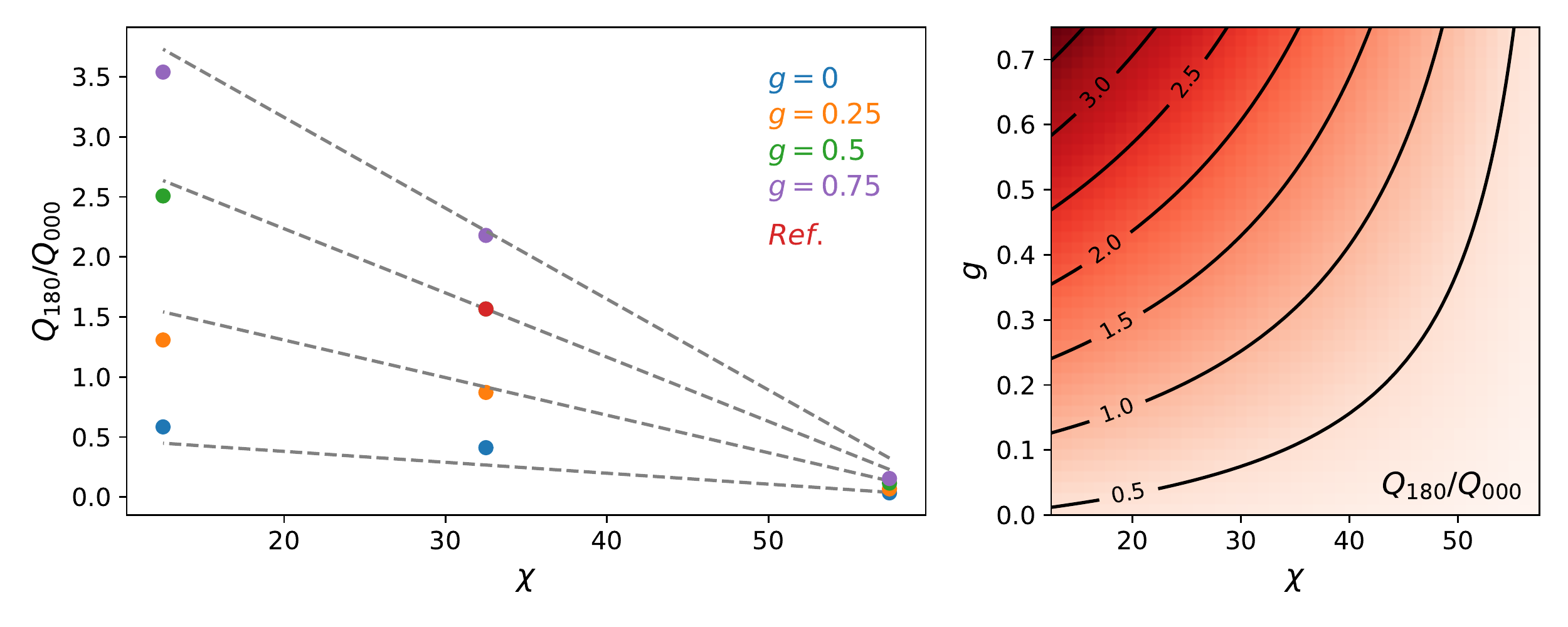}}
    \caption{Quadrant polarization ratios $Q_{180}/Q_{000}$ as a function of asymmetry $g$ and disk wall slope $\chi$. The gray dashed lines represent $g=0, 0.25, 0.5,$ and $0.75$ for the reference parameters $X_r: \omega=0.5, p_{max}=0.5, i=32.5^\circ$.}
    \label{fig:DiagQuadq}
\end{figure}

The quadrant polarization parameters given in Table~\ref{tab:diskresults} describe the azimuthal dependence of the polarized radiation. A combination of different quadrant parameters could be investigated \citep{Schmid21}, but we restrict our discussion to the front-to-back ratio $Q_{180}/Q_{000}$ and the corresponding intensity ratio $I_{180}/I_{000}$ shown in Fig. ~\ref{fig:DiagAzi} for different $i$ and as functions of $\omega$, $g$, $\chi$, and $\alpha$. The ratios are almost independent of $p_{\rm max}$ and therefore this is not shown. Nearly face-on disks ($i=7.5^\circ$) are close to axisymmetric and the ratios are close to one for all model parameters. The front--backside asymmetries become obvious for moderately inclined disks ($i=32.5^\circ$ and $42.5^\circ$) and they are rather similar for the polarization ratio $Q_{180}/Q_{000}$ and the intensity ratio $I_{180}/I_{000}$.
The values for larger inclinations are not shown, because the front side 
disappears and the $Q_{180}$ and $I_{180}$ values go to zero. For the wall
slope $\chi=57.5^\circ$ this happens already for $i=32.5^\circ$ and $42.5^\circ$.

For inclined disks, the dependence of the front-to-back ratios is particularly strong with respect to the scattering asymmetry parameter $g$ and the disk slope $\chi$ and the ratio $Q_{180}/Q_{000}$ can be used as diagnostics.
For example, for $i=32^\circ$, $\omega=0.5,$ and $\alpha=10^\circ$, the dependence can be fitted with the linear relation:
\begin{equation}
    \dfrac{Q_{180}/Q_{000}}{(Q_{180}/Q_{000})_r} \approx \Big( 1+ A_g\dfrac{g-g_r}{g_r}\Big) \Big( 1+ A_\chi\dfrac{\chi-\chi_r}{\chi_r}\Big)
\end{equation}
where $A_g = 0.83\pm 0.04$ and $A_\chi = -1.11 \pm 0.08$ and the relation is plotted in Fig.~\ref{fig:DiagQuadq}.
Much brighter front sides, as seen in many moderately inclined disks, indicate that the dust favors forward scattering $g>0.25$ and the illuminated inner disk wall on the front side is flat enough to be well visible. If the back side is brighter, then the illuminated wall on the front side becomes hardly visible. 
Other quadrant polarization ratios, such as $Q_{090}/Q_{000}$, can also be used to determine $g$ and $\chi$ using the numerical results presented in Table~\ref{tab:diskresults}. Results can be verified with the investigation of different quadrant ratios and additional effects, such as deviations of the disk geometry from rotational symmetry, the presence of shadows from hot dust near the star, or significant contributions from diffuse dust located above the disk surface, and other effects may be identified.

\subsection{Comparison of the scattered and thermal disk emission}

The spectral energy distributions of transition disk systems show  a well-defined far-IR emission excess which is explained by the thermal emission from the dust in the disk, which is heated mainly by the absorption of stellar radiation. It was further shown by \cite{Garufi17} that for transition disks (Meeus group I disks) there exists  a strong correlation between this far-IR excess and the relative polarized intensity $F({\rm fIR})/F_\star \approx 35 \times Q_\varphi/I_\star$ which are both quantities normalized to the emission of the stars. Therefore, the comparison between the scattered intensities $I/I_\star$ or polarized intensity $Q_\varphi/I_\star$ and the IR excess $F_{\rm IR}/F_\star$ can be used to obtain rough constraints on  the disk surface reflectivity and therefore also the single scattering albedo $\omega$.

For our scattering models, we can assume that the reflecting dust of the inner disk wall is also strongly absorbing and as discussed for a flat surface, the absorbed radiation ${\cal{K}}$ is directly related to the integrated surface reflectivity or surface albedo ${\cal{R}}$ according to ${\cal{K}} = 1- {\cal{R}}$. The absorbed radiation integrated over all wavelengths heats the dust which produces thermal emission measurable in the spectral energy distribution of the system as IR excess $F_{\rm IR}/F_\star$. 
 
We use the double ratios
\begin{equation}
    \Lambda_I = \frac{\langle \overline{I}(i)/I_\star \rangle}{F_{\rm IR}(i)/F_\star}\quad {\rm and} \quad
    \Lambda_p = \frac{\langle \overline{Q}_\varphi(i)/I_\star \rangle}{F_{\rm IR}(i)/F_\star}\,,
\label{Eq:albedos}
\end{equation}
which can be considered as an apparent disk scattering albedo and polarized albedo \citep[according to][]{Tschudi21}, and one can expect that these quantities depend strongly on the single scattering albedo $\omega$ of the dust.

In this approach, the relationship between measurable radiation parameters and the dust scattering parameters is less close. A major complication is that the IR excess $F_{\rm IR}(i)/F_\star$ provides a "bolometric" measure of the absorbed radiation energy integrated over all wavelengths, while the scattered intensity $\overline{I}(i)/I_\star$ is typically only measured for one or a few wavelengths. 
Thus, one needs to derive or predict the wavelength dependence of the scattered radiation $\langle \overline{I}(i)/I_\star \rangle$ or $\langle \overline{Q}_\varphi(i)/I_\star \rangle$ and estimate the entire energy input contributing to the far-IR excess. This includes the stellar radiation, energetic radiation from the disk accretion, near-IR radiation from hot dust located near the star, the galactic radiation field, and cosmic rays.
In most cases, the absorption of stellar radiation will be the dominant source of energy for the far-IR emission of transition disks.

In this work, as a first approximation, we neglect   the wavelength dependencies for the scattering and the absorption by the disk. We assume that results from such simple models will probably introduce uncertainties of a factor of a few for an estimate on $\omega$ from the apparent disk albedo $\Lambda_I$ or polarized albedo $\Lambda_p$ determinations. First estimates by \cite{Garufi17} for a large sample or the detailed determination by \cite{Tschudi21} yield $\Lambda_p\approx 0.03$, which is compatible with the $\omega=0.5$ of the reference model. 
Most likely, the apparent disk albedo determination and the corresponding constraints on $\omega$ could be improved with detailed investigations of wavelength-dependent effects and the derivation of "bolometric" correction factors, but this is beyond the scope of this paper.
In this section, we provide  only a simplified assessment of how the apparent albedo is expected to depend on the parameters of our disk models.

\begin{figure}[]
    \centering
    \includegraphics[scale=0.47]{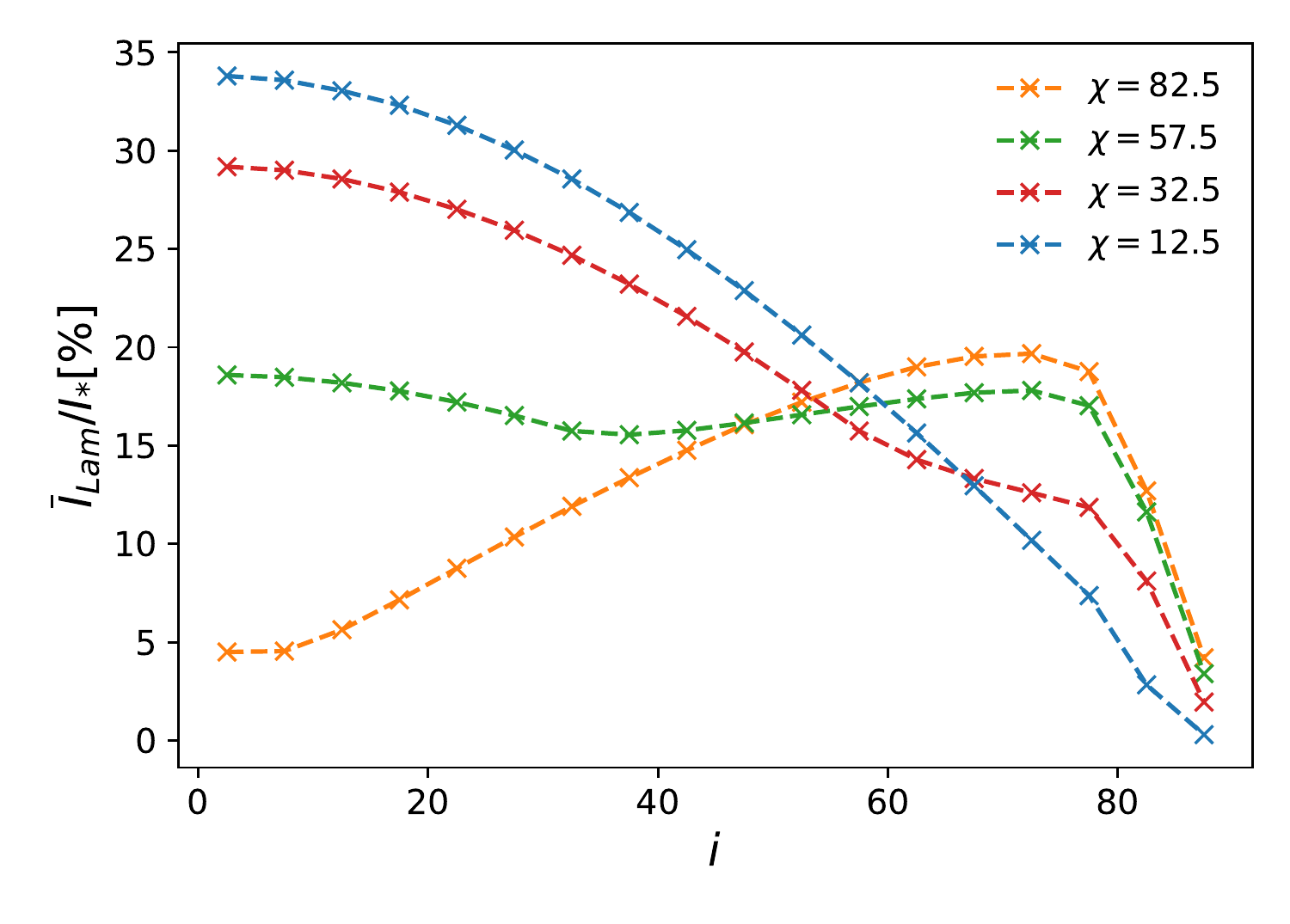}
    \caption{Reflected emission $\overline{I}_{\rm Lam}(i,\chi)/I_\star[\%]$    for a scattering disk with a perfect Lambert surface ${\cal{R}}=1$ and a wall height of $\alpha=10^\circ$.}
    \label{fig:i_lam_chi}
\end{figure}

\begin{table}[]
    \centering
    \caption{Reflected emission $\overline{I}_{\rm Lam}(i,\chi)/I_\star[\%]$ for a perfect Lambert disk surface
    }
    \begin{tabular}{c c c c c c}
    \hline\hline
    $i$ & $\chi=$  & $12.5^{\circ}$ & $32.5^{\circ}$ &  $57.5^{\circ}$ & $82.5^{\circ}$\\
    \hline
    $7.5^{\circ}$ & & 33.6 & 29.1 & 18.5 & 4.6\\
    $32.5^{\circ}$& & 28.6 & 24.7 & 15.8 & 11.9\\
    $57.5^{\circ}$& & 18.2 & 15.8 & 17.0 & 18.2\\
    $77.5^{\circ}$& & 7.4 & 11.9 & 17.1 & 18.8\\
    \hline
    \end{tabular}
    \tablefoot{Numerical results of the reflected emission at different inclinations $i$ for disks with different wall slopes $\chi$ and fixed wall height $\alpha=10^\circ$, by a Lambert surface with ${\cal{R}}=1$, as in Fig.~\ref{fig:i_lam_chi}.}
    \label{lambert_ref}
\end{table}

\subsubsection{Derivation of the IR excess}

There exist very sophisticated models for the thermal emission of the dust in circumstellar disks which calculate the detailed spectral energy distribution based on hydrodynamical disk geometries and considering thermo-chemical processes of the gas and the dust \citep[e.g.,][]{Woitke16}.
These models are characterized by ten or more parameters and they are not appropriate for our simple models. We use a rough approximation for the integrated IR excess $\overline{F}_{\rm IR}/F_\star$ which is compatible with our six-parameter transition disk models.

Therefore,  for the calculation of the IR excess of the disk
$\overline{F}_{\rm IR}/F_\star$ , we adopt the same disk geometry as for the scattered radiation, assuming that the disk is also optically thick for infrared light similar to the model of \citet{DAlessio05}. The emitted infrared energy from point $S(r,h)$ is assumed to be equal to the energy of absorbed photons at $S(r,h)$ as calculated by the scattering models and each point on the disk $S(r,h)$ emits in the same way as a Lambert surface $I_{\rm Lam}(\theta) \propto \cos \theta$ (see Fig~\ref{fig:05model}). We only consider the total energy of the emitted thermal radiation from the transition disk and disregard the spectral energy distribution. 

The thermal emission for this kind of model can be calculated with a scattering model which assumes that all stellar light is absorbed and converted into IR thermal emission. In this case, the thermal emission is equal to the results of the scattering model for a disk with a perfectly reflecting Lambert surface,
\begin{equation}
    \frac{F_{\rm IR}(\omega=0)}{F_\star} =\frac{\overline{I}_{\rm Lam}({\cal{R}}=1)}{I_\star} \,.
\end{equation}
The reflected intensity of such a disk $\overline{I}_{\rm Lam}(i,\chi,\alpha)/I_\star$ depends on the disk geometry $i$, $\alpha$, and $\chi$ only. 
Figure~\ref{fig:i_lam_chi} shows $\overline{I}_{\rm Lam}(i)/I_\star$ for $\alpha=10^\circ$ as a function of inclination for different wall slopes $\chi$ and  assuming perfect reflection ${\cal{R}}=1,$ and Table~\ref{lambert_ref} gives numerical values for four inclinations. 
The Lambertian disk $\overline{I}_{\rm Lam}(i)/I_\star$ shows strong differences in the $i$-dependence compared to the $g=0.5$ multiple-scattering model $\overline{I}(i,\chi)/I_\star$ in Fig.~\ref{fig:integrate-param} because the latter produces a strong forward scattering peak for inclined disks with small $\chi=12.5^\circ$ and $32.5^\circ$.

The disk with $\alpha=10^\circ$ intercepts 17.6~\% of the stellar light and the perfect Lambert surface produces for small $\chi$ an intensity of more than $\overline{I}_{\rm Lam}/I_\star>25~\%$ for polar directions. 
It should be noted that the results in Fig.~\ref{fig:i_lam_chi} or Table~\ref{lambert_ref} consider only the radiation scattered at one location on the disk. Not considered is for example the reflected light from the back side which would contribute significantly for high ${\cal{R}}$ to the illumination of the inner wall on the front side for a high-albedo disk surface as can be seen by the drop in $\overline{I}_{\rm Lam}(i)/I_\star$ for $i>90^\circ-\alpha=80^\circ$. This neglected second-order effect would add some light to the reflected light $\overline{I}_{\rm Lam}/I_\star$ for smaller $i$. This effect is very small for low-albedo dust.
The dependence of the thermal radiation on the disk height $\alpha$ can be described approximately by combining the results in Fig.~\ref{fig:i_lam_chi} and Table~\ref{lambert_ref} with the scaling factor $\sin \alpha/\sin \alpha_r$ (where $\alpha_r=10^\circ$), which accounts for the amount of stellar radiation intercepted by the disk. 

 For the calculation of the thermal emission, we adopt the description for the scattered light for a disk with a perfect Lambert surface $\overline{I}_{\rm Lam}(i,\chi,\alpha)/I_\star$ and simply scale with ${\cal{K}}= 1- {\cal{R}}$ for the absorbed radiation: 
\begin{equation}
    \frac{F_{\rm IR}(i,\chi,\alpha,\omega,g)}{F_\star}\approx
    \frac{\overline{I}_{\rm Lam}(i,\chi,\alpha_r)}{I_\star}\,
    \frac{\sin\alpha}{\sin \alpha_r}\,(1-{\cal{R}}(\chi,\omega,g))\,.
\label{Eq:Fir}
\end{equation}
The disk-integrated reflectivity ${\cal{R}}(\chi,\omega,g)$ is a function of the scattering parameters $\omega$ and $g$, and also depends on the wall slope $\chi$ as given in Table~\ref{tab:emiss} for 36 parameter combinations. 
The ${\cal{R}}$-values vary by up to a factor of 100 between ${\cal{R}}\approx 0.43$ for high-scattering albedo $\omega=0.8$, close to isotropic scattering $g=0.25$, and flat wall slope $\chi=12.5^\circ$ and ${\cal{R}}\approx 0.04$ for low albedo $\omega=0.8$, strong forward scattering $g=0.75$, and steep wall slopes $\chi=82.5^\circ$. The impact of the model parameters is much less dramatic for the absorption ${\cal{K}} = 1- {\cal{R}}$ and the thermal emission, because the extreme values are ${\cal{K}}\approx 0.57$ and 1.0 which differ by less than a factor of two. 

It is obvious that most of the radiation interacting with the disk is absorbed for the parameters selected for our model grid and the values for the reference model are about ${\cal{R}}\approx 0.10$ for the reflectivity or ${\cal{K}}\approx 0.9$ for the absorption.This yields together with $\overline{I}_{\rm Lam}(i=32.5^\circ,\chi=32.5^\circ)/I_\star=24.7~\%$ for the reference model an IR excess which is consistent with the typical results $(F_{mid-IR}+F_{far-IR})/F_{\star} \approx  20\% \pm 10\%$ derived for a larger disk sample studied within the DIANA project \citep{Woitke19}.

\begin{table}
    \caption{Integrated surface reflectivities ${\cal{R}}(\chi,\omega,g)$
    } 
    \label{tab:emiss}
    \begin{tabular}{c c c c c c }
    \hline\hline
        \noalign{\smallskip}
    $\omega$ & $g$ & $\chi= 12.5^\circ$& $32.5^\circ$ & $57.5^\circ$ & $82.5^\circ$ \\
        \noalign{\smallskip}
    \hline
    \noalign{\smallskip}
    0.8 & 0.25 & 0.4287 &  0.3095 & 0.2312 & 0.1951 \\ 
    0.8 & 0.5  & 0.4011 &  0.2575 & 0.1741 & 0.1393 \\ 
    0.8 & 0.75 & 0.3408 &  0.1706 & 0.0969 & 0.0712 \\ 
    \noalign{\smallskip}
    0.5 & 0.25 & 0.2058 &  0.1277 & 0.0859 & 0.0690 \\
    0.5 & 0.5  & 0.1885 &  {\it 0.0968} & 0.0558 & 0.0415 \\
    0.5 & 0.75 & 0.1497 &  0.0537 & 0.0252 & 0.0172 \\
    \noalign{\smallskip}
    0.2 & 0.25 & 0.0696 &  0.0391 & 0.0245 & 0.0190 \\
    0.2 & 0.5  & 0.0630 &  0.0278 & 0.0145 & 0.0103 \\
    0.2 & 0.75 & 0.0476 &  0.0139 & 0.006 & 0.004 \\
    \noalign{\smallskip}
    \hline
    \end{tabular}
    \tablefoot{
    The values are derived for $p_{\rm max}=0.5$ and $\alpha=10^\circ$, but they are essentially independent of $p_{\rm max}$ and can be scaled with $\alpha$. The 0.5-model is given with italic fonts.
    }
\end{table}

\subsubsection{Apparent disk scattering albedos}
\begin{figure}
    \centering
    \resizebox{\hsize}{!}{\includegraphics{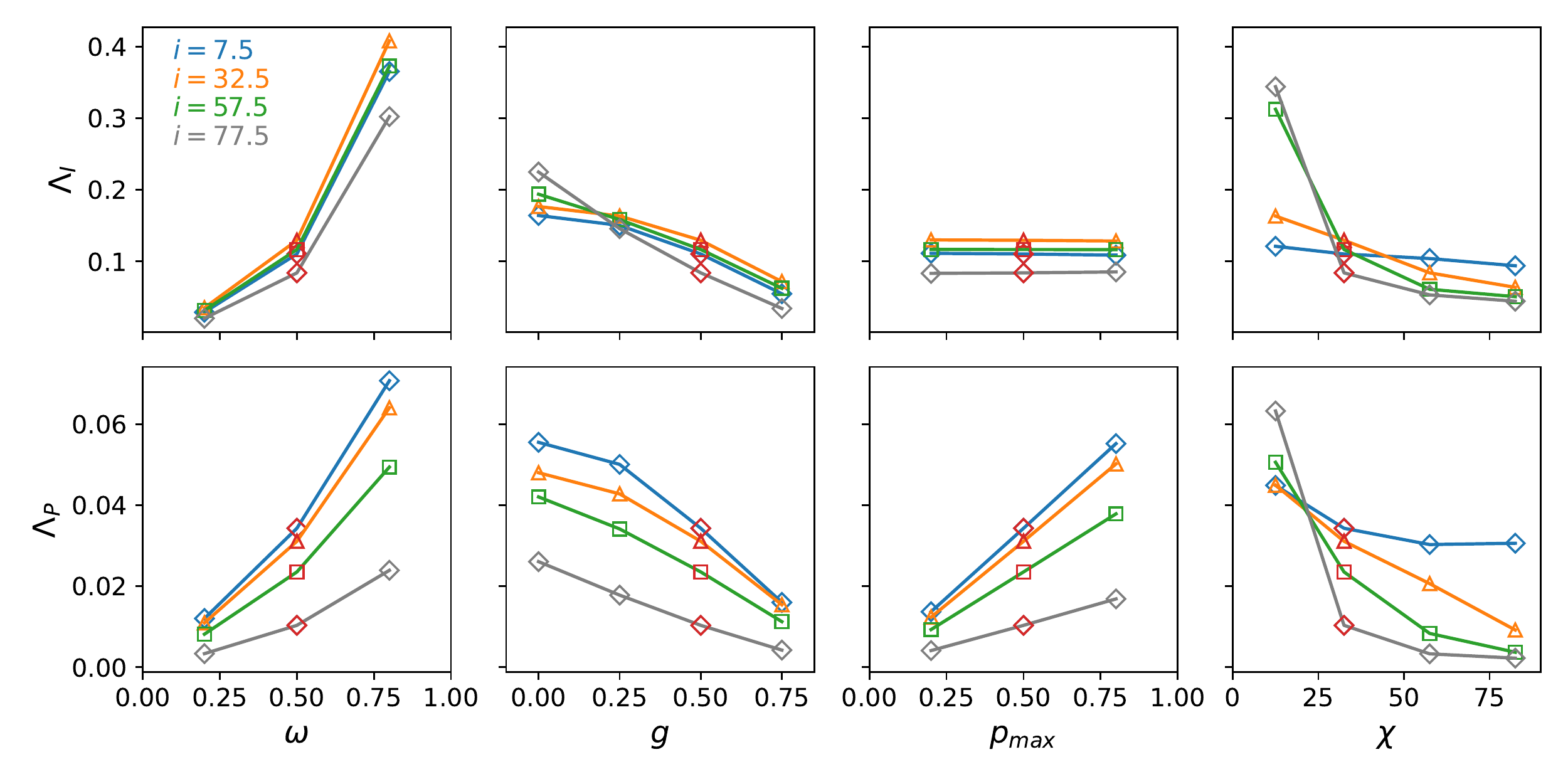}}
    \caption{Dependencies of the apparent disk scattering albedo $\Lambda_I$ and polarized albedo $\Lambda_P$ on the scattering parameters $\omega, g,$ and $p_{max}$ and geometric parameters $i$ and $\chi$. For parameters not indicated in the panel, the reference parameters $\omega,\,g,p_{\rm max}=0.5,$ and $\chi=32.5^\circ$ are used. 
    The values of the reference model for $i=7.5^\circ$, $32.5^\circ$, $57.5^\circ$ , and $77.5^\circ$ are plotted with open red symbols.}
    \label{fig:diag-sca-ir}
\end{figure}

The approximation for the IR excess $F_{\rm IR}/I_\star$, together with the disk integrated scattered intensity $\overline{I}/I_\star$ or polarized intensity $\overline{Q}_\varphi/I_\star$ , provides the apparent disk scattering albedo $\Lambda_I$ or polarized albedo $\Lambda_P$ defined in Eq.~(\ref{Eq:albedos}).
The albedo is a relative quantity which does not depend on the fraction of stellar photons interacting with the disk. Therefore, the dependence of the albedo on $\alpha$ is very small for our models. On the other side, the apparent disk albedo is a wavelength-averaged quantity and this introduces issues with the $\lambda$-dependence of the scattering parameters which should be known for the range covered by the stellar radiation.

The dependencies of the apparent disk albedos $\Lambda_I(i)$ and $\Lambda_p(i)$ on the scattering parameters $\omega$, $g$, $p_{\rm max}$, and the wall slope $\chi$ are shown in Fig.~\ref{fig:diag-sca-ir} for different inclination $i$. This shows that $\Lambda_I(i)$ correlates very strongly with the single scattering albedo $\omega$ and anti-correlates with the asymmetry parameter $g$ as expected from the relative dependencies $\overline{I}(\omega)/\overline{I}_r$ and $\overline{I}(g)/\overline{I}_r$ given in Fig.~\ref{fig:DiagRelative}.
However, for the albedos, the effect of higher reflectivies ${\cal{R}}$ for large $\omega$ and low $g$ is further enhanced by the reduced absorption and the overall disk albedo dependence is roughly $\Lambda \propto {\cal{R}}/(1-{\cal{R}})$.

For small disk inclinations $i=7.5^\circ$ or $i=32.5^\circ$, the dependence of $\Lambda_I$ on $i$ and the wall slope $\chi$ is very small and $\Lambda_I$ is therefore mainly a function of the two scattering parameters $\omega$ and $g$ and is very useful for constraining the dust properties for disks seen close to pole-on. This is of particular value because disk slopes and scale heights are difficult to derive for disks 
seen close to pole-on \citep[e.g.,][]{Rich21}.

The polarized scattered intensity $\overline{Q}_\varphi/I_\star$ is often easier to measure than $\overline{I}/I_\star$ with high-contrast disk observations \citep{Schmid21} and this provides the polarized scattering albedo $\Lambda_P$. The dependencies for $\Lambda_P$ are more complex because the polarization introduces additional strong effects related to $i$, $p_{\rm max}$.
Also, the dependence on the wall slope is larger for $\Lambda_P$ than for $\Lambda_I$, but still quite small for $i$ close to pole-on ($i=7.5^\circ,\, 32.5^\circ$). 

The typical $\Lambda_P$-values in Fig.~\ref{fig:diag-sca-ir} are in surprisingly good agreement with the study of \citet{Garufi17}, who,  for a sample of transition disks, find a strong correlation between IR excess and polarized light contrast of $F_{\rm IR}/F_\star \approx 35 \times \overline{Q}_\varphi/I_\star$ or $\Lambda_P \approx 0.029$ while the tabulated values (Tables~\ref{tab:diskresults},\ref{lambert_ref},\ref{tab:emiss}) for the reference model with $i=32.5^\circ$ yield $\Lambda_P\approx 0.0069/(0.247\,(1-0.0968))=0.031$ using Eqs.~(\ref{Eq:albedos}) and (\ref{Eq:Fir}). An agreement within a factor of a few indicates that the model calculations in this work accurately represent the overall budget between scattering and absorption of real transition disks. The almost perfect match between the reference model and the relation of \citet{Garufi17} is a coincidence when considering the model parameter dependencies of $\Lambda_P$ and the large dispersion and uncertainties in the compiled observational values.
The intensity albedo of the reference model is $\Lambda_I\approx \Lambda_p/\langle p_\varphi \rangle = 0.13$ and this value also matches well with the measured value for the inner wall (inner ring) of HD~169142 by \cite{Tschudi21}.

\section{Discussion and conclusions}
\label{Sect:discus}

This paper presents model simulations of the intensity and polarization of the scattered stellar light from the inner wall of transition disks. Results for many model parameter combinations are calculated in order to quantify the dependencies of the "observable" radiation parameters on the disk geometry and the dust scattering properties.
The results can be readily compared with measurements and used to constrain the properties of transition disks; in particular, the light scattering parameters of the dust. 

The presented transition disk models are very simple and fully characterized by only six parameters. This confers the important advantage that the dependencies on each parameter can be easily understood and investigated in detail and that the very basic relations describing the disk properties can be easily quantified. 

Indeed, real transition disks are much more complex than the selected description of an axisymmetric inner wall described by the parameters inclination $i$, wall height $\alpha,$ and slope $\chi,$ and the dust scattering parameters albedo $\omega$, asymmetry $g$, and polarization $p_{\rm max}$. In addition, the dust scattering in the disk is treated in the same way as multiple scattering on a semi-infinite (optically thick) plane parallel surface, assuming that the dust properties are the same everywhere in the disk.

When using the presented results as tools for the interpretation of observational data, it is important to be aware of these simplifications and to therefore compare model results only to measurements that do not depend strongly on system properties not considered in the simple model. In the following, we review our main simulation results and critically discuss  their application for the interpretation of observations.

\subsection{Dust scattering parameters}
The description of the dust scattering presented in Sect.~\ref{Sect:plpar} explains the most important properties of the radiation reflected from a multi-scattering dusty surface. We also find a strong correlation between the single scattering albedo $\omega$ and the total surface reflectivity ${\cal{R}}$, an anti-correlation between ${\cal{R}}$ and the forward scattering parameter $g$, a higher ${\cal{R}}$ for grazing incidence, and the highest fractional polarization for scattering angles around $\theta_S\approx 90^\circ$ , as described in the literature \citep[e.g.,][]{Chandrasekhar60,Abhyankar70,Dlugach74,vandeHulst80}. In addition, we add many results for the polarization of the reflected light for forward scattering dust which are specifically useful for the investigation of the reflection from optically thick circumstellar disks.
A particularly interesting result is the obtained maximum fractional polarization parameter $max(p(\theta_{pp}))$  for the reflected polarization, which shows a very strong dependence on the scattering albedo $\omega$ and polarization $p_{\rm max}$ but very little dependence on the scattering asymmetry $g$ and very moderate dependence on the incidence angle $\theta_0$.
This is very useful for tightly constraining  $\omega$ and $p_{\rm max}$
from fractional polarization measurements of disks.

Describing the scattering in the disk surface using only three dust parameters is certainly a simplification. However, which extension to the dust scattering model is required to achieve a significant improvement for the interpretation of disk observations  seems to be unclear. The adoption of a Rayleigh-scattering-like polarization phase function $p_{\rm sca}(\theta_S)$, which is symmetric with respect to $\theta_S=90^\circ$, could introduce a significant bias effect for the interpretation of the azimuthal distribution of the polarization signal in very well observed disks. The dust in debris disks seems to have a different $p_{\rm sca}(\theta_S)$ function \citep[e.g.,][]{Milli19,Arriaga20} and this possibility should also be investigated for transition disks.

For a given model, we adopt the same dust scattering parameters everywhere in the disk. This could be a strong over-simplification as dust particles in disks are known to have different size distributions at different locations and this is probably also the case for vertical or radial directions within the scattering surface of disks. For example, there could exist a significant vertical stratification with small particles with Rayleigh scattering-like properties in the uppermost layers and larger, more forward-scattering particles in deeper layers. This could produce reflected radiation  $I_\lambda(\theta,\phi)$, $p_\lambda(\theta,\phi)$ from a surface with wavelength dependencies which are not compatible with one homogeneously distributed dust-particle population. An investigation of such stratification effects would be very interesting, because they could produce signatures that might be clearly recognizable in high-quality data.

\subsection{Scattering in a plane parallel surface layer}

Our disk models adopt a plane-parallel surface geometry everywhere on the disk, where the scattered photon escapes from the same location as the incident photon. This is certainly a reasonable assumption for surface areas $S(r,h)$ in the "middle" of an extended flat wall (at $r_0<r<r_w$) where diffuse dust above the surface behaves like the uppermost layers of a plane-parallel surface. However, the plane-parallel surface assumption breaks down for optically thin, diffuse dust inside of $r_0$, for diffuse dust above the rim of the wall, and for illuminated dust above the outer flat surface of the disk, because scattered photons may escape for directions $\theta>90^\circ$ for which escape is not possible in the plane-parallel geometry adopted for our transition disk models. 
For example, diffuse dust above the disk at $h(r)/r>\tan\alpha$ will produce optically thin forward scattering and this could significantly enhance the intensity brightness of the front side for highly inclined disks. Therefore, the calculated diagnostic parameters for the intensity and polarized intensity ratios between the front and back side $I_{180}/I_{000}$ and $Q_{180}/Q_{000}$, respectively, should only be used very cautiously for the interpretation of strongly inclined disks $i>45^\circ$. 

The model results for highly inclined disks are still useful because they provide an estimate or upper limit for the expected scattered light produced by the optically thick disk wall. Any additional signal above this level could then be attributed to scattering occurring above the disk surface \citep{Glauser08,Villenave20} and this could be used to quantify the diffuse dust above the disk surface.

For small disk inclinations $i,$ or for the reflected radiation from the back wall, the forward scattering effect by the diffuse dust is much less important and the signal from the quasi plane-parallel disk wall surface is more visible and therefore dominant. For all these cases, the results from our simple model should provide a good description of the scattered intensity and polarization.

\subsection{Disk geometry}
For the geometry of the transition disk, the simulations adopt  a rotationally symmetric inner wall with a fixed wall slope $\chi$ and an angular wall height $\alpha$ and no variations in the azimuthal direction. 
Real disks are often much more complicated, showing a large diversity of structures for transition disks including spirals \citep[e.g.,][]{Garufi13,Benisty15}, eccentric rings \citep{Rameau12}, circular gaps \citep{Quanz13}, irregular small-scale structures \citep{Ginski21}, outer low-surface-brightness regions for extended flaring disks \citep{Ginski16}, and shadows from hot dust near the star \citep{Marino15,Pinilla18}. 
All these features are not captured by our simple models and the comparison of calculated results with such observations must consider the impact of the more complex geometry. One can study only disks that are close to centro-symmetric, but this would strongly restrict the available sample for the detailed investigation of the dust in circumstellar disks. For investigations of a larger sample, it is useful to study the uncertainties introduced by the not-so-well-defined disk geometry on the observational parameters and the derived dust properties.

Very significant discrepancies between observations and the results of the axisymmetric models can occur for disks with large-scale azimuthal structures which have a significant impact on the disk-integrated $\overline{I},\, \overline{Q}_\varphi$ or disk-averaged $\langle p_\varphi\rangle$ radiation parameters. Less critical are small-scale structures, such as small spiral features, localized shadows, or bumps in the disk surface because their effect will at least partly average out, meaning that comparisons of measured values $\overline{I}$ and $\overline{Q}_\varphi$ with axisymmetric models are still useful. Also, more complicated radial disk structures, such as a bright inner disk wall or faint but extended contributions from the outer regions of a flaring disk could at first approximation be compared with a model with one wall slope $\chi_{\rm model}$, intermediate between the steep slope of the inner wall $\chi_{\rm inner}>\chi_{\rm model}$ and the flat slope of the outer flaring disk surface $\chi_{\rm outer}<\chi_{\rm model}$. 

Azimuthal inhomogeneities of disks can be measured as brightness differences between the left and right quadrant polarization parameters with respect to the projected minor disk axis as
\begin{equation}
    \Delta = \frac{|Q_{090}|-|Q_{270}|}{|Q_{090}|+|Q_{270}|}\, , 
\end{equation}
and the equivalent for $|U_{045}|$ and $|U_{315}|$ or $|U_{135}|$ and $|U_{225}|$ \citep{Schmid21}. This gives a rough measure of the strength of the azimuthal feature from which one can estimate its impact on the disk-integrated or disk-averaged radiation parameters when comparing observations of non-symmetric disks to the results of the symmetric disk models. 

\subsection{Reflected radiation and diagnostic parameters}
The presented model calculations provide two-dimensional images for the scattered intensity $I(x,y)$, azimuthal polarization $Q_\varphi(x,y)$, and other radiation parameters. It is useful for the comparison with observations to deduce disk-integrated radiation parameters  from
these model images that are scaled to the stellar intensity, as in $\overline{I}(i)/I_\star$, $\overline{Q}_\phi(i)/I_\star$, the disk-averaged fractional polarization $\langle p_\varphi \rangle$, or quadrant polarization values $Q_{xxx}(i)/I_\star$ or $U_{xxx}(i)/I_\star$ which can also be deduced from observations without introducing significant ambiguities by the diversity of disk morphologies. We note that the measured values must be corrected for instrumental effects, in particular for the signal convolution with the instrumental PSF, which can introduce significant cancellation of the disk signal \citep{Schmid06, Avenhaus14, 
Avenhaus17, Heikamp19, Tschudi21}.
Unfortunately, available observational data with high accuracy and quantified uncertainties are still very limited and often only some of the radiation parameters can be determined for a given disk. Therefore, the derivation of disk parameters from the comparison with model results can be rather ambiguous. 
The model results presented in this work allow us to carry out a detailed investigation of the important parameter ambiguities involved in the interpretation of observations of transition disks and provide diagnostic relations that can be used to constrain  key model parameters for different types of observational data. For example, Fig.~\ref{fig:integrate-param} shows that, for a given inclination
$i$, $\overline{I}/I_\star$, $\overline{Q}_\varphi/I_\star$, and $\overline{Q}/I_\star$ all strongly  correlate with the angular wall height $\alpha$ and single scattering albedo $\omega$, and anti-correlate with wall slope $\chi$ and scattering asymmetry parameter $g$. In addition, the polarization parameters $\overline{Q}_\varphi/I_\star$ and $\overline{Q}/I_\star$ also correlated with $p_{\rm max}$. If only the scattered intensity $\overline{I}/I_\star$ or only the polarized intensity is measured then it is almost impossible to constrain individual disk parameters without additional information. Therefore, it is important to select and obtain more observational information and better diagnostic parameters with more diagnostic power. 

Strong ambiguities can be partly broken if both $\overline{I}/I_\star$ and $\overline{Q}_\varphi/I_\star$ or $\langle p_\varphi \rangle=\overline{Q}_\varphi/\overline{I}$ are known. The fractional polarization correlates with $p_{\rm max}$, but anti-correlates with $\omega$, while the dependencies are small for the parameter $g$, the wall slope, if $\chi\leq 57.5^\circ$, and zero for the wall height $\alpha$, while $i$ and the relative dependence $\langle p_\varphi(i) \rangle$ are typically well known. Essentially, the impact of uncertainties in the disk geometry on the resulting disk-averaged fractional polarization $\langle p_\varphi(i) \rangle$ or the maximum fractional polarization max($p_\varphi$) are strongly reduced, these parameters are mainly used to probe the dust-scattering parameters $\omega$, $p_{\rm max}$, and $g$. 

Similarly, the measurement of the wavelength dependence of the radiation parameters $(\overline{I}/I_\star)_\lambda$ or $(\overline{Q}_\varphi/I_\star)_\lambda$ does not depend strongly on the relatively poorly understood disk geometry because one can assume that the scattering geometry for the radiation with different wavelengths is very similar. This provides well-defined constraints on the wavelength dependence of the scattering parameters $\omega(\lambda)$, $g(\lambda),$ and $p_{\rm max}(\lambda),$ which can be compared with the prediction of different dust models.

More constraints on the dust-scattering parameters, in particular on the scattering asymmetry parameter, can be gained from derivations of the azimuthal dependence of the scattered radiation in inclined disks using for example ratios for the front-to-back intensity ratio $I_{180}/I_{000}$ or ratios of quadrant polarization parameters $Q_{180}/Q_{000}$. 
In addition, one may constrain the single scattering albedo $\omega$ from derivations of the effective disk albedos $\Lambda_I$ or $\Lambda_p$. These are less direct methods which might be strongly affected by the not-well-known scattering geometry. However, observations seem to confirm general trends, such as $g>0.25$ from the often-observed brighter front side in inclined disks, and $\omega\leq 0.75$ from the correlation between scattered and thermal emission from resolved transition disks $\Lambda_p\approx 0.03$ as derived by \citet{Garufi17}. These findings can at least exclude dust with a close-to-isotropic scattering phase function $g\nless 0.25$ and very high scattering albedos of $\omega\ngtr 0.9$. 

\subsection{Dust properties in transition disks}
Ultimately, we are interested in deriving dust properties for transition disks, such as the particle size distribution $n(a)$ where $a$ is the particle radius, the particle composition, and structure. These parameters are strongly linked to the used dust scattering parameters $\omega$, $g$, and $p_{\rm max}$ but there exist no well-established relations between dust properties and dust scattering parameters for propto-planetary disks. Detailed dust models are required for a translation between these two types of dust description. 

Mie theory describes scattering for homogeneous spherical particles and is useful for a first rough characterization.  For wavelengths much longer than the size of the particles $\lambda > 2\pi a$, this theory predicts a fast decrease in the absorption cross-section with wavelengths, roughly as $\kappa(\lambda)\propto \lambda^{-2}$ and an even faster decrease for the scattering $\sigma(\lambda)\propto \lambda^{-4}$. Therefore, the extinction cross-section is very small and mostly due to absorption. Therefore, according to Eq.~(\ref{EqOmega}),  $\omega$ is also small. Thus, for long wavelengths, the expected scattering parameters are $\omega\leq 0.5$, $g<0.5$, and $p_{\rm max}>0.5$ as expected for scattering of small particles in the Rayleigh regime such as interstellar dust at $\lambda>2~\mu$m \citep{Weingartner01}. For short wavelengths $\lambda \leq 2\pi \alpha$, the cross-sections $\sigma$ and $\kappa$ are roughly independent of $\lambda$ and are of the same order as the particle size $\approx \pi\,a^2$. The scattering parameters are therefore $\omega>0.5$, $g>0.5,$ and $p_{\rm max}<0.5,$ as expected from scattering and Fraunhofer diffraction by large particles similar to the scattering of cometary dust at $\lambda < 1~\mu$m \citep[e.g.,][]{Lasue09}.

Derived scattering parameters, for example for the disk around HD~142527, but also for other transition disks, seem to indicate a higher reflectivity and therefore higher $\omega$ for longer wavelengths, a roughly constant asymmetry parameter from $0.7~\mu$m to $2.2~\mu$m, and an increasing max$(p_\varphi)$ with wavelength \citep{Hunziker21}. This is not easily compatible with a simple dust model based on Mie theory. Usually, dust models based on Mie theory seem to be in qualitative agreement with observations of the thermal and scattered emission if only the reflected light at one wavelength is considered. However, we are not aware of a detailed study using Mie theory that compares models and observations of the thermal emission, the wavelength dependence of the reflected radiation, and the fractional polarization of a protoplanetary disk in detail. This would provide the dust scattering parameters $\sigma(\lambda,a)$, $\omega(\lambda,a)$, $g(\lambda,a)$, and $p_{\rm max}(\lambda,a)$ as functions of wavelength.
Alternative dust models based on large porous grains or aggregates $a>\lambda$ with a lot of small-scale structures could quite naturally explain $\omega>0.5$ at $2~\mu$m, a $g>0.5$ over a broad wavelength range, and still a quite high scattering polarization $p_{\rm max}\approx 0.3$ at $0.7~\mu$m. These produce a high scattering polarization by the interaction of the light with small substructures, and strong forward scattering and high scattering albedo because of the strong forward diffraction introduced by the large particles \citep[e.g.,][]{Kolokolova2010,Min2016,Tazaki18}. 
Model calculations for disks with such models are presented by \citet{Tazaki19}, who investigate in detail the wavelength dependence and the fractional polarization of the disk radiation parameters $(I/I_\star)_\lambda$ and $(\langle p_\varphi \rangle)_\lambda$ on the properties of dust aggregates. 
Further calculations of this kind using detailed dust models for the derivations of the scattering parameters $\sigma(\lambda,a)$, $\omega(\lambda,a)$, $g(\lambda,a)$, and $p_{\rm max}(\lambda,a)$, or even the radiation parameters of disks considering the effects of the radiative transfer and the disk geometry are required in addition to accurate measurements in order to derive the nature of the dust particles in protoplanetary disks. 

There will most likely always exist different (competing) dust models that can explain the observationally derived parameters for the dust scattering of a circumstellar disk. Therefore, the characterization of the dust scattering phase matrix, or a simple description thereof with the three parameters $\omega$, $g$, and $p_{\rm max}$, seems to be a useful interface between dust models and disk observations to constrain the dust properties. The model results presented in this work show that there exist well-defined and relatively universal dependencies between these dust scattering parameters and the reflected radiation from disk that are measurable with high-resolution observations. It is important to collect quantitative measurements for more disks because additional constraints on the dust can be derived from circumstantial evidence based on systematic effects found for disks with different physical conditions, like dust temperatures, stellar irradiation, mass accretion rates, dust-to-gas ratios, and different complementary dust characteristics like silicon oxide(SiO) or polycyclic aromatic hydrocarbons (PAHs) emission features in the mid-IR. This will provide a more comprehensive picture of the particle properties and clarify the evolution of the dust in protoplanetary disks. 


\begin{acknowledgements}
We thank the anonymous referee for the valuable comments on the manuscript. This work has been carried out within the frame of the National Centre for Competence in Research PlanetS supported by the Swiss National Science Foundation (SNSF).
\end{acknowledgements}

\bibliographystyle{aa} 
\bibliography{example}

\begin{thebibliography}{83}
\expandafter\ifx\csname natexlab\endcsname\relax\def\natexlab#1{#1}\fi

\bibitem[{{Abhyankar} \& {Fymat}(1970)}]{Abhyankar70}
{Abhyankar}, K.~D. \& {Fymat}, A.~L. 1970, \aap, 4, 101

\bibitem[{{Abhyankar} \& {Fymat}(1971)}]{Abhyankar71}
{Abhyankar}, K.~D. \& {Fymat}, A.~L. 1971, \apjs, 23, 35

\bibitem[{{Andrews}(2015)}]{Andrews15}
{Andrews}, S.~M. 2015, \pasp, 127, 961

\bibitem[{{Ardila} {et~al.}(2007){Ardila}, {Golimowski}, {Krist}, {Clampin},
  {Ford}, \& {Illingworth}}]{Ardila07}
{Ardila}, D.~R., {Golimowski}, D.~A., {Krist}, J.~E., {et~al.} 2007, \apj, 665,
  512

\bibitem[{{Arriaga} {et~al.}(2020){Arriaga}, {Fitzgerald}, {Duch{\^e}ne},
  {Kalas}, {Millar-Blanchaer}, {Perrin}, {Chen}, {Mazoyer}, {Ammons}, {Bailey},
  {Barman}, {Bulger}, {Chilcote}, {Cotten}, {De Rosa}, {Doyon}, {Esposito},
  {Follette}, {Gerard}, {Goodsell}, {Graham}, {Greenbaum}, {Hibon}, {Hom},
  {Hung}, {Ingraham}, {Konopacky}, {Macintosh}, {Maire}, {Marchis}, {Marley},
  {Marois}, {Metchev}, {Nielsen}, {Oppenheimer}, {Palmer}, {Patience},
  {Poyneer}, {Pueyo}, {Rajan}, {Rameau}, {Rantakyr{\"o}}, {Ruffio},
  {Savransky}, {Schneider}, {Sivaramakrishnan}, {Song}, {Soummer}, {Thomas},
  {Wang}, {Ward-Duong}, \& {Wolff}}]{Arriaga20}
{Arriaga}, P., {Fitzgerald}, M.~P., {Duch{\^e}ne}, G., {et~al.} 2020, \aj, 160,
  79

\bibitem[{{Avenhaus} {et~al.}(2018){Avenhaus}, {Quanz}, {Garufi}, {Perez},
  {Casassus}, {Pinte}, {Bertrang}, {Caceres}, {Benisty}, \&
  {Dominik}}]{Avenhaus18}
{Avenhaus}, H., {Quanz}, S.~P., {Garufi}, A., {et~al.} 2018, \apj, 863, 44

\bibitem[{{Avenhaus} {et~al.}(2014){Avenhaus}, {Quanz}, {Meyer}, {Brittain},
  {Carr}, \& {Najita}}]{Avenhaus14}
{Avenhaus}, H., {Quanz}, S.~P., {Meyer}, M.~R., {et~al.} 2014, \apj, 790, 56

\bibitem[{{Avenhaus} {et~al.}(2017){Avenhaus}, {Quanz}, {Schmid}, {Dominik},
  {Stolker}, {Ginski}, {de Boer}, {Szul{\'a}gyi}, {Garufi}, {Zurlo},
  {Hagelberg}, {Benisty}, {Henning}, {M{\'e}nard}, {Meyer}, {Baruffolo},
  {Bazzon}, {Beuzit}, {Costille}, {Dohlen}, {Girard}, {Gisler}, {Kasper},
  {Mouillet}, {Pragt}, {Roelfsema}, {Salasnich}, \& {Sauvage}}]{Avenhaus17}
{Avenhaus}, H., {Quanz}, S.~P., {Schmid}, H.~M., {et~al.} 2017, \aj, 154, 33

\bibitem[{{Bastien} \& {Menard}(1988)}]{Bastien88}
{Bastien}, P. \& {Menard}, F. 1988, \apj, 326, 334

\bibitem[{{Bazzon} {et~al.}(2014){Bazzon}, {Schmid}, \& {Buenzli}}]{Bazzon14}
{Bazzon}, A., {Schmid}, H.~M., \& {Buenzli}, E. 2014, \aap, 572, A6

\bibitem[{{Benisty} {et~al.}(2015){Benisty}, {Juhasz}, {Boccaletti},
  {Avenhaus}, {Milli}, {Thalmann}, {Dominik}, {Pinilla}, {Buenzli}, {Pohl},
  {Beuzit}, {Birnstiel}, {de Boer}, {Bonnefoy}, {Chauvin}, {Christiaens},
  {Garufi}, {Grady}, {Henning}, {Huelamo}, {Isella}, {Langlois}, {M{\'e}nard},
  {Mouillet}, {Olofsson}, {Pantin}, {Pinte}, \& {Pueyo}}]{Benisty15}
{Benisty}, M., {Juhasz}, A., {Boccaletti}, A., {et~al.} 2015, \aap, 578, L6

\bibitem[{{Benisty} {et~al.}(2017){Benisty}, {Stolker}, {Pohl}, {de Boer},
  {Lesur}, {Dominik}, {Dullemond}, {Langlois}, {Min}, {Wagner}, {Henning},
  {Juhasz}, {Pinilla}, {Facchini}, {Apai}, {van Boekel}, {Garufi}, {Ginski},
  {M{\'e}nard}, {Pinte}, {Quanz}, {Zurlo}, {Boccaletti}, {Bonnefoy}, {Beuzit},
  {Chauvin}, {Cudel}, {Desidera}, {Feldt}, {Fontanive}, {Gratton}, {Kasper},
  {Lagrange}, {LeCoroller}, {Mouillet}, {Mesa}, {Sissa}, {Vigan}, {Antichi},
  {Buey}, {Fusco}, {Gisler}, {Llored}, {Magnard}, {Moeller-Nilsson}, {Pragt},
  {Roelfsema}, {Sauvage}, \& {Wildi}}]{Benisty17}
{Benisty}, M., {Stolker}, T., {Pohl}, A., {et~al.} 2017, \aap, 597, A42

\bibitem[{{Bertrang} {et~al.}(2018){Bertrang}, {Avenhaus}, {Casassus},
  {Montesinos}, {Kirchschlager}, {Perez}, {Cieza}, \& {Wolf}}]{Bertrang18}
{Bertrang}, G.~H.~M., {Avenhaus}, H., {Casassus}, S., {et~al.} 2018, \mnras,
  474, 5105

\bibitem[{{Calvet} {et~al.}(2005){Calvet}, {D'Alessio}, {Watson},
  {Franco-Hern{\'a}ndez}, {Furlan}, {Green}, {Sutter}, {Forrest}, {Hartmann},
  {Uchida}, {Keller}, {Sargent}, {Najita}, {Herter}, {Barry}, \&
  {Hall}}]{Calvet05}
{Calvet}, N., {D'Alessio}, P., {Watson}, D.~M., {et~al.} 2005, \apjl, 630, L185

\bibitem[{{Canovas} {et~al.}(2015){Canovas}, {M{\'e}nard}, {de Boer}, {Pinte},
  {Avenhaus}, \& {Schreiber}}]{Canovas15}
{Canovas}, H., {M{\'e}nard}, F., {de Boer}, J., {et~al.} 2015, \aap, 582, L7

\bibitem[{{Canovas} {et~al.}(2013){Canovas}, {M{\'e}nard}, {Hales},
  {Jord{\'a}n}, {Schreiber}, {Casassus}, {Gledhill}, \& {Pinte}}]{Canovas13}
{Canovas}, H., {M{\'e}nard}, F., {Hales}, A., {et~al.} 2013, \aap, 556, A123

\bibitem[{{Chandrasekhar}(1960)}]{Chandrasekhar60}
{Chandrasekhar}, S. 1960, {Radiative transfer}

\bibitem[{{D'Alessio} {et~al.}(2005){D'Alessio}, {Hartmann}, {Calvet},
  {Franco-Hern{\'a}ndez}, {Forrest}, {Sargent}, {Furlan}, {Uchida}, {Green},
  {Watson}, {Chen}, {Kemper}, {Sloan}, \& {Najita}}]{DAlessio05}
{D'Alessio}, P., {Hartmann}, L., {Calvet}, N., {et~al.} 2005, \apj, 621, 461

\bibitem[{{de Juan Ovelar} {et~al.}(2013){de Juan Ovelar}, {Min}, {Dominik},
  {Thalmann}, {Pinilla}, {Benisty}, \& {Birnstiel}}]{deJuanOvelar13}
{de Juan Ovelar}, M., {Min}, M., {Dominik}, C., {et~al.} 2013, \aap, 560, A111

\bibitem[{{Debes} {et~al.}(2013){Debes}, {Jang-Condell}, {Weinberger},
  {Roberge}, \& {Schneider}}]{Debes13}
{Debes}, J.~H., {Jang-Condell}, H., {Weinberger}, A.~J., {Roberge}, A., \&
  {Schneider}, G. 2013, \apj, 771, 45

\bibitem[{{Dlugach} \& {Yanovitskij}(1974)}]{Dlugach74}
{Dlugach}, J.~M. \& {Yanovitskij}, E.~G. 1974, \icarus, 22, 66

\bibitem[{{Dong} {et~al.}(2012){Dong}, {Rafikov}, {Zhu}, {Hartmann}, {Whitney},
  {Brandt}, {Muto}, {Hashimoto}, {Grady}, {Follette}, {Kuzuhara}, {Tanii},
  {Itoh}, {Thalmann}, {Wisniewski}, {Mayama}, {Janson}, {Abe}, {Brandner},
  {Carson}, {Egner}, {Feldt}, {Goto}, {Guyon}, {Hayano}, {Hayashi}, {Hayashi},
  {Henning}, {Hodapp}, {Honda}, {Inutsuka}, {Ishii}, {Iye}, {Kandori}, {Knapp},
  {Kudo}, {Kusakabe}, {Matsuo}, {McElwain}, {Miyama}, {Morino}, {Moro-Martin},
  {Nishimura}, {Pyo}, {Suto}, {Suzuki}, {Takami}, {Takato}, {Terada}, {Tomono},
  {Turner}, {Watanabe}, {Yamada}, {Takami}, {Usuda}, \& {Tamura}}]{Dong12}
{Dong}, R., {Rafikov}, R., {Zhu}, Z., {et~al.} 2012, \apj, 750, 161

\bibitem[{{Dong} {et~al.}(2016){Dong}, {Zhu}, {Fung}, {Rafikov}, {Chiang}, \&
  {Wagner}}]{Dong16}
{Dong}, R., {Zhu}, Z., {Fung}, J., {et~al.} 2016, \apjl, 816, L12

\bibitem[{{Draine}(2003)}]{Draine03}
{Draine}, B.~T. 2003, \apj, 598, 1017

\bibitem[{{Draine} \& {Lee}(1984)}]{Draine84}
{Draine}, B.~T. \& {Lee}, H.~M. 1984, \apj, 285, 89

\bibitem[{{Espaillat} {et~al.}(2014){Espaillat}, {Muzerolle}, {Najita},
  {Andrews}, {Zhu}, {Calvet}, {Kraus}, {Hashimoto}, {Kraus}, \&
  {D'Alessio}}]{Espaillat14}
{Espaillat}, C., {Muzerolle}, J., {Najita}, J., {et~al.} 2014, in Protostars
  and Planets VI, ed. H.~{Beuther}, R.~S. {Klessen}, C.~P. {Dullemond}, \&
  T.~{Henning}, 497

\bibitem[{{Fischer} {et~al.}(1994){Fischer}, {Henning}, \& {Yorke}}]{Fischer94}
{Fischer}, O., {Henning}, T., \& {Yorke}, H.~W. 1994, \aap, 284, 187

\bibitem[{{Garufi} {et~al.}(2017){Garufi}, {Benisty}, {Stolker}, {Avenhaus},
  {de Boer}, {Pohl}, {Quanz}, {Dominik}, {Ginski}, {Thalmann}, {van Boekel},
  {Boccaletti}, {Henning}, {Janson}, {Salter}, {Schmid}, {Sissa}, {Langlois},
  {Beuzit}, {Chauvin}, {Mouillet}, {Augereau}, {Bazzon}, {Biller}, {Bonnefoy},
  {Buenzli}, {Cheetham}, {Daemgen}, {Desidera}, {Engler}, {Feldt}, {Girard},
  {Gratton}, {Hagelberg}, {Keller}, {Keppler}, {Kenworthy}, {Kral}, {Lopez},
  {Maire}, {Menard}, {Mesa}, {Messina}, {Meyer}, {Milli}, {Min}, {Muller},
  {Olofsson}, {Pawellek}, {Pinte}, {Szulagyi}, {Vigan}, {Wahhaj}, {Waters}, \&
  {Zurlo}}]{Garufi17}
{Garufi}, A., {Benisty}, M., {Stolker}, T., {et~al.} 2017, The Messenger, 169,
  32

\bibitem[{{Garufi} {et~al.}(2013){Garufi}, {Quanz}, {Avenhaus}, {Buenzli},
  {Dominik}, {Meru}, {Meyer}, {Pinilla}, {Schmid}, \& {Wolf}}]{Garufi13}
{Garufi}, A., {Quanz}, S.~P., {Avenhaus}, H., {et~al.} 2013, \aap, 560, A105

\bibitem[{{Ginski} {et~al.}(2021){Ginski}, {Gratton}, {Bohn}, {Dominik},
  {Jorquera}, {Chauvin}, {Milli}, {Rodriguez}, {Benisty}, {Launhardt},
  {Mueller}, {Cugno}, {van Holstein}, {Boccaletti}, {Muro-Arena}, {Desidera},
  {Keppler}, {Zurlo}, {Sissa}, {Henning}, {Janson}, {Langlois}, {Bonnefoy},
  {Cantalloube}, {D'Orazi}, {Feldt}, {Hagelberg}, {Segransan}, {Lagrange},
  {Lazzoni}, {Meyer}, {Romero}, {Schmidt}, {Vigan}, {Petit}, {Roelfsema},
  {Pragt}, \& {Weber}}]{Ginski21}
{Ginski}, C., {Gratton}, R., {Bohn}, A., {et~al.} 2021, arXiv e-prints,
  arXiv:2111.11077

\bibitem[{{Ginski} {et~al.}(2016){Ginski}, {Stolker}, {Pinilla}, {Dominik},
  {Boccaletti}, {de Boer}, {Benisty}, {Biller}, {Feldt}, {Garufi}, {Keller},
  {Kenworthy}, {Maire}, {M{\'e}nard}, {Mesa}, {Milli}, {Min}, {Pinte}, {Quanz},
  {van Boekel}, {Bonnefoy}, {Chauvin}, {Desidera}, {Gratton}, {Girard},
  {Keppler}, {Kopytova}, {Lagrange}, {Langlois}, {Rouan}, \&
  {Vigan}}]{Ginski16}
{Ginski}, C., {Stolker}, T., {Pinilla}, P., {et~al.} 2016, \aap, 595, A112

\bibitem[{{Glauser} {et~al.}(2008){Glauser}, {M{\'e}nard}, {Pinte},
  {Duch{\^e}ne}, {G{\"u}del}, {Monin}, \& {Padgett}}]{Glauser08}
{Glauser}, A.~M., {M{\'e}nard}, F., {Pinte}, C., {et~al.} 2008, \aap, 485, 531

\bibitem[{{Grady} {et~al.}(2005){Grady}, {Woodgate}, {Bowers}, {Gull}, {Sitko},
  {Carpenter}, {Lynch}, {Russell}, {Perry}, {Williger}, {Roberge}, {Bouret}, \&
  {Sahu}}]{Grady05}
{Grady}, C.~A., {Woodgate}, B.~E., {Bowers}, C.~W., {et~al.} 2005, \apj, 630,
  958

\bibitem[{{Hashimoto} {et~al.}(2012){Hashimoto}, {Dong}, {Kudo}, {Honda},
  {McClure}, {Zhu}, {Muto}, {Wisniewski}, {Abe}, {Brandner}, {Brandt},
  {Carson}, {Egner}, {Feldt}, {Fukagawa}, {Goto}, {Grady}, {Guyon}, {Hayano},
  {Hayashi}, {Hayashi}, {Henning}, {Hodapp}, {Ishii}, {Iye}, {Janson},
  {Kandori}, {Knapp}, {Kusakabe}, {Kuzuhara}, {Kwon}, {Matsuo}, {Mayama},
  {McElwain}, {Miyama}, {Morino}, {Moro-Martin}, {Nishimura}, {Pyo}, {Serabyn},
  {Suenaga}, {Suto}, {Suzuki}, {Takahashi}, {Takami}, {Takato}, {Terada},
  {Thalmann}, {Tomono}, {Turner}, {Watanabe}, {Yamada}, {Takami}, {Usuda}, \&
  {Tamura}}]{Hashimoto12}
{Hashimoto}, J., {Dong}, R., {Kudo}, T., {et~al.} 2012, \apjl, 758, L19

\bibitem[{{Heikamp} \& {Keller}(2019)}]{Heikamp19}
{Heikamp}, S. \& {Keller}, C.~U. 2019, \aap, 627, A156

\bibitem[{{Henyey} \& {Greenstein}(1941)}]{Henyey41}
{Henyey}, L.~G. \& {Greenstein}, J.~L. 1941, \apj, 93, 70

\bibitem[{{Hunziker} {et~al.}(2021){Hunziker}, {Schmid}, {Ma}, {Menard},
  {Avenhaus}, {Boccaletti}, {Beuzit}, {Chauvin}, {Dohlen}, {Dominik}, {Engler},
  {Ginski}, {Gratton}, {Henning}, {Langlois}, {Milli}, {Mouillet}, {Tschudi},
  {van Holstein}, \& {Vigan}}]{Hunziker21}
{Hunziker}, S., {Schmid}, H.~M., {Ma}, J., {et~al.} 2021, \aap, 648, A110

\bibitem[{{Jang-Condell}(2017)}]{JangCondell17}
{Jang-Condell}, H. 2017, \apj, 835, 12

\bibitem[{{Jang-Condell} \& {Turner}(2013)}]{JangCondell13}
{Jang-Condell}, H. \& {Turner}, N.~J. 2013, \apj, 772, 34

\bibitem[{{Keppler} {et~al.}(2018){Keppler}, {Benisty}, {M{\"u}ller},
  {Henning}, {van Boekel}, {Cantalloube}, {Ginski}, {van Holstein}, {Maire},
  {Pohl}, {Samland }, {Avenhaus}, {Baudino}, {Boccaletti}, {de Boer},
  {Bonnefoy}, {Chauvin}, {Desidera}, {Langlois}, {Lazzoni}, {Marleau},
  {Mordasini}, {Pawellek}, {Stolker}, {Vigan}, {Zurlo}, {Birnstiel},
  {Brandner}, {Feldt}, {Flock}, {Girard}, {Gratton}, {Hagelberg}, {Isella},
  {Janson}, {Juhasz}, {Kemmer}, {Kral}, {Lagrange}, {Launhardt}, {Matter},
  {M{\'e}nard}, {Milli}, {Molli{\`e}re}, {Olofsson}, {P{\'e}rez}, {Pinilla},
  {Pinte}, {Quanz}, {Schmidt}, {Udry}, {Wahhaj}, {Williams}, {Buenzli},
  {Cudel}, {Dominik}, {Galicher}, {Kasper}, {Lannier}, {Mesa}, {Mouillet},
  {Peretti}, {Perrot}, {Salter}, {Sissa}, {Wildi}, {Abe}, {Antichi},
  {Augereau}, {Baruffolo}, {Baudoz}, {Bazzon}, {Beuzit}, {Blanchard}, {Brems},
  {Buey}, {De Caprio}, {Carbillet}, {Carle}, {Cascone}, {Cheetham}, {Claudi},
  {Costille}, {Delboulb{\'e}}, {Dohlen}, {Fantinel}, {Feautrier}, {Fusco},
  {Giro}, {Gluck}, {Gry}, {Hubin}, {Hugot}, {Jaquet}, {Le Mignant}, {Llored},
  {Madec}, {Magnard}, {Martinez}, {Maurel}, {Meyer}, {M{\"o}ller-Nilsson},
  {Moulin}, {Mugnier}, {Orign{\'e}}, {Pavlov}, {Perret}, {Petit}, {Pragt},
  {Puget}, {Rabou}, {Ramos}, {Rigal}, {Rochat}, {Roelfsema}, {Rousset}, {Roux},
  {Salasnich}, {Sauvage}, {Sevin}, {Soenke}, {Stadler}, {Suarez}, {Turatto}, \&
  {Weber}}]{Keppler18}
{Keppler}, M., {Benisty}, M., {M{\"u}ller}, A., {et~al.} 2018, \aap, 617, A44

\bibitem[{{Kolokolova} \& {Kimura}(2010)}]{Kolokolova2010}
{Kolokolova}, L. \& {Kimura}, H. 2010, \aap, 513, A40

\bibitem[{{Lasue} {et~al.}(2009){Lasue}, {Levasseur-Regourd}, {Hadamcik}, \&
  {Alcouffe}}]{Lasue09}
{Lasue}, J., {Levasseur-Regourd}, A.~C., {Hadamcik}, E., \& {Alcouffe}, G.
  2009, \icarus, 199, 129

\bibitem[{{Marino} {et~al.}(2015){Marino}, {Perez}, \& {Casassus}}]{Marino15}
{Marino}, S., {Perez}, S., \& {Casassus}, S. 2015, \apjl, 798, L44

\bibitem[{{Mayama} {et~al.}(2012){Mayama}, {Hashimoto}, {Muto}, {Tsukagoshi},
  {Kusakabe}, {Kuzuhara}, {Takahashi}, {Kudo}, {Dong}, {Fukagawa}, {Takami},
  {Momose}, {Wisniewski}, {Follette}, {Abe}, {Akiyama}, {Brandner}, {Brandt},
  {Carson}, {Egner}, {Feldt}, {Goto}, {Grady}, {Guyon}, {Hayano}, {Hayashi},
  {Hayashi}, {Henning}, {Hodapp}, {Ishii}, {Iye}, {Janson}, {Kand ori}, {Kwon},
  {Knapp}, {Matsuo}, {McElwain}, {Miyama}, {Morino}, {Moro-Martin},
  {Nishimura}, {Pyo}, {Serabyn}, {Suto}, {Suzuki}, {Takato}, {Terada},
  {Thalmann}, {Tomono}, {Turner}, {Watanabe}, {Yamada}, {Takami}, {Usuda}, \&
  {Tamura}}]{Mayama12}
{Mayama}, S., {Hashimoto}, J., {Muto}, T., {et~al.} 2012, \apjl, 760, L26

\bibitem[{{McLean} {et~al.}(2017){McLean}, {Stam}, {Bagnulo}, {Borisov},
  {Devog{\`e}le}, {Cellino}, {Rivet}, {Bendjoya}, {Vernet}, {Paolini}, \&
  {Pollacco}}]{McLean17}
{McLean}, W., {Stam}, D.~M., {Bagnulo}, S., {et~al.} 2017, \aap, 601, A142

\bibitem[{{Milli} {et~al.}(2019){Milli}, {Engler}, {Schmid}, {Olofsson},
  {M{\'e}nard}, {Kral}, {Boccaletti}, {Th{\'e}bault}, {Choquet}, {Mouillet},
  {Lagrange}, {Augereau}, {Pinte}, {Chauvin}, {Dominik}, {Perrot}, {Zurlo},
  {Henning}, {Beuzit}, {Avenhaus}, {Bazzon}, {Moulin}, {Llored},
  {Moeller-Nilsson}, {Roelfsema}, \& {Pragt}}]{Milli19}
{Milli}, J., {Engler}, N., {Schmid}, H.~M., {et~al.} 2019, \aap, 626, A54

\bibitem[{{Min} {et~al.}(2009){Min}, {Dullemond}, {Dominik}, {de Koter}, \&
  {Hovenier}}]{Min09}
{Min}, M., {Dullemond}, C.~P., {Dominik}, C., {de Koter}, A., \& {Hovenier},
  J.~W. 2009, \aap, 497, 155

\bibitem[{{Min} {et~al.}(2016){Min}, {Rab}, {Woitke}, {Dominik}, \&
  {M{\'e}nard}}]{Min2016}
{Min}, M., {Rab}, C., {Woitke}, P., {Dominik}, C., \& {M{\'e}nard}, F. 2016,
  \aap, 585, A13

\bibitem[{{Monnier} {et~al.}(2017){Monnier}, {Harries}, {Aarnio}, {Adams},
  {Andrews}, {Calvet}, {Espaillat}, {Hartmann}, {Hinkley}, {Kraus}, {McClure},
  {Oppenheimer}, {Perrin}, \& {Wilner}}]{Monnier17}
{Monnier}, J.~D., {Harries}, T.~J., {Aarnio}, A., {et~al.} 2017, \apj, 838, 20

\bibitem[{{Monnier} {et~al.}(2019){Monnier}, {Harries}, {Bae}, {Setterholm},
  {Laws}, {Aarnio}, {Adams}, {Andrews}, {Calvet}, {Espaillat}, {Hartmann},
  {Kraus}, {McClure}, {Miller}, {Oppenheimer}, {Wilner}, \& {Zhu}}]{Monnier19}
{Monnier}, J.~D., {Harries}, T.~J., {Bae}, J., {et~al.} 2019, \apj, 872, 122

\bibitem[{{Mulders} {et~al.}(2013){Mulders}, {Min}, {Dominik}, {Debes}, \&
  {Schneider}}]{Mulders13}
{Mulders}, G.~D., {Min}, M., {Dominik}, C., {Debes}, J.~H., \& {Schneider}, G.
  2013, \aap, 549, A112

\bibitem[{{Murakawa}(2010)}]{Murakawa10}
{Murakawa}, K. 2010, \aap, 518, A63

\bibitem[{{Perrin} {et~al.}(2009){Perrin}, {Schneider}, {Duchene}, {Pinte},
  {Grady}, {Wisniewski}, \& {Hines}}]{Perrin09}
{Perrin}, M.~D., {Schneider}, G., {Duchene}, G., {et~al.} 2009, \apjl, 707,
  L132

\bibitem[{{Pinilla} {et~al.}(2018){Pinilla}, {Benisty}, {de Boer}, {Manara},
  {Bouvier}, {Dominik}, {Ginski}, {Loomis}, \& {Sicilia Aguilar}}]{Pinilla18}
{Pinilla}, P., {Benisty}, M., {de Boer}, J., {et~al.} 2018, \apj, 868, 85

\bibitem[{{Pinte} {et~al.}(2006){Pinte}, {M{\'e}nard}, {Duch{\^e}ne}, \&
  {Bastien}}]{Pinte06}
{Pinte}, C., {M{\'e}nard}, F., {Duch{\^e}ne}, G., \& {Bastien}, P. 2006, \aap,
  459, 797

\bibitem[{{Pinte} {et~al.}(2008){Pinte}, {Padgett}, {M{\'e}nard},
  {Stapelfeldt}, {Schneider}, {Olofsson}, {Pani{\'c}}, {Augereau},
  {Duch{\^e}ne}, {Krist}, {Pontoppidan}, {Perrin}, {Grady}, {Kessler-Silacci},
  {van Dishoeck}, {Lommen}, {Silverstone}, {Hines}, {Wolf}, {Blake}, {Henning},
  \& {Stecklum}}]{Pinte08}
{Pinte}, C., {Padgett}, D.~L., {M{\'e}nard}, F., {et~al.} 2008, \aap, 489, 633

\bibitem[{{Pohl} {et~al.}(2015){Pohl}, {Pinilla}, {Benisty}, {Ataiee},
  {Juh{\'a}sz}, {Dullemond}, {Van Boekel}, \& {Henning}}]{Pohl15}
{Pohl}, A., {Pinilla}, P., {Benisty}, M., {et~al.} 2015, \mnras, 453, 1768

\bibitem[{{Quanz} {et~al.}(2013){Quanz}, {Avenhaus}, {Buenzli}, {Garufi},
  {Schmid}, \& {Wolf}}]{Quanz13}
{Quanz}, S.~P., {Avenhaus}, H., {Buenzli}, E., {et~al.} 2013, \apjl, 766, L2

\bibitem[{{Quanz} {et~al.}(2011){Quanz}, {Schmid}, {Geissler}, {Meyer},
  {Henning}, {Brandner}, \& {Wolf}}]{Quanz11}
{Quanz}, S.~P., {Schmid}, H.~M., {Geissler}, K., {et~al.} 2011, \apj, 738, 23

\bibitem[{{Rameau} {et~al.}(2012){Rameau}, {Chauvin}, {Lagrange},
  {Th{\'e}bault}, {Milli}, {Girard}, \& {Bonnefoy}}]{Rameau12}
{Rameau}, J., {Chauvin}, G., {Lagrange}, A.~M., {et~al.} 2012, \aap, 546, A24

\bibitem[{{Rapson} {et~al.}(2015){Rapson}, {Kastner}, {Andrews}, {Hines},
  {Macintosh}, {Millar-Blanchaer}, \& {Tamura}}]{Rapson15}
{Rapson}, V.~A., {Kastner}, J.~H., {Andrews}, S.~M., {et~al.} 2015, \apjl, 803,
  L10

\bibitem[{{Rich} {et~al.}(2021){Rich}, {Teague}, {Monnier}, {Davies}, {Bosman},
  {Harries}, {Calvet}, {Adams}, {Wilner}, \& {Zhu}}]{Rich21}
{Rich}, E.~A., {Teague}, R., {Monnier}, J.~D., {et~al.} 2021, \apj, 913, 138

\bibitem[{{Schmid}(1992)}]{Schmid92}
{Schmid}, H.~M. 1992, \aap, 254, 224

\bibitem[{{Schmid}(2021)}]{Schmid21}
{Schmid}, H.~M. 2021, \aap, 655, A83

\bibitem[{{Schmid}(2022 in press)}]{Schmid22}
{Schmid}, H.~M. 2022 in press, IAU Symp.

\bibitem[{{Schmid} {et~al.}(2011){Schmid}, {Joos}, {Buenzli}, \&
  {Gisler}}]{Schmid11}
{Schmid}, H.~M., {Joos}, F., {Buenzli}, E., \& {Gisler}, D. 2011, \icarus, 212,
  701

\bibitem[{{Schmid} {et~al.}(2006){Schmid}, {Joos}, \& {Tschan}}]{Schmid06}
{Schmid}, H.~M., {Joos}, F., \& {Tschan}, D. 2006, \aap, 452, 657

\bibitem[{{Shen} {et~al.}(2009){Shen}, {Draine}, \& {Johnson}}]{Shen09}
{Shen}, Y., {Draine}, B.~T., \& {Johnson}, E.~T. 2009, \apj, 696, 2126

\bibitem[{{Stolker} {et~al.}(2016){Stolker}, {Dominik}, {Avenhaus}, {Min}, {de
  Boer}, {Ginski}, {Schmid}, {Juhasz}, {Bazzon}, {Waters}, {Garufi},
  {Augereau}, {Benisty}, {Boccaletti}, {Henning}, {Langlois}, {Maire},
  {M{\'e}nard}, {Meyer}, {Pinte}, {Quanz}, {Thalmann}, {Beuzit}, {Carbillet},
  {Costille}, {Dohlen}, {Feldt}, {Gisler}, {Mouillet}, {Pavlov}, {Perret},
  {Petit}, {Pragt}, {Rochat}, {Roelfsema}, {Salasnich}, {Soenke}, \&
  {Wildi}}]{Stolker16}
{Stolker}, T., {Dominik}, C., {Avenhaus}, H., {et~al.} 2016, \aap, 595, A113

\bibitem[{{Takami} {et~al.}(2014){Takami}, {Hasegawa}, {Muto}, {Gu}, {Dong},
  {Karr}, {Hashimoto}, {Kusakabe}, {Chapillon}, {Tang}, {Itoh}, {Carson},
  {Follette}, {Mayama}, {Sitko}, {Janson}, {Grady}, {Kudo}, {Akiyama}, {Kwon},
  {Takahashi}, {Suenaga}, {Abe}, {Brandner}, {Brandt}, {Currie}, {Egner},
  {Feldt}, {Guyon}, {Hayano}, {Hayashi}, {Hayashi}, {Henning}, {Hodapp},
  {Honda}, {Ishii}, {Iye}, {Kandori}, {Knapp}, {Kuzuhara}, {McElwain},
  {Matsuo}, {Miyama}, {Morino}, {Moro-Martin}, {Nishimura}, {Pyo}, {Serabyn},
  {Suto}, {Suzuki}, {Takato}, {Terada}, {Thalmann}, {Tomono}, {Turner},
  {Wisniewski}, {Watanabe}, {Yamada}, {Takami}, {Usuda}, \&
  {Tamura}}]{Takami14}
{Takami}, M., {Hasegawa}, Y., {Muto}, T., {et~al.} 2014, \apj, 795, 71

\bibitem[{{Tazaki} {et~al.}(2021){Tazaki}, {Murakawa}, {Muto}, {Honda}, \&
  {Inoue}}]{Tazaki21}
{Tazaki}, R., {Murakawa}, K., {Muto}, T., {Honda}, M., \& {Inoue}, A.~K. 2021,
  \apj, 910, 26

\bibitem[{{Tazaki} \& {Tanaka}(2018)}]{Tazaki18}
{Tazaki}, R. \& {Tanaka}, H. 2018, \apj, 860, 79

\bibitem[{{Tazaki} {et~al.}(2019){Tazaki}, {Tanaka}, {Muto}, {Kataoka}, \&
  {Okuzumi}}]{Tazaki19}
{Tazaki}, R., {Tanaka}, H., {Muto}, T., {Kataoka}, A., \& {Okuzumi}, S. 2019,
  \mnras, 485, 4951

\bibitem[{{Tschudi} \& {Schmid}(2021)}]{Tschudi21}
{Tschudi}, C. \& {Schmid}, H.~M. 2021, \aap, 655, A37

\bibitem[{{van de Hulst}(1980)}]{vandeHulst80}
{van de Hulst}, H.~C. 1980, {Multiple light scattering. Vols.\_1\_and\_2.}

\bibitem[{{Villenave} {et~al.}(2020){Villenave}, {M{\'e}nard}, {Dent},
  {Duch{\^e}ne}, {Stapelfeldt}, {Benisty}, {Boehler}, {van der Plas}, {Pinte},
  {Telkamp}, {Wolff}, {Flores}, {Lesur}, {Louvet}, {Riols}, {Dougados},
  {Williams}, \& {Padgett}}]{Villenave20}
{Villenave}, M., {M{\'e}nard}, F., {Dent}, W.~R.~F., {et~al.} 2020, \aap, 642,
  A164

\bibitem[{{Weingartner} \& {Draine}(2001)}]{Weingartner01}
{Weingartner}, J.~C. \& {Draine}, B.~T. 2001, \apj, 548, 296

\bibitem[{{Whitney} \& {Hartmann}(1992)}]{Whitney92}
{Whitney}, B.~A. \& {Hartmann}, L. 1992, \apj, 395, 529

\bibitem[{{Whitney} \& {Hartmann}(1993)}]{Whitney93}
{Whitney}, B.~A. \& {Hartmann}, L. 1993, \apj, 402, 605

\bibitem[{{Whitney} {et~al.}(2013){Whitney}, {Robitaille}, {Bjorkman}, {Dong},
  {Wolff}, {Wood}, \& {Honor}}]{Whitney13}
{Whitney}, B.~A., {Robitaille}, T.~P., {Bjorkman}, J.~E., {et~al.} 2013, \apjs,
  207, 30

\bibitem[{{Witt}(1977)}]{Witt77}
{Witt}, A.~N. 1977, \apjs, 35, 1

\bibitem[{{Woitke} {et~al.}(2019){Woitke}, {Kamp}, {Antonellini}, {Anthonioz},
  {Baldovin-Saveedra}, {Carmona}, {Dionatos}, {Dominik}, {Greaves},
  {G{\"u}del}, {Ilee}, {Liebhardt}, {Menard}, {Min}, {Pinte}, {Rab}, {Rigon},
  {Thi}, {Thureau}, \& {Waters}}]{Woitke19}
{Woitke}, P., {Kamp}, I., {Antonellini}, S., {et~al.} 2019, \pasp, 131, 064301

\bibitem[{{Woitke} {et~al.}(2016){Woitke}, {Min}, {Pinte}, {Thi}, {Kamp},
  {Rab}, {Anthonioz}, {Antonellini}, {Baldovin-Saavedra}, {Carmona}, {Dominik},
  {Dionatos}, {Greaves}, {G{\"u}del}, {Ilee}, {Liebhart}, {M{\'e}nard},
  {Rigon}, {Waters}, {Aresu}, {Meijerink}, \& {Spaans}}]{Woitke16}
{Woitke}, P., {Min}, M., {Pinte}, C., {et~al.} 2016, \aap, 586, A103

\end{thebibliography}

\begin{appendix}
\section{Results for plane parallel surfaces}
\label{AppSurface}

Our models calculate the reflected photon intensity $I$, the linearly polarized intensity $P$ (or Stokes $Q$ and $U$), and the fractional polarization $p$ from a plane parallel surface as two-dimensional functions of the polar and azimuthal angles $\theta$ and $\phi$ (see Fig.~\ref{fig:coordinate}). All our models assume that the incident radiation is unpolarized. Model parameters are the polar incidence angle $\theta_0$, and the dust parameters scattering albedo $\omega$, scattering asymmetry $g$, and fractional scattering polarization $p_{\rm max}$. 
The main text of this paper describes in Sect.~\ref{section: reference model} results for the principal plane ($\phi=0^\circ$ or $180^\circ$) for the 0.5-model with parameters $\omega=0.5$, $g=0.5$, $p_{\rm max}=0.5$, and $\theta_0=57.5^\circ$. These results are plotted in this Appendix with red color as references. We present in Appendix~\ref{AppSurfacePhi} additional results for the 0.5-model for the radiation escaping in other directions than in the principal plane. Appendix~\ref{AppSurfacePara} describes how the reflected radiation in the principal plane depends on the model parameters $\omega$, $g$, $p_{\rm max}$, and $\theta_0$.

\begin{figure*}[]
    \centering
    \includegraphics[width = \hsize]{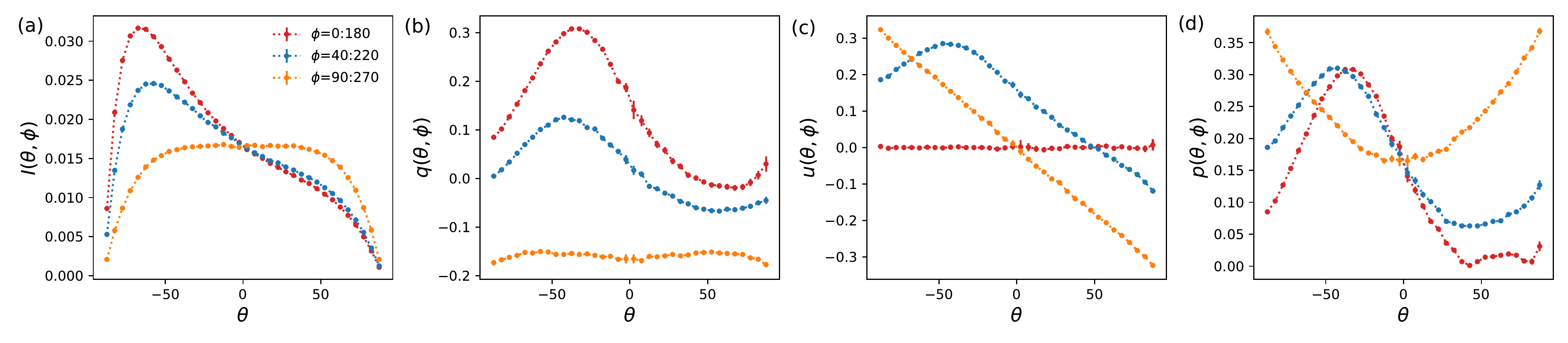}
    \vspace{-0.8cm}
    \caption{Reflected intensity $I$, Stokes parameters $q, u$ and fractional polarization $p$ for the 0.5-model for the principal plane (red) and the two vertical planes for the $\phi$-angles $40^\circ\colon220^\circ$ (blue) and $90^\circ\colon 270^\circ$ (yellow)}
    \label{fig:05model2}
\end{figure*}

\subsection{Dependencies on the azimuth angle}
\label{AppSurfacePhi}
\paragraph{Intensity.} 
The principal plane picks only a special case $\phi=0^\circ\colon180^\circ$ of the reflected intensity $I(\theta,\phi)$ of a plane parallel surface. The first panel in Fig.~\ref{fig:05model2} also plots the reflected intensities for planes perpendicular to the surface with $\phi$-angles $40^\circ\colon220^\circ$ and $90^\circ\colon270^\circ$ using the same $\pm\theta$-convention as for $\theta_{\rm pp}$ for the principal plane (Sect.~\ref{section: reference model}. 
The scattering problem is symmetric with respect to the principal plane, meaning that the reflected intensity for $I(\pm\theta,40^\circ\colon220^\circ)$ is equal to $I(\mp\theta,140^\circ\colon320^\circ)$ and $I(\theta,90^\circ)=I(-\theta,270^\circ)$. All intensity curves go through the same value $I(\theta=0^\circ)$ for the photon escape direction perpendicular to the surface. 
 
Figure~\ref{fig:05model2}(a) shows that the forward peaked scattering phase function creates the strongest $I(\pm\theta)$-asymmetry in the principal plane, with a gradual evolution for other $\phi$-angles towards the symmetric case $I(\pm\theta,90^\circ\colon270^\circ)$. Therefore the reflected intensity in the principal plane $\phi=0^\circ\colon180^\circ$ can be used for guessing the intensity at other $\phi$-angles.

\paragraph{Polarization.}
The fractional polarization of the scattered radiation given in Fig.~\ref{fig:05model2}(b) and (c) needs to be described by $q=Q/I$ and $u=U/I$ for arbitrary $\phi$ angles, because there are not only perpendicular and parallel Stokes $q$ components with respect to the vertical plane but also nonzero Stokes $u$ polarization components in the $45^\circ$ and $135^\circ$ direction. Only in the principal plane the $u$-component is zero for all angles $\theta$, and the deviations of the red curve from zero in panel (c) illustrate very well the statistical uncertainties of the Monte Carlo simulations. For the principal plane there is also $p=(q^2+u^2)^{1/2}=|q|$ as can be inferred from panel (b) and (d).

For all other azimuth angles the polarization signals are more complex. For example the yellow curves in Fig.~\ref{fig:05model2}(b) and (c) for the scattered polarization for the $\phi=90^\circ,270^\circ$ plane has for all $\theta$ a negative $q$-component and a $u$-component which changes from strongly positive to strongly negative. This corresponds to a rotation of the polarization angle from $57.5^\circ$ at $\theta=-90^\circ$ for photons escaping to the one side just along the surface to an angle $122.5^\circ$ at $\theta=+90^\circ$ for photons escaping to the opposite side along the surface, both with scattering angles of $90^\circ$. The polarization angle is $90^\circ$ or parallel to the vertical plane for photons escaping vertically at $\theta=0^\circ$ where the scattering angle is about $32.5^\circ$. All these polarization position angles are roughly perpendicular to the propagation direction of the incoming photon and such a perpendicular polarization is also expected for single scattering. Multiple scattering events add small contributions which lead to small second-order deviations from the perpendicular polarization angles.

The perpendicular scattering polarization would be more obvious, if we would show the polarization relative to the scattering plane defined by the propagation directions of the incoming $\theta_0,0^\circ$ and escaping photon $\theta,\phi$, instead of the polarization relative to the perpendicular planes as used in Fig.~\ref{fig:05model2}. However, the geometry of the scattering plane depends on the incidence angle $\theta_0$ and this complicates comparisons with the red curve for scattering in the principal plane.

The yellow curve for the fractional polarization $p$ in Fig.~\ref{fig:05model2}(d) for the $\phi=90^\circ,270^\circ$ plane has maxima for the right angle scattering at $\theta=\pm 90^\circ$ and a minimum for $\theta=0^\circ$. For intermediate $\phi$-angles the fractional polarization $p$ is (very) roughly between the red and yellow curve. 

\subsection{Dependencies on model parameters}
\label{AppSurfacePara}

\paragraph{Scattering albedo.}
\begin{figure*}[]
    \centering
    \captionsetup{margin=0cm}
    \includegraphics[width=\hsize]{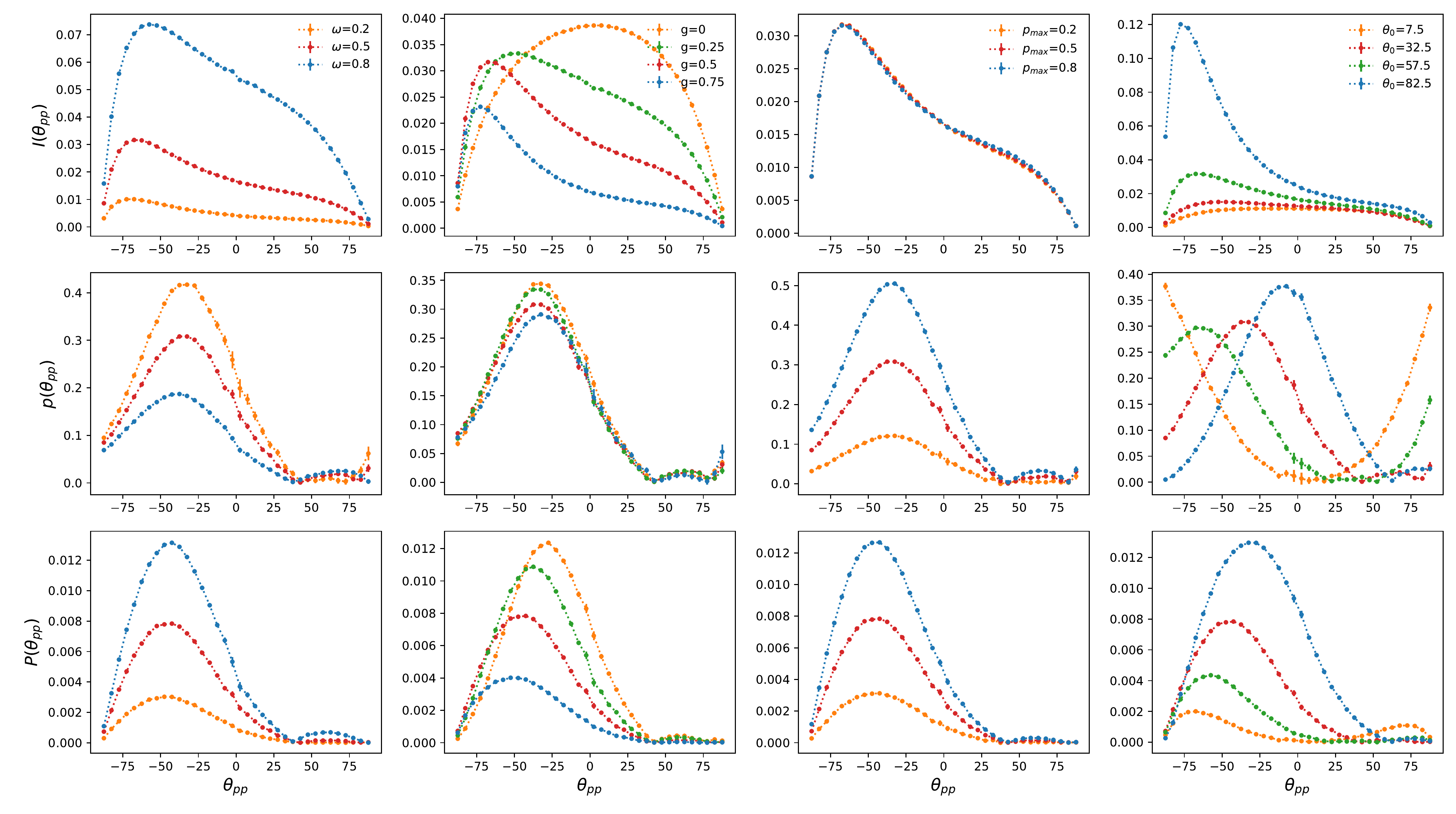}
    \vspace{-0.7cm}
    \caption{Reflected intensity $I(\theta_{\rm pp})$, fractional polarization $p(\theta_{\rm pp})$, and polarized intensity $P(\theta_{\rm pp})$ in the principal plane for a scattering surface with a HG$_{\rm pol}$ phase function. The red curves is the standard model with $\theta_0=57.5^\circ$, $\omega=0.5$, $g=0.5$, and $p_{\rm max} = 0.5$ and in each column one of these parameters is varied as indicated in the top panel.} 
    \label{fig:hg_dep}
\end{figure*}

The single scattering albedo $\omega$ has a very important impact on the amount of reflected intensity as can be seen in Fig.~\ref{fig:hg_dep} for the principal plane $I(\theta_{\rm pp})$. For our standard model parameters $\theta_0=57.5^{\circ}$, $g=0.5$ and $p_{\rm max}=0.5$, the peak reflectivity max$(I(\theta_{\rm pp}))$ decreases for standard model parameters from 0.074, to 0.031 and 0.010, if the single scattering albedo is reduced from 0.8 to 0.5 and 0.2. These characteristic values are also given in Table~\ref{tab:param} together with the angle $\theta_{\rm pp}^{\rm maxI}$ where this maximum occurs. 

The reflected polarized flux $P(\theta_{\rm pp})$ shows also a correlation with $\omega$, but with a less steep gradient when compared to the intensity. A low single scattering albedo  reduces the multiple-scattered photons---which contribute more "randomized" or unpolarized reflected
light--- more significantly,  while the strongly polarizing single-scattered photons are less suppressed. Therefore, the maximum fractional polarization of the reflected light max($p(\theta_{\rm pp})$) is high for low albedo and increases from about 19~\% to 31~\% and 43~\% for $\omega$ decreasing from 0.8, to 0.5 and 0.2, respectively (see Table~\ref{tab:param}). 

\paragraph{Scattering asymmetry parameter.}
The scattering asymmetry parameter $g$ can significantly change the strength and the angular distribution of the reflected intensity $I(\theta_{\rm pp})$ as shown in Fig.~\ref{fig:hg_dep} for $\theta_0=57.5^\circ$, $\omega=0.5,$ and $p_{\rm max}=0.5$. 
For isotropic scattering $g\approx 0,$ about half of the incident photons propagate after the first scattering upwards and have a high probability of escaping, while for strongly forward scattering particles $g\approx 0.5$ or 0.75, the photons propagate after the scattering predominantly further into the dusty disk and they have a high probability of absorption and the reflectivity decreases strongly with the $g$-parameter. 
Also, the peak of the $I(\theta_{\rm pp})$-curves shifts from perpendicular escape directions $\theta_{\rm pp}^{\rm maxI}\approx 0^\circ$ for isotropic scattering to large scattering angles $\theta_{\rm pp}^{\rm maxI}$ for larger $g$ according to Table.~\ref{tab:param}. Interestingly, the fractional polarization of the scattered radiation $p(\theta_{\rm pp})$ in Fig.~\ref{fig:hg_dep} depends only slightly on the scattering asymmetry parameter. The polarization $p$ follows from the ratio between escaping single scattered and multiple scattered photons, and this ratio clearly does not significantly change for different $g$-values. 
Independently of $g$, the maximum $p$-values occurs near $\theta_{\rm pp}^{\rm maxp}\approx -35^\circ$ or approximately near right-angle scattering $\theta_{\rm pp}^{\rm maxp}-\theta_0\approx 90^\circ$.

\paragraph{Scattering polarization.}
The maximum fractional polarization $p_{\rm max}$ of the polarization phase function has a strong effect on $p(\theta,\phi)$ of the reflected radiation. For the models with $\theta_0=57.5^\circ$, $\omega=0.5,$ and $g=0.5$ shown in Fig.~\ref{fig:hg_dep}, there exists a well-defined relation for the maximum fractional polarization of the reflected radiation max$(p(\theta_{\rm pp}))\approx 0.62\cdot p_{\rm max}$ (see Table.~\ref{tab:param}). This tight relation is helpful for deriving the dust parameter $p_{\rm max}$ from observations. However, $p(\theta,\phi)$ depends also on $\omega$ and this must be considered for the derivation of $p_{\rm max}$ (see Sect.~\ref{section: fractional polarization}).

The impact of different $p_{\rm max}$-parameter on the reflected intensity $I(\theta_{\rm pp})$ is hardly visible in Fig.~\ref{fig:hg_dep}. There are small differences because the higher order scatterings depend on the polarization of the scattered photons as described in \citet{Chandrasekhar60} for Rayleigh scattering, but this effect is not important when compared to the strong dependencies caused by the parameters $\omega$ and $g$.

\paragraph{Incidence angle.}
The incidence angle $\theta_0$ defines for a given viewing direction the scattering angle. Forward scattering dust $g>0$ and an incidence close to perpendicular $\theta_0\approx 0^\circ$ will scatter the photons predominantly into deeper layers and the surface reflectivity is low. For grazing incidence $\theta_0\rightarrow 90^\circ$ the forward scattering dust will produce a high reflectivity in the forward direction $\theta_{\rm pp}\approx -70^\circ$. This produces the frequently observed intensity maximum for the front side surface of inclined disks. This effect is further amplified by the fact that the first interaction for photons with grazing incidence takes place often close to the surface, and if scattering occurs, then the probability for photon escape is relatively high. 

The fractional polarization $p(\theta_{\rm pp})$ has a strong maximum near angles $\theta_{\rm pp}^{\rm maxp}\approx \theta_0 -90^\circ$ corresponding to right-angle scattering. For the polarized intensity $P$, the angle for the maximum reflectivity is intermediate between the peaks for the fractional polarization $p$ and the intensity $I$. 

\begin{table*}[]
    \caption{Key values for the reflected radiation in the plane parallel surface models}
    \label{tab:param}
    \centering
    \begin{tabular}{c c c c c c c c c c c c}
    \hline\hline
    \noalign{\smallskip}
    Parm.  & Ref. & \multispan{2}{\hfil $\omega$-dep.\hfil} 
                         & \multispan{3}{\hfil$g$-dep.\hfil}
                                 &\multispan{2}{\hfil$p_{\rm max}$-dep.\hfil}
                                        & \multispan{3}{\hfil$\theta_0$-dep.\hfil} \\
                                        
 $\omega$ & $\omega_r$=$0.5$ & 0.20 & 0.80 & 
 $\omega_r$ & $\omega_r$ & $\omega_r$ & $\omega_r$ & $\omega_r$ & $\omega_r$ & $\omega_r$ & $\omega_r$  \\ 
 $g$      & $g_r=0.5$ & $g_r$ & $g_r$ & 0.00 & 0.25 & 0.75 & $g_r$ & $g_r$ 
            & $g_r$ & $g_r$ & $g_r$ \\
 $p_{\rm max}$      
            & $p_r=0.5$ & $p_r$ & $p_r$ & $p_r$ & $p_r$ & $p_r$ & 0.20 & 0.80 & 
               $p_r$ & $p_r$ & $p_r$  \\ 
 $\theta_0$  & $\theta_{0,r}$=57.5$^\circ$ & $\theta_{0,r}$ & $\theta_{0,r}$ & $\theta_{0,r}$ & $\theta_{0,r}$ & $\theta_{0,r}$ &
 $\theta_{0,r}$ & $\theta_{0,r}$ & $7.5^\circ$ & $32.5^\circ$ & $82.5^\circ$\\
 \hline
 \noalign{\smallskip}
 ${\cal{R}}$                          & 0.091 & 0.025 & 0.253 & 0.155 & 0.126 & 0.047 & 0.091 & 0.091 & 0.048 & 0.058 & 0.194 \\
 \noalign{\smallskip}
 ${\rm max}(I(\theta_{\rm pp}))$      & 0.031 & 0.010 & 0.074 & 0.039 & 0.033 & 0.023 & 0.032 & 0.032 & 0.011 & 0.015 & 0.120 \\
 $\theta_{\rm pp}^{\rm maxI}[^\circ]$ & -62.5 & -67.5 & -57.5 & 7.5.  & -52.5 & -72.5 & -67.5 & -67.5 & -2.5  & -47.5 & -77.5 \\
 \noalign{\smallskip}
 ${\rm max}(p(\theta_{\rm pp}))$      & 0.31  & 0.43  & 0.19  & 0.35  & 0.33  & 0.30  & 0.12  & 0.50  & 0.37  & 0.30  & 0.39  \\
 $\theta_{\rm pp}^{\rm maxp}[^\circ]$ & -32.5 & -32.5 & -37.5 & -37.5 & -37.5 & -27.5 & -37.5 & -37.5 & -87.5 & -67.5 & -12.5 \\

    \hline
    \end{tabular}
\addtolength{\tabcolsep}{1.2pt}
\tablefoot{Given values include integrated reflectivity ${\cal{R}}$, peak reflectivity ${\rm max}(I(\theta_{\rm pp}))$ and peak fractional polarization ${\rm max}(p(\theta_{\rm pp}))$ for the principal plane and the corresponding angle bin $\theta_{\rm pp}^{\rm maxI}$ and $\theta_{\rm pp}^{\rm maxp}$ respectively, as discussed in Fig.~\ref{fig:hg_dep}. Numerical uncertainties for the calculated model results are at a level of about one or two units for the last digit of the given values.}
\end{table*}

\section{Additional results for the disk models}
The main text of this paper provides quantitative results for the radiation parameters for the reference disk model in Fig.~\ref{fig:fluxvsiref} and lists in Table~\ref{tab:diagnostic} corresponding values for four inclinations and for many different model parameters.  
This appendix provides for most disk models in Table~\ref{tab:diagnostic} the plots of radiation parameters as functions of the disk inclination which should be helpful for a detailed assessment of the model results.

\label{section: inclination-dep}
\subsection{Disk integrated and disk averaged radiation parameters}
\begin{figure*}
    \centering
    \begin{subfigure}[b]{\hsize}
        \centering
        \includegraphics[width=17cm]{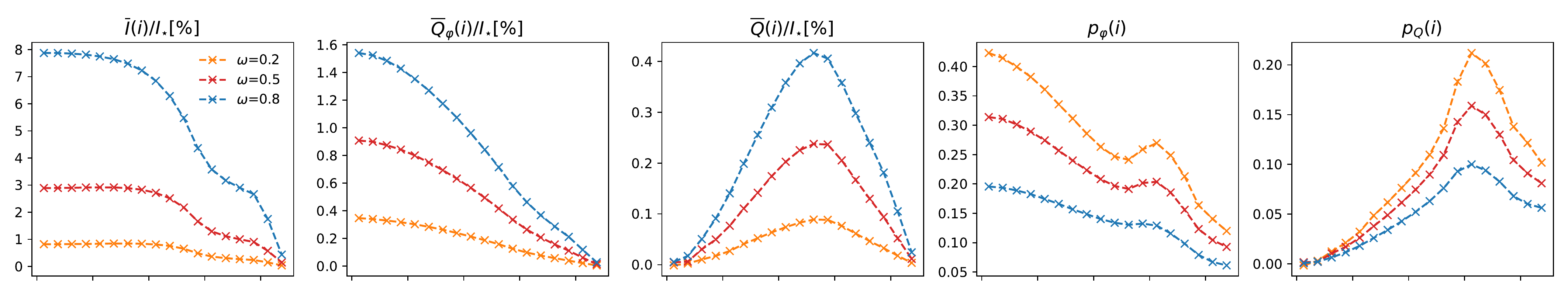}
    \end{subfigure}
    \begin{subfigure}[b]{\hsize}
        \centering
        \includegraphics[width=17cm]{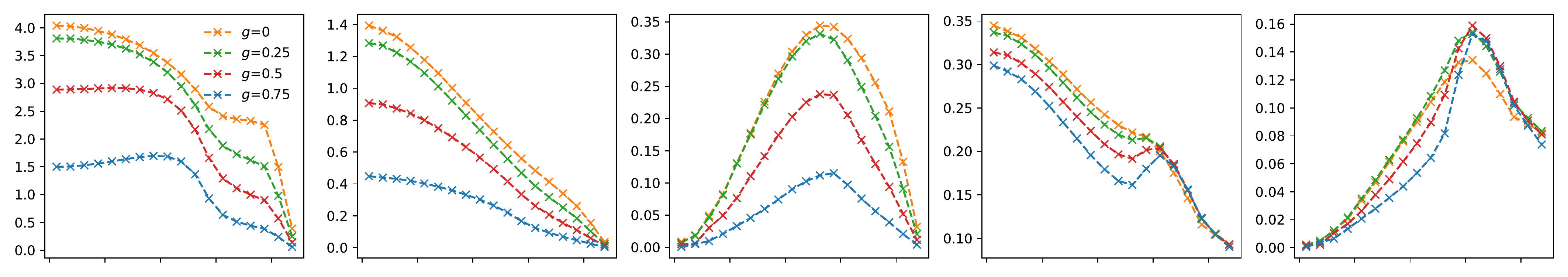}
    \end{subfigure}
    \begin{subfigure}[b]{\hsize}
        \centering
        \includegraphics[width=17cm]{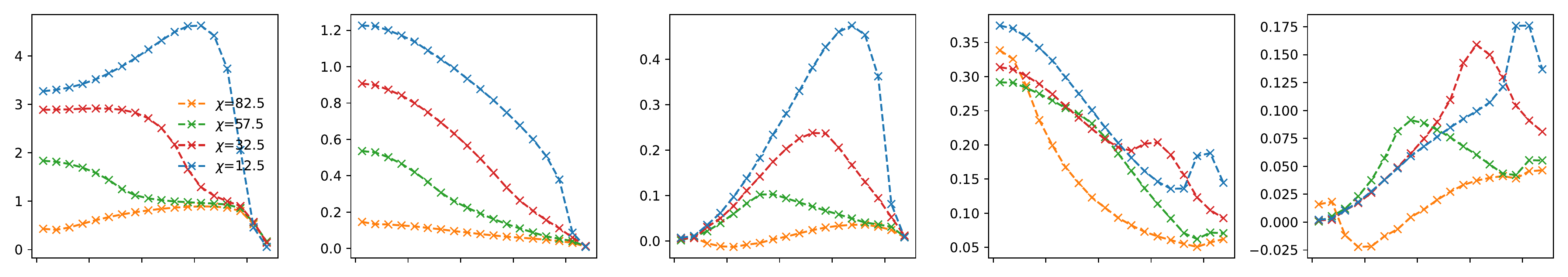}
    \end{subfigure}
    \begin{subfigure}[b]{\hsize}
        \centering
        \includegraphics[width=17cm]{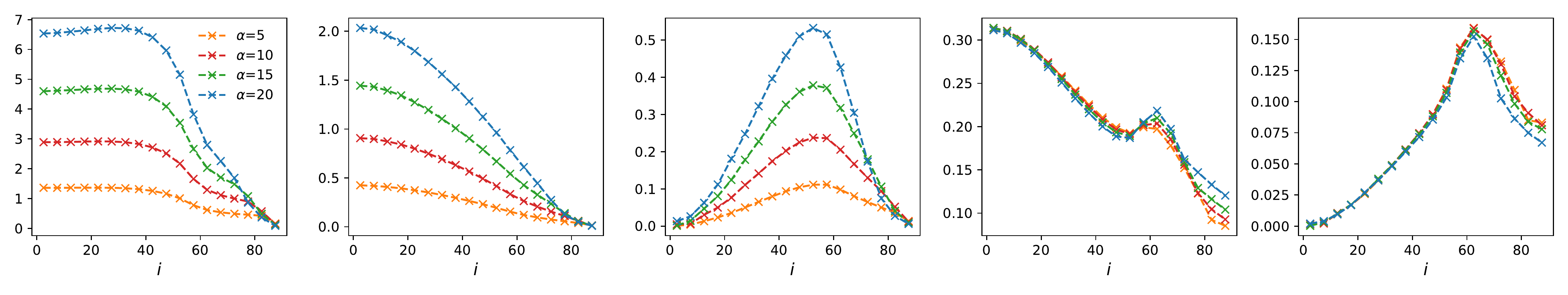}
    \end{subfigure}
    
    \caption{Integrated disk intensity $\overline{I}(i)/I_\star$, azimuthal polarization $\overline{Q}_\varphi(i)/I_\star$, Stokes polarization $\overline{Q}(i)/I_\star$ and the disk averaged fractional polarizations for the azimuthal polarization $\langle p_\varphi\rangle=\overline{Q_\varphi}/\overline{I}$ and the Stokes $Q$ polarization $\langle p_Q\rangle=\overline{Q}/\overline{I}$ as functions of disk inclination $i$. Model parameters are $\omega=0.5$, $g=0.5$ and $\chi=32.5^\circ$ unless other values are given in the leftmost panel. Fixed model parameters are $\alpha=10^\circ$, and $p_{\rm max}=0.5$. The red curve in each panel is the reference model.}
    \label{fig:integrate-param}
\end{figure*}

Figure~\ref{fig:integrate-param} gives an overview of the model parameter dependence for the integrated reflected intensity $\overline{I}(i)/I_\star$, the azimuthal polarization $\overline{Q}_\varphi(i)/I_\star$, the Stokes parameter $\overline{Q}(i)/I_\star$, and the disk averaged fractional polarizations $\langle p_\varphi\rangle=\overline{Q}_\varphi/\overline{I}$ and $\langle p_Q\rangle=\overline{Q}/\overline{I}$. Each panel includes the result for the reference model plotted in red. 

A general feature of the transition disk model is that they are bright in intensity $\overline{I}(i)$ and polarized intensity $\overline{Q}_\phi(i)$ for pole-on viewing angles ($i=0^\circ$), when the illuminated inner wall is well visible. The signal decrease for larger inclination and become zero for edge-on disks ($i=90^\circ$) when the illuminated regions are hidden. 

The four model parameters $\omega$, $g$, $\chi$, and $\alpha$ investigated in Fig.~\ref{fig:integrate-param} have a strong influence on the overall level of the $\overline{I}(i)$, $\overline{Q}_{\varphi}(i),$ and $\overline{Q}(i)$ curves. The parameter $p_{\rm max}$ is not included because it only changes ---to first order---  the polarization level, as discussed for the plane-parallel case and as illustrated below.

The relative inclination dependence for $\overline{I}(i)$, $\overline{Q}_\varphi(i)$, and $\overline{Q}(i)$ or the shape of the curves are not strongly changed by $\omega$, $g,$ or $\alpha$. According to the numerical values in Table~\ref{tab:diskresults} for $i=7.5^\circ,\,32.5^\circ,\,57.5^\circ,$ and $77.5^\circ$, the reflected intensities $\overline{I}(i)/I_\star$ for $\omega=0.2$ are simply scaled by a factor of $0.28\pm 0.01$, and for $\omega=0.8$ by a factor of $2.7\pm 0.2$ relative to the reference case $\omega=0.5$. As described for the plane parallel case, a low $\omega$ produces scattered light with less multiple scattering and therefore a higher fractional polarization. For the asymmetry parameter $g$, both the intensity and polarized intensity are enhanced for low $g$, while the fractional polarization signal stays roughly the same. The angular wall height $\alpha$ scales the integrated radiation parameters approximately by a factor $\alpha/\alpha_r$ except for large $i$ when the front wall can hide illuminated disk regions on the back side and therefore introduce more complex $\alpha$-dependencies.
The wall slope $\chi$ has important effects on both the level and the shape of the $\overline{I}(i)$ and $\overline{Q}_{\varphi}(i)$ curves. Disks with flat walls, equivalent to small $\chi$, are significantly brighter when compared to disks with steep walls. For flat walls the incidence angle $\theta_0$ of the incoming light is large and close to grazing incidence and therefore the photons have a high probability to escape from the disk surface. Contrary to this, photons penetrate deeply into the disk for steep incidence angles (small $\theta_0$) and for strongly forward scattering dust, the probability for photon absorption is high. The shape of the curves for the fractional polarizations $\langle p_\varphi(i) \rangle$ and $\langle p_Q(i) \rangle$ show the same behaviour as the reference case  for many disk parameters.

Most model parameters hardly change the "normal" $i$-dependence for $p_{\varphi}(i)$ and $p_Q(i)$ except for the wall slope $\chi$, which has a strong impact. The curves show for small $i\ll 90^\circ -\chi$ the expected decrease with $i$ for $p_{\varphi}(i)$ or increase with $i$ for $p_{Q}(i)$ as long as the illuminated inner wall on the front side is well visible. However, when $i$ approaches $\chi$ the typically lower $p$, negative $p_Q$ front side component disappears and the overall $p_Q$ shows a bump around $i\approx 90^\circ -\chi$. This bump is strong and narrow for small $\chi$ and weak and broad for large $\chi$ and for $\chi=82.5^\circ$ it is around $i\approx 5^\circ$ and therefore hardly visible. For $i>90^\circ -\chi$ various effects play a role which depend on details of the scattering geometry, like hiding of parts of the back wall, grazing photon emergence from the lower or upper back side, negative polarization because of multiple scattering under $\theta_s \gtrsim  150^\circ$ back-scattering conditions. 
These special effects, which become important for cases $i \gtrsim 90^\circ - \chi$, are beyond the scope of our general description of model parameter dependencies for transition disks. 
\subsection{Quadrant polarization parameters}
\begin{figure}
    \centering
    \includegraphics[width =0.49\textwidth]{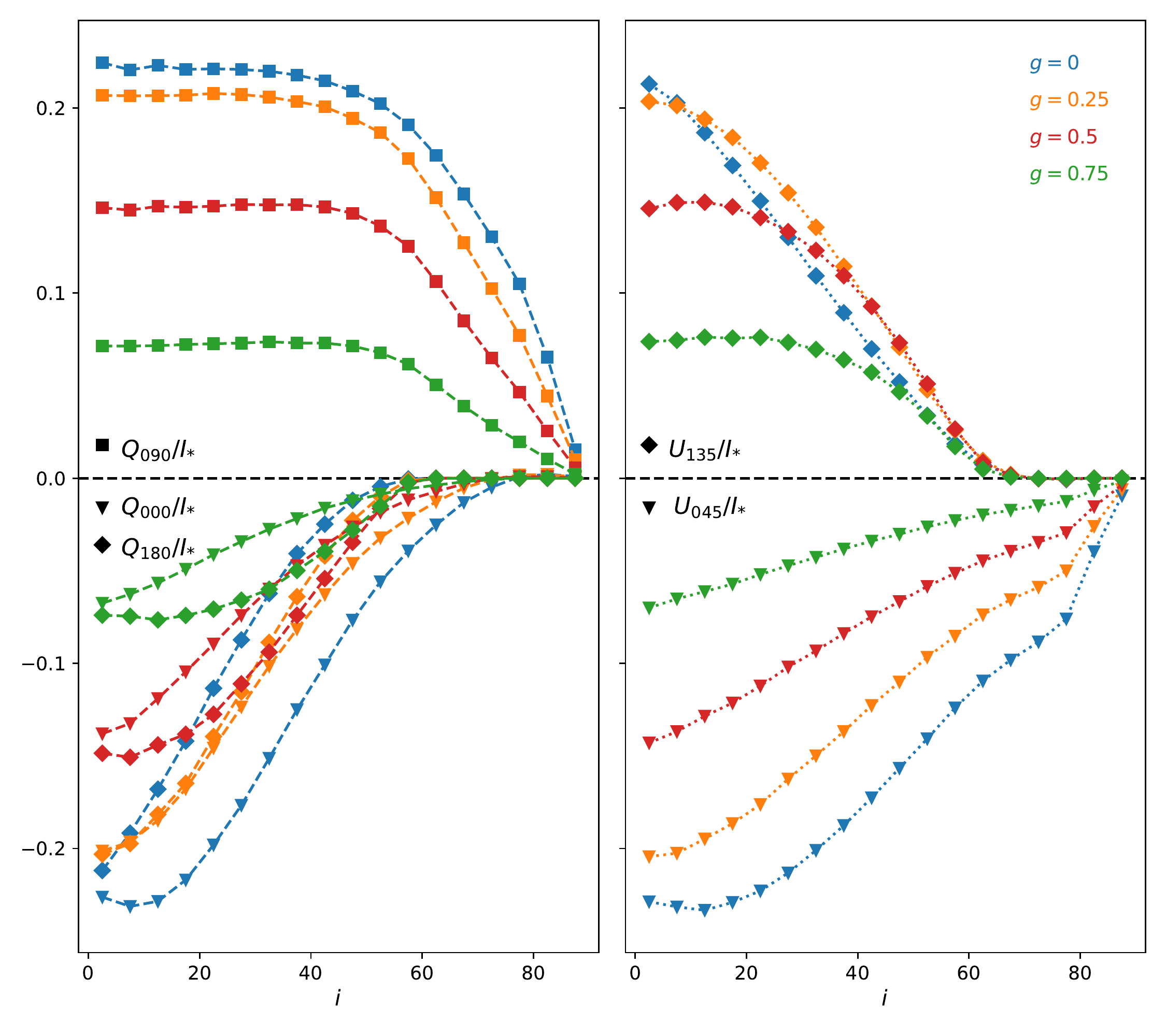}
    \caption{Quadrant polarization values $Q_{xxx}(i)/I_\star$ (left) and $U_{xxx}(i)/I_\star$ (right) for the scattering asymmetry parameters $g=0.0,\, 0.25,\,0.5$ and $0.75$, and fixed parameter $\omega=0.5$, $p_{\rm max}=0.5$, $\chi=32.5^\circ$ and $\alpha=10^\circ$. The red curves is the reference model. 
    }
    \label{fig:disk-g-dep}
\end{figure}

\begin{figure}
    \centering
    \includegraphics[width =0.49\textwidth]{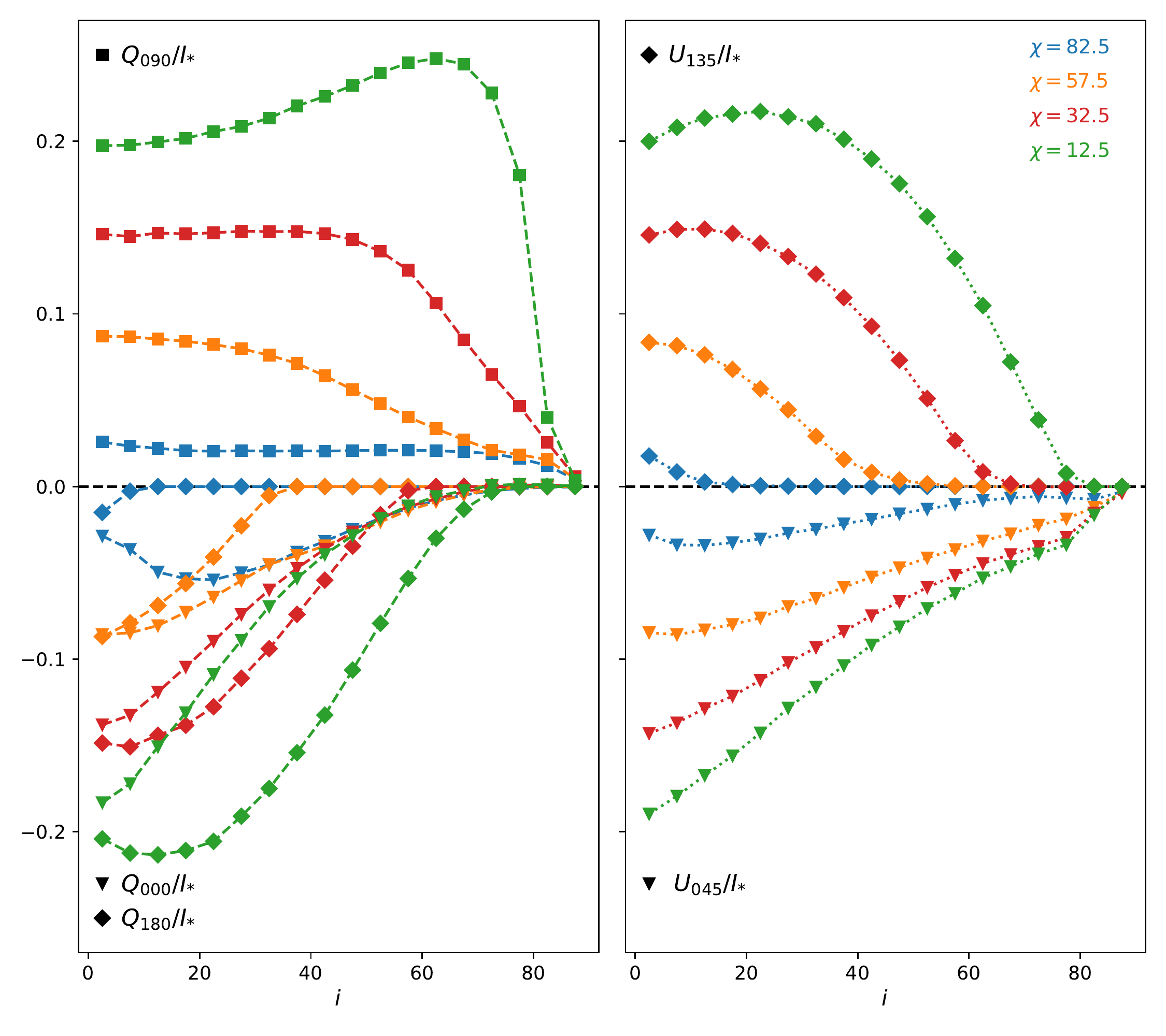}
    \caption{$Q_{xxx}(i)/I_\star$ (left) and $U_{xxx}(i)/I_\star$
    (right) for the wall slopes $\chi=7.5^\circ,\, 32.5^\circ,\, 57.5^\circ$ and $82.5^\circ$ and fixed parameter $\omega=0.5$, $g=0.5$, $p_{\rm max}=0.5$ and $\alpha=10^\circ$.  The red curves is the reference model.}
    \label{fig:disk-chi-dep}
\end{figure}

The azimuthal polarization depends significantly on the scattering asymmetry parameter $g$ and the wall slope $\chi$ as illustrated in Figs.~\ref{fig:disk-morphology}(a) and \ref{fig:disk-morphology}(b). These dependencies can be quantified by the polarization quadrant parameters as illustrated in Figs.~\ref{fig:disk-g-dep} and \ref{fig:disk-chi-dep}.
In general, high reflectivity and therefore also strong signals for the quadrant parameters $Q_{xxx}/I_\star$ and $U_{xxx}/I_\star$ are obtained for low $g=0$ and flat wall slopes. 

The quadrant polarization parameters $Q_{xxx}/I_\star$ and $U_{xxx}/I_\star$ also depend strongly on the model parameters $\omega$, $p_{\rm max}$, and $\alpha$ which enhance or lower the overall level of the scattering polarization for the entire disk, almost independently of the azimuthal angle.
Therefore, the values $Q_{xxx}/I_\star$ and $U_{xxx}/I_\star$ have a limited diagnostic value. More useful for the investigation of the scattering asymmetry $g$ or the wall slope $\chi$ are relative quadrant values $Q_{xxx}/\overline{Q}_\varphi$ and $U_{xxx}/\overline{Q}_\varphi$ as given in Table~\ref{tab:diskresults} or ratios between quadrant polarization intensities like $Q_{180}/Q_{000}$ or $U_{135}/U_{045}$, because these ratios depend mainly on $g$ and $\chi$ but are almost independent of $\omega$, $p_{\rm max}$ and $\alpha$. 

The interpretation of the front-to-back disk brightness ratio $Q_{180}/Q_{000}$ should also consider the corresponding intensity effect. Therefore, we also list in Table~\ref{tab:diskresults} the corresponding front-to-back intensity ratio $I_{180}/I_{000}$ derived in exactly the same way as $Q_{180}/Q_{000}$, but from the $I(x,y)$-image instead of the $Q(x,y)$-image. The diagnostic potential of these ratios for the interpretation of data is briefly discussed in Sect.~\ref{section: azimuthal distribution}.

\end{appendix}
\end{document}